\documentclass[11pt]{article}
\usepackage[a4paper]{geometry}
\usepackage{amsmath,amsfonts,amssymb}
\usepackage{authblk}
\usepackage{alltt}
\usepackage{bm,bbm,bbold}
\usepackage{booktabs}
\usepackage{comment}
\usepackage{dcolumn}
\usepackage{empheq} 
\usepackage{enumitem}
\usepackage{float}  
\usepackage{graphicx}
\usepackage{ltablex}  
\usepackage[scr=boondox]{mathalpha}
\usepackage{mathtools}
\usepackage{physics}
\usepackage{relsize}
\usepackage{setspace}
\usepackage{txfonts}
\usepackage{xcolor}
\usepackage[most]{tcolorbox} 
\usepackage{upgreek}
\usepackage{slashed}
\numberwithin{equation}{section}
\usepackage[sort&compress,numbers,merge]{natbib}
\usepackage[colorlinks=true,breaklinks=true]{hyperref}
\hypersetup{
allcolors=[rgb]{0.0 0.0 0.6},
linkcolor=[rgb]{0.0 0.0 1}
}
\usepackage[capitalise]{cleveref}
\crefformat{pluralequation}{#2{\color{black} Eqs.~(}#1{\color{black} )}#3}
\Crefformat{pluralequation}
{#2{\color{black} Equations~(}#1{\color{black} )}#3}
\newcommand{\aB}{\alpha_{\mathsmaller{\cal B}}}

\def\arad{\alpha_{\rm rad}}

\def\aS{\alpha_{\mathsmaller{\cal S}}}
\def\BSF{\mathsmaller{\rm BSF}}
\def\cc{\rm c.c.}

\def\elas{{\rm elas}}
\def\ellB{\ell_{\mathsmaller{\cal B}}}
\def\ellS{\ell_{\mathsmaller{\cal S}}}
\def\ES{E_{\mathsmaller{\cal S}}}

\def\hc{\rm h.c.}
\def\ID{\mathbbm{1}}
\def\ii{\im}
\def\im{\mathbbm{i}}
\def\inel{{\rm inel}}
\def\kappaB{\kappa_{\mathsmaller{\cal B}}}
\def\kappan{\kappa_{n}}

\def\kappaS{\kappa_{\mathsmaller{\cal S}}}

\def\mB{m_{\mathsmaller{\cal B}}}
\def\mmedB{m_{{\rm med},\mathsmaller{\cal B}}}
\def\mmedS{m_{{\rm med},\mathsmaller{\cal S}}}
\def\mrad{m_{\rm rad}}
\def\mS{m_{\mathsmaller{\cal S}}}
\def\mT{m_{\mathsmaller{T}}}
\def\mX{m_{\mathsmaller{X}}}

\def\nS{n_{\mathsmaller{\cal S}}}

\def\NN{\mathbbm{N}}
\def\nn{\nonumber}

\def\prad{p^{\rm rad}}
\def\reg{{\rm reg}}

\def\rM{r^{\mathsmaller{\cal M}}}
\def\rMstar{r^{\mathsmaller{\cal M}\,\star}}

\def\RR{\mathbbm{R}}
\def\sym{\mathscr{c}}
\def\subsqrts{{\scalebox{0.8}{$\scriptscriptstyle\sqrt{s}$}}}

\def\twoPI{\mathsmaller{\rm 2PI}}
\def\unreg{{\rm unreg}}
\def\VB{V_{\mathsmaller{\cal B}}}
\def\vrel{v_{\rm rel}}
\def\VS{V_{\mathsmaller{\cal S}}}
\def\WW{\mathbbm{W}}
\def\xB{x_{\mathsmaller{\cal B}}}
\def\xS{x_{\mathsmaller{\cal S}}}
\def\yrad{y_{\rm rad}}

\def\zetaS{\zeta_{\mathsmaller{\cal S}}}
\def\zetaB{\zeta_{\mathsmaller{\cal B}}}
\def\zetan{\zeta_n}
\def\zetanp{\zeta_{n'}}
\def\zS{z_{\mathsmaller{\cal S}}}
\def\zB{z_{\mathsmaller{\cal B}}}
\def\zn{z_{n}}
\def\znp{z_{n'}}
\def\ZZ{\mathbbm{Z}}
\DeclareMathOperator{\atantwo}{atan2}

\DeclareMathOperator{\arcsinh}{arcsinh}
\DeclareMathOperator{\arccosh}{arccosh}

\DeclareMathOperator{\Arg}{Arg}

\DeclareMathOperator{\Log}{Log}
\DeclareMathOperator{\sgn}{sgn}
\newcommand{\myshadebox}[1]{%
\tcboxmath[
colback=gray!20, 
colframe=gray!20, 
boxrule=0pt,      
arc=0pt,          
left=0.5pt, right=0.5pt, top=0.5pt, bottom=0.5pt
]{#1}%
}
\colorlet{lightblue}{blue!30}
\colorlet{lightred}{red!30}

\usepackage{tikz}
\usetikzlibrary{arrows,shapes}
\usetikzlibrary{trees,patterns}
\usetikzlibrary{matrix,arrows} 				
\usetikzlibrary{positioning}				  
\usetikzlibrary{calc,through}				  
\usetikzlibrary{decorations.pathreplacing}  
\usepackage{pgffor}							

\usetikzlibrary{decorations.pathmorphing}	
\usetikzlibrary{decorations.markings}
\tikzset{
>=stealth', 
vector/.style={decorate, decoration={snake}, draw},
provector/.style={decorate, decoration={snake,amplitude=2.5pt}, draw},
antivector/.style={decorate, decoration={snake,amplitude=-2.5pt}, draw},
bigvector/.style={decorate, decoration={snake,amplitude=4pt}, draw},
fermion/.style={draw=black, postaction={decorate}, 
	decoration={markings,mark=at position .55 with {\arrow[draw=black]{>}}}},
fermionbar/.style={draw=black, postaction={decorate},
    decoration={markings,mark=at position .55 with {\arrow[draw=black]{<}}}},
fermionnoarrow/.style={draw=black},
doublefermion/.style={draw=black,double, postaction={decorate},
	decoration={markings,mark=at position .57 with {\arrow[draw=black]{>}}}},
doublefermionbar/.style={draw=black,double, postaction={decorate},
	decoration={markings,mark=at position .57 with {\arrow[draw=black]{<}}}},
doublefermionnoarrow/.style={draw=black,double},
gluon/.style={decorate, draw=black,
    decoration={coil,amplitude=4pt, segment length=5pt}},
scalar/.style={dashed,draw=black, postaction={decorate},
	decoration={markings,mark=at position .55 with {\arrow[draw=black]{>}}}},
scalarbar/.style={dashed,draw=black, postaction={decorate},
    decoration={markings,mark=at position .55 with {\arrow[draw=black]{<}}}},
scalarnoarrow/.style={dashed,draw=black},
momentum/.style={draw=black, postaction={decorate},
    decoration={markings,mark=at position 1 with {\arrow[draw=black]{>}}}},
antimomentum/.style={draw=black, postaction={decorate},
    decoration={markings,mark=at position 0.1 with {\arrow[draw=black]{<}}}}
}

\tikzstyle{block} = [draw, rectangle, minimum height=3em, minimum width=6em]
    
\title{\bf Unitarity violation and restoration \\ 
in radiative bound-state formation}
\author[1,2]{\Large
Marcos M. Flores\footnote{marcos.flores@fys.uio.no}} 
\author[1]{Kalliopi Petraki\footnote{kalliopi.petraki@phys.ens.fr}
}

\affil[1]{\large\it
Laboratoire de Physique de l'École Normale Supérieure, 
ENS, 
Université PSL, 
CNRS, 
Sorbonne Université, 
Université Paris Cité,
Paris, F-75005, France}

\affil[2]{\large\it 
Department of Physics, 
University of Oslo, 
Box 1048, N-0316 Oslo, Norway}

\begin{document}

\maketitle
\begin{abstract}
\noindent
State-of-the-art calculations motivated by dark matter exhibit severe violation of partial-wave unitarity in the non-relativistic regime in radiative bound-state-formation processes. It has been recently shown, in a model-independent fashion, that unitarity is restored by the proper resummation of the inelastic contributions to the self-energy of the incoming state.  In this work, we first derive Kramers-like formulae for individual partial waves, demonstrating that existing calculations of bound-state formation severely violate unitarity. We then discuss how unitarity is restored through the resummation of the absorptive contributions to the incoming-state self-energy,  generated by bound-state formation processes, taking into account their analytic structure in the complex momentum plane. Our results can be generalized to a variety of theories and employed in phenomenological studies, such as dark-matter freeze-out, indirect detection and self-interactions. 
\end{abstract}

\clearpage
\noindent\rule{\textwidth}{0.4pt}
\setcounter{tocdepth}{2}
\tableofcontents
\noindent\rule{\textwidth}{0.4pt}

\clearpage
\section{Introduction \label{sec:Intro}}

Understanding inelasticity in the non-relativistic regime is important for a wide range of new physics scenarios, such as those pertaining to the cosmology and galactic dynamics of dark matter (DM). In this context, work in recent years on \emph{long-range interactions} has brought into focus the importance of \emph{bound states} in weakly-coupled theories. Bound-state formation (BSF) has emerged as a dominant type of inelasticity at low velocities when the interacting particles are much heavier than the force mediators to which they couple~\cite{vonHarling:2014kha}. This includes models with multi-TeV particles coupled to the Weak interactions of the Standard Model (SM), that have been the major focus of new-physics searches over recent decades.  However, the exploration of long-range dynamics has extended well beyond the SM forces, deepening our understanding of inelasticity at low energies, revealing new effects and exposing limitations of existing computational approaches.

A notable development is the connection between long-range interactions and the unitarity limits on partial-wave elastic and inelastic cross-sections. The momentum dependence of the unitarity limits suggests that they can be approached or attained in the non-relativistic regime, across a continuum of energies, only by interactions that are (sufficiently) long-ranged and attractive~\cite{Flores:2024sfy,Baldes:2017gzw}. Attractive long-range interactions imply also the existence of bound states. In minimal QED-like theories, BSF occurs via emission of an Abelian gauge boson or a neutral scalar. The non-relativistic potential of the incoming scattering state, being the same as that of the bound state, is then both attractive and long-ranged. These processes can thus saturate the unitarity limit on inelastic cross-sections across a range of non-relativistic momenta, provided that the relevant couplings are sufficiently large~\cite{vonHarling:2014kha,Petraki:2015hla,Petraki:2016cnz,Baldes:2017gzw}.

More complex theories give rise to a variety of radiative transitions, including BSF via emission of a scalar or gauge boson that carries away a conserved charge.  The emission of a charged boson typically alters the non-relativistic potential between the scattering and bound states. This results in large overlaps of the incoming and outgoing wavefunctions~\cite{Oncala:2019yvj}. Remarkably, it can lead to apparent violation of the inelastic unitarity limit, \emph{for arbitrarily small couplings at sufficiently low velocities}. Such theories include Abelian and non-Abelian gauge theories where BSF can occur via emission of a charged scalar~\cite{Oncala:2019yvj}, such as the Higgs doublet~\cite{Oncala:2021tkz,Oncala:2021swy}, or through the emission of a non-Abelian gauge boson~\cite{Binder:2023ckj}. The issue arises also when the emitted charge is global~\cite{Oncala:2019yvj}.

The severity of unitarity violation by these processes underscores an important limitation in the validity of the methods employed to compute them. It is reasonable to expect that in weakly-coupled theories, it suffices to consider only long-range or finite-range interactions in the non-relativistic potential, characterized by low (soft) momentum exchanges that can significantly distort wavefunctions even for small couplings. In contrast, contact interactions, characterized by large (hard) momentum exchanges, have a smaller impact, unless their coupling is strong. The computations referenced above indeed consider Coulomb-like potentials. However, if an inelastic amplitude appears to be so large as to approach (or even exceed) the unitarity bound -- whether this is due to a large coupling, a resonance, or a large wavefunction overlap -- then the underlying interaction may contribute appreciably to the non-relativistic potential regardless of its range.  Upon squaring, inelastic processes indeed generate irreducible contributions to the self-energy kernels that must be resummed to obtain the scattering-state wavefunctions. Taking these contributions into account is necessary to formally retain consistency with unitarity. It has been previously employed in nuclear physics (see e.g.~\cite{Carbonell1993}), and for the purpose of regulating resonances in the context of DM~\cite{Blum:2016nrz}. However, these implementations have not been sufficiently formal or general. 

More recently, Ref.~\cite{Flores:2024sfy} deduced the anti-Hermitian potential generated by inelastic processes in terms of the irreducible inelastic amplitudes, i.e.~the inelastic amplitudes with all incoming-state self-energy factors amputated, using the unitarity relation that underlies the optical theorem. It is worth emphasizing that the deduced form of the anti-Hermitian potential \emph{predicts} the correct sign for the resulting imaginary phase shift that can unitarize the inelastic cross-sections. Moreover, being \emph{non-local but separable}, it allows for a compact analytical solution of the Schr\"odinger equation.  This, in turn, allows relating the cross-sections corrected by this effect to those that neglect it, yielding a simple prescription that can be used in phenomenological studies. The results apply to any partial wave, and are independent of the details of the underlying interactions that may cause tension with unitarity, up to certain analyticity and convergence conditions on the complex momentum plane~\cite{Flores:2024sfy}. (For another approach tailored to unitarizing contact-type inelastic interactions, see~\cite{Parikh:2024mwa}, while for previous ans\"atze that improve perturbative amplitudes, see Refs.~\cite{Aydemir:2012nz,Kamada:2022zwb}.)

Reference~\cite{Flores:2025uoh} extended the methodology and theoretical foundations of Ref.~\cite{Flores:2024sfy}. It derived the same anti-Hermitian potential in two additional ways: by integrating out inelastic channels via Feshbach projection~\cite{Feshbach:1958nx,Feshbach:1962nra}, and by combining the continuity equation with LSZ reduction for two-particle states. The uniqueness and completeness of the anti-Hermitian potential imply a \emph{unique and complete unitarization scheme.} Reference~\cite{Flores:2025uoh} further developed a framework for the systematic treatment of the cases where the analyticity and convergence assumptions of the simple prescription of Ref.~\cite{Flores:2024sfy} do not hold. Here we use this extended framework to unitarize BSF, whose inelastic amplitudes are ultrasoft, thereby convergent at large momenta, but exhibit rich non-analytic structure on the complex momentum plane.

The aim of the present work is two-fold: (i) to quantify the severity of unitarity violation in existing BSF computations, and (ii) to apply the formal solution of Refs.~\cite{Flores:2024sfy,Flores:2025uoh} to BSF processes, considering the specific analyticity and convergence conditions of the latter. To this end, we consider monopole capture processes via emission of a charged scalar, in the context of which the severe unitarity violation was first pointed out~\cite{Oncala:2019yvj,Oncala:2021swy,Oncala:2021tkz}. Monopole transitions, characterized by the simple $\Delta\ell = 0$ angular-momentum selection rule, offer easy bookkeeping of the different partial waves. Moreover, being parametrized by general coupling strengths for the  scattering and bound states, they emulate the dynamics of other types of transitions, such as dipole transitions via emission of gauge bosons in Abelian and non-Abelian theories. 

For point (i), we derive Kramers-like formulae for individual partial waves, summing over all BSF cross-sections of a given angular mode. We show that, depending on the scattering-to-bound coupling ratio, and owing to excited bound levels, these sums may demonstrate unphysical growth at low velocities leading to apparent violation of unitarity for arbitrarily small couplings.  (See also \cite{Beneke:2024nxh} for an analysis of the scaling of the BSF cross-sections, and a semi-classical interpretation of the unitarity violation in BSF.)
For point (ii), we compute the contributions to the regularization scheme arising from singularities of the monopole BSF amplitudes in the complex momentum plane. As we argue, our methodology is applicable to radiative BSF of any type. We find that these singularities play a crucial role in the unitarization of BSF processes, and we discuss the resulting features in detail.

Beyond formal consistency, the proper unitarization of BSF processes has broad phenomenological implications. For DM freeze-out, the severe unitarity violation can seemingly prevent freeze-out~\cite{Binder:2023ckj}; resumming BSF restores unitarity and ensures freeze-out, yielding reliable predictions for the parameters of the theory~\cite{Petraki:2025zvv}. For DM indirect detection, unitarized BSF is likewise required for reliable predictions of both high-energy signals from the decay of metastable bound states and low-energy emission associated with BSF, in symmetric and asymmetric DM scenarios (see e.g.~\cite{Pearce:2013ola,Pearce:2015zca,Asadi:2016ybp,Cirelli:2016rnw,Baldes:2017gzu,Cirelli:2018iax,Baldes:2020hwx}). The present work is thus essential for quantitatively reliable phenomenology.

This paper is organized as follows: In \cref{sec:Models}, we introduce monopole BSF processes, provide examples of particle physics models in which they occur, and review their amplitudes and cross-sections. In \cref{sec:UnitarityViol}, we examine the scaling of the cross-sections, derive Kramers-like formulae analytically and numerically, and delineate the unitarity violation. In \cref{sec:UnitarityRestore}, we discuss the unitarity restoration in BSF processes. We conclude in \cref{sec:Concl}.

\clearpage
\section{Monopole transitions in particle physics theories \label{sec:Models}}

\subsection{Models \label{sec:Models_Lagrangians}}

The capture of unbound particles into bound states necessitates dissipation of energy that can happen radiatively. The angular-momentum selection rules for radiative transition processes depend on the spin of the emitted particles. It is well known that the emission of gauge bosons gives rise at leading order to dipole transitions, $|\Delta \ell|=1$, where $\Delta\ell$ accounts for the spin of the emitted vector. 
In contrast, emission of neutral scalars by a particle-antiparticle pair or a pair of two identical particles requires $\Delta \ell =$~even. However, the monopole modes ($\Delta \ell = 0$) vanish due to the incoming and outgoing wavefunctions being orthogonal, rendering the quadrupole modes ($|\Delta\ell| = 2$) dominant.\footnote{For a pair of particles whose coupling-to-mass ratio is different, the transitions via emission of a neutral scalar are dipole at leading order, $|\Delta \ell|=1$~\cite{Petraki:2015hla,Petraki:2016cnz}.} 
While these have been the two most commonly considered cases in the context of both SM and beyond-SM physics, a significant development has emerged fairly recently: if the emitted scalar is charged under a symmetry, the potentials of the incoming and outgoing states differ, thereby allowing for monopole transitions~\cite{Oncala:2019yvj,Ko:2019wxq}.

For Coulomb-like potentials, the angular-momentum selection rule introduces a suppression factor $\sim(p/\mu)^{2|\Delta\ell|}\sim \aB^{2|\Delta\ell|}$ on the cross-section, with $\mu$ being the reduced mass of the interacting particles, $\aB$ the coupling strength in the bound state, and $p\sim \mu \aB$ the typical momentum exchange (see e.g.~\cite{Petraki:2016cnz}). This scaling implies that monopole transitions can be very fast processes~\cite{Oncala:2019yvj,Ko:2019wxq}. 

In addition, the overlaps of wavefunctions of \emph{different potentials} exhibit non-monotonic dependence on the momenta, and can become very large, even if only for a finite momentum range~\cite{Oncala:2019yvj}. Besides monopole transitions, this feature occurs in other commonly considered processes, such as dipole transitions via emission of a non-Abelian gauge boson~\cite{Binder:2023ckj,Harz:2018csl}, even as the latter are suppressed by extra powers of the relevant couplings due to the momentum dependence of the radiative vertex. Noting that the only known fundamental scalar in Nature, the Higgs doublet, is charged under two gauge symmetries, the importance of monopole transitions becomes particularly evident. In the following, we provide example models where monopole transitions occur. These models can be invoked in the context of DM and beyond; a complete phenomenological analysis is, however, beyond the scope of this work.

\subsubsection{Abelian model with doubly-charged scalar mediator}
We consider a heavy complex scalar or Dirac fermion $X$, of mass $\mX$, that couples to a dark Abelian gauge force $U(1)_{\mathsmaller{D}}$. $X$ couples also to a light complex scalar $\varrho$ of mass $m_\varrho$, that is doubly charged under $U(1)_{\mathsmaller{D}}$, as well as to a light real scalar $\varphi$ of mass $m_\varphi$ that is neutral under $U(1)_{\mathsmaller{D}}$. The interaction Lagrangian, for the case of $X$ being scalar, is
\begin{align}
{\cal L} =
&-\dfrac{1}{4} F_{\mu\nu} F^{\mu\nu}
+(D_\mu X)^\dagger (D^\mu X)
+(D_\mu \varrho)^\dagger (D^\mu \varrho)
+\dfrac{1}{2}(\partial_\mu \varphi) (\partial^\mu \varphi)
-m_X^2 |X|^2 
-m_\varrho^2 |\varrho|^2 
-\dfrac{1}{2} m_\varphi^2 \varphi^2 
\nn \\
&
-\frac{y_{X\varrho} \, m_X}{2}~\left( X^2 \varrho^\dagger  + {X^{\dagger}}^2 \varrho \right) 
-y_{X\varphi} \, m_X~|X|^2 \varphi
-(y_{\varrho\varphi} \, m_\varphi)~|\varrho|^2 \varphi
-\dfrac{y_{\varphi} \, m_\varphi}{3}~\varphi^3
\label{eq:L_Abelian}
\\
&
- \frac{\lambda_X}{4} |X|^4 
- \frac{\lambda_\varrho}{4} |\varrho|^4 
- \frac{\lambda_\varphi}{4!} \varphi^4 
- \lambda_{X\varrho} |X|^2|\varrho|^2  
- \dfrac{\lambda_{X\varphi}}{2} |X|^2 \varphi^2
- \dfrac{\lambda_{\varrho\varphi}}{2} |\varrho|^2 \varphi^2 ,
\nn 
\end{align}
with 
$F^{\mu\nu} \equiv \partial^\mu V^\nu -\partial^\nu V^\mu$ and 
$D_j^\mu \equiv \partial^\mu +\im q_j g V^{\mu}$, 
where $V$ is the gauge boson, and the index $j$ denotes the particle of charge $q_j$. The charges are $q_X = 1$ and $q_\varrho = 2$ for the $X$ and $\varrho$ fields respectively. The quartic terms stabilize the scalar potential at large field values. We define for convenience,  
\begin{align}
\alpha_{V} \equiv g^2/(4\pi), \quad    
\alpha_{\varrho} \equiv y_{X\varrho}^2/(16\pi), \quad
\alpha_{\varphi} \equiv y_{X\varphi}^2/(16\pi) .
\end{align}
It is possible that the dark sector couples also to the SM, via biquadratic couplings of the new scalars to the Higgs doublet, and/or kinetic mixing of $V$ with the Hypercharge gauge boson. The Lagrangian \eqref{eq:L_Abelian} extends the model of Ref.~\cite{Oncala:2019yvj} by the neutral scalar $\varphi$, in order to allow for a larger range of non-relativistic potentials for the $XX$, $X^\dagger X^\dagger$ and $XX^\dagger$ pairs, as we shall see shortly.

In this model, $X$ is stabilized by an accidental $\mathbb{Z}_2$ symmetry, and can play the role of DM. Its relic density is determined by the strength of its annihilation processes, including the formation and decay of metastable bound states that we discuss next. 
The light species, $\varrho$, $\varphi$ and $V^\mu$, redshift as radiation until late times, and, depending on their relative mass hierarchies, may decay into each other and/or into SM particles, thereby eventually attaining negligible relic densities; we refer to \cite{Oncala:2019yvj} for more detailed phenomenological considerations.

\begin{figure}[t!]
\centering
\begin{tikzpicture}[line width=1pt, scale=1]
\begin{scope}[shift={(0,1)}]
\begin{scope}[shift={(-1.9,0)}]
\node at (-1.3,1){$X$};
\node at (-1.3,0){$X^{\dagger}$};
\draw (-1,1) -- (1,1);\draw[fermion] 	(-1,1) -- (-0.4,1);\draw[fermion] 	(0.7,1) -- (1,1);
\draw (-1,0) -- (1,0);\draw[fermionbar] (-1,0) -- (-0.4,0);\draw[fermionbar](0.7,0) -- (1,0);
\node at (1.3,1){$X$};
\node at (1.3,0){$X^{\dagger}$};
\draw[fill=lightblue,shift={(0,0.5)}] (-0.5,-0.5) rectangle (0.5,0.5);
\node at (0,0.5){${\cal K}_{\mathsmaller{XX^{\dagger}}}^{\twoPI}$};
\end{scope}
\node at (0,0.5){$=$};
\begin{scope}[shift={(1.9,0)}]
\node at (-1.3,1){$X$};
\node at (-1.3,0){$X^{\dagger}$};
\draw[fermion] 		(-1,1) -- (0,1);
\draw[fermionbar] 	(-1,0) -- (0,0);	
\draw[vector]	(0,0) -- (0,1);
\node at (0.5,0.5){$V$};
\draw[fermion] 		(0,1) -- (1,1);
\draw[fermionbar] 	(0,0) -- (1,0);	
\node at (1.3,1){$X$};
\node at (1.3,0){$X^{\dagger}$};
\end{scope}
\node at (3.8,0.5){$+$};
\begin{scope}[shift={(5.7,0)}]
\node at (-1.3,1){$X$};
\node at (-1.3,0){$X^{\dagger}$};
\draw[fermion] 		(-1,1) -- (0,1);
\draw[fermionbar] 	(-1,0) -- (0,0);	
\draw[scalarnoarrow](0,1) -- (0,0);
\node at (0.5,0.5){$\varphi$};
\draw[fermion]   	(0,1) -- (1,1);
\draw[fermionbar] 	(0,0) -- (1,0);	
\node at (1.3,1){$X$};
\node at (1.3,0){$X^{\dagger}$};
\end{scope}
\node at (7.6,0.5){$+$};
\begin{scope}[shift={(9.5,0)}]
\node at (-1.3,1){$X$};
\node at (-1.3,0){$X^{\dagger}$};
\draw[fermion] 		(-1,1) -- (0,1);
\draw[fermionbar] 	(-1,0) -- (0,0);	
\draw[scalar]	(0,1) -- (0,0);
\node at (0.5,0.5){$\varrho$};
\draw[fermionbar] 	(0,1) -- (1,1);
\draw[fermion] 		(0,0) -- (1,0);	
\node at (1.3,1){$X^{\dagger}$};
\node at (1.3,0){$X$};
\end{scope}
\end{scope}
%
%
\begin{scope}[shift={(0,-1)}]
\begin{scope}[shift={(-1.9,0)}]
\node at (-1.3,1){$X$};
\node at (-1.3,0){$X$};
\draw (-1,1) -- (1,1);\draw[fermion] (-1,1) -- (-0.4,1);\draw[fermion] (0.7,1) -- (1,1);
\draw (-1,0) -- (1,0);\draw[fermion] (-1,0) -- (-0.4,0);\draw[fermion] (0.7,0) -- (1,0);
\node at (1.3,1){$X$};
\node at (1.3,0){$X$};
\draw[fill=lightred,shift={(0,0.5)}] (-0.5,-0.5) rectangle (0.5,0.5);
\node at (0,0.5){${\cal K}_{\mathsmaller{XX}}^{\twoPI}$};
\end{scope}
\node at (0,0.5){$=$};
\begin{scope}[shift={(1.9,0)}]
\node at (-1.3,1){$X$};
\node at (-1.3,0){$X$};
\draw[fermion]	(-1,1) -- (0,1);
\draw[fermion]	(-1,0) -- (0,0);	
\draw[vector]	(0,1) -- (0,0);
\node at (0.5,0.5){$V$};
\draw[fermion] 	(0,1) -- (1,1);
\draw[fermion] 	(0,0) -- (1,0);	
\node at (1.3,1){$X$};
\node at (1.3,0){$X$};
\end{scope}
\node at (3.8,0.5){$+$};
\begin{scope}[shift={(5.7,0)}]
\node at (-1.3,1){$X$};
\node at (-1.3,0){$X$};
\draw[fermion] 		(-1,1) -- (0,1);
\draw[fermion]   	(-1,0) -- (0,0);	
\draw[scalarnoarrow](0,1) -- (0,0);
\node at (0.5,0.5){$\varphi$};
\draw[fermion]   	(0,1) -- (1,1);
\draw[fermion]  	(0,0) -- (1,0);	
\node at (1.3,1){$X$};
\node at (1.3,0){$X$};
\end{scope}
\end{scope}
\end{tikzpicture}
\caption{\label{fig:2PI} 
The 2PI kernels generating long- and finite-range potentials for $XX^{\dagger}$ pairs (upper) and $XX$ or $X^{\dagger}X^{\dagger}$ pairs (lower), in the model of \cref{eq:L_Abelian}. The arrows denote the flow of the $U(1)_D$ charge. Figure adapted from Ref.~\cite{Oncala:2019yvj}.}
\end{figure}
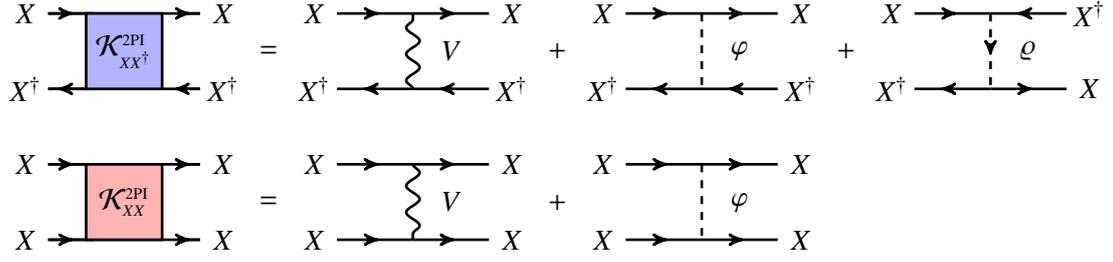
\begin{figure}[t!]
\centering
\begin{tikzpicture}[line width=1pt, scale=1]
\begin{scope}[shift={(0,0.5)}]
\node at (-4.3,1){$X$};
\node at (-4.3,0){$X$};
\draw[fermion]	(-2.1,1) -- (-2,1);
\draw[fermion]	(-2.1,0) -- (-2,0);
\draw (-4,1) -- (0,1);
\draw (-4,0) -- (0,0);
%
\draw[fill=lightred,shift={(-3,0.5)}] (-0.5,-0.5) rectangle (0.5,0.5);
\draw[fill=lightred,shift={(-1,0.5)}] (-0.5,-0.5) rectangle (0.5,0.5);
\node at (-3,0.5){${\cal K}_{\mathsmaller{XX}}^{\twoPI}$};
\node at (-1,0.5){${\cal K}_{\mathsmaller{XX}}^{\twoPI}$};
\node at (-2,0.5){$\cdots$};
\draw[scalar]	(0,0) -- (0.6,-0.6);
\node at (0.75,-0.5){$\varrho$};
\node at (4.3,1){$X$};
\node at (4.3,0){$X^{\dagger}$};
\draw[fermion]		(2,1) -- (2.1,1);
\draw[fermionbar]	(2,0) -- (2.1,0);
\draw (0,1) -- (4,1);
\draw (0,0) -- (4,0);
%
\draw[fill=lightblue,shift={(3,0.5)}] (-0.5,-0.5) rectangle (0.5,0.5);
\draw[fill=lightblue,shift={(1,0.5)}] (-0.5,-0.5) rectangle (0.5,0.5);
\node at (3,0.5){${\cal K}_{\mathsmaller{XX^{\dagger}}}^{\twoPI}$};
\node at (1,0.5){${\cal K}_{\mathsmaller{XX^{\dagger}}}^{\twoPI}$};
\node at (2,0.5){$\cdots$};
\draw[fill=none,gray,line width=1.5pt] (2.1,0.5) ellipse (1.8 and 0.75);
\node at (4.6,0.5){${\cal B}$};
\end{scope}
%
\node at (0,0){$+$};
%
\begin{scope}[shift={(0,-1.5)}]
\node at (-4.3,1){$X$};
\node at (-4.3,0){$X$};
\draw[fermion]	(-2.1,1) -- (-2,1);
\draw[fermion]	(-2.1,0) -- (-2,0);
\draw (-4,0) -- (-3.5,1);\draw (-3.5,1) -- (0,1);
\draw (-4,1) -- (-3.5,0);\draw (-3.5,0) -- (0,0);
%
\draw[fill=lightred,shift={(-3,0.5)}] (-0.5,-0.5) rectangle (0.5,0.5);
\draw[fill=lightred,shift={(-1,0.5)}] (-0.5,-0.5) rectangle (0.5,0.5);
\node at (-3,0.5){${\cal K}_{\mathsmaller{XX}}^{\twoPI}$};
\node at (-1,0.5){${\cal K}_{\mathsmaller{XX}}^{\twoPI}$};
\node at (-2,0.5){$\cdots$};
\draw[scalar]	(0,0) -- (0.6,-0.6);
\node at (0.75,-0.5){$\varrho$};
\node at (4.3,1){$X$};
\node at (4.3,0){$X^{\dagger}$};
\draw[fermion]		(2,1) -- (2.1,1);
\draw[fermionbar]	(2,0) -- (2.1,0);
\draw (0,1) -- (4,1);
\draw (0,0) -- (4,0);
%
\draw[fill=lightblue,shift={(3,0.5)}] (-0.5,-0.5) rectangle (0.5,0.5);
\draw[fill=lightblue,shift={(1,0.5)}] (-0.5,-0.5) rectangle (0.5,0.5);
\node at (3,0.5){${\cal K}_{\mathsmaller{XX^{\dagger}}}^{\twoPI}$};
\node at (1,0.5){${\cal K}_{\mathsmaller{XX^{\dagger}}}^{\twoPI}$};
\node at (2,0.5){$\cdots$};
\draw[fill=none,gray,line width=1.5pt] (2.1,0.5) ellipse (1.8 and 0.75);
\node at (4.6,0.5){${\cal B}$};
\end{scope}
\end{tikzpicture}
\caption{
\label{fig:BSF_monopole_XXtoXXdagger_FeynDiag}
Bound-state formation via emission of a charged scalar, $X+ X \to {\cal B} (XX^{\dagger}) + \varrho$, in the model of \cref{eq:L_Abelian}. The arrows denote the flow of the $U(1)_D$ charge. Figure adapted from Ref.~\cite{Oncala:2019yvj}.} 
\end{figure}
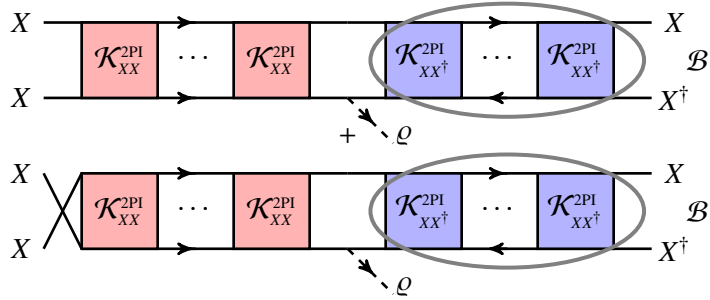

The exchange of the massless $V^\mu$ and light $\varrho$ and $\varphi$ bosons, shown in \cref{fig:2PI}, generates long- and short-range potentials~\cite{Petraki:2015hla,Petraki:2016cnz,Oncala:2019yvj},
\begin{subequations}
\label{eq:Potentials}
\label[pluralequation]{eqs:Potentials}
\begin{align}
V_{XX^\dagger} (r) &= -\dfrac{1}{r}  
\left(
+\alpha_V 
+\alpha_\varphi e^{-m_\varphi r} 
+ (-1)^\ell \alpha_\varrho e^{-m_\varrho r} 
\right),
\label{eq:Potential_XXdagger}
\\
V_{XX} (r) = V_{X^\dagger X^\dagger} (r) &= 
-
\dfrac{1+(-1)^\ell}{2}
\dfrac{1}{r}  
\left(-\alpha_V + \alpha_\varphi e^{-m_\varphi r} \right),
\label{eq:Potential_XX}
\end{align}
\end{subequations}
where the dependence of \cref{eq:Potential_XXdagger} on the angular mode $\ell$ is due to the $u$-channel $\varrho$ exchange~\cite[appendix~A]{Oncala:2019yvj}, and in \cref{eq:Potential_XX} we have taken into account that the interacting particles are identical~\cite[appendix~A]{Oncala:2021tkz}. It is clear that even in the limit of the local symmetry becoming global ($\alpha_V \to 0$), the potentials \eqref{eq:Potential_XXdagger} and \eqref{eq:Potential_XX} are different, as long as $\alpha_\varrho \neq 0$. 
For $\ell=$~even (which are the relevant modes for monopole transitions if $X$ is scalar), 
the $XX^\dagger$ potential is attractive and supports bound states. On the other hand, the $XX$ and $X^\dagger X^\dagger$ potential is repulsive at large distances, but may be either attractive or repulsive at shorter distances, depending on the relative strength of the $\alpha_V$ and $\alpha_\varphi$ couplings. The condition for $XX$ and $X^\dagger X^\dagger$ supporting bound states can be determined numerically (see~\cite[Fig.~6 and Eq.~(3.30)]{Harz:2019rro}).

In this setup, $XX^\dagger$ bound states can form from $XX^\dagger$ scattering states with emission of a $V^\mu$ or $\varphi$ boson; these are dipole and quadrupole transitions, respectively~\cite{Petraki:2016cnz}. $XX^\dagger$ bound states may also form from $XX$ scattering states via emission of a $\varrho$ scalar,
\begin{align}
X+X \to {\cal B} (XX^\dagger) + \varrho ,   
\label{eq:BSF_monopole_XXtoXXdagger}
\end{align}
as shown in \cref{fig:BSF_monopole_XXtoXXdagger_FeynDiag}, or its conjugate process. These are monopole transitions~\cite{Oncala:2019yvj}. Equivalent considerations can be extended to the formation of $XX$ and $X^\dagger X^\dagger$ bound states, if they exist.

\subsubsection{Higgs-portal models}

We consider scalar or fermionic electroweak multiplets $X_N$, with $N$ denoting their multiplicity under ${\rm SU}_L(2)$. Fields whose multiplicities differ by 1 may interact with each other via trilinear couplings to the Higgs doublet. For fermionic $X_N$, the SM Lagrangian is extended by
\begin{align}
{\delta \cal L} &= \sum_N \left[
\bar{X}_N (\im \slashed{D} -m_N )X_N
-(y_{N} \bar{X}_{N+1} H X_N + \hc)
\right],
\label{eq:Lagrangian}
\end{align}
where the sum over $N$ extends schematically over the multiplets that are introduced in a given model. The Hypercharge assignments of the $X_N$ multiplets must be such that the interaction terms are gauge invariant. 

Such interactions arise in supersymmetric and non-supersymmetric frameworks that have long served as archetypal models for WIMP DM (see e.g.~\cite{Lopez-Honorez:2017ora} for a classification and related phenomenological considerations).
After electroweak symmetry breaking, multiplets coupled to each other via $H$ acquire mass mixing terms that, upon diagonalization, generate mass splittings for the various components. The lightest component, typically stabilized by a $\mathbb{Z}_2$ symmetry, may constitute the DM of the universe. 

The processes that annihilate the DM multiplets and affect the DM relic density include capture into metastable bound states. If the masses of the DM multiplets are sufficiently large, their chemical decoupling from the primordial plasma begins, and may even complete, before the electroweak phase transition. In this case, bound states may form with emission of a Higgs doublet, in rapid monopole transitions, for example~\cite{Oncala:2021tkz,Oncala:2021swy} 
\begin{align}
\bar{X}_N X_{N+1} \to {\cal B} (\bar{X}_{N}X_{N}) +H.     
\end{align} 
Note that the leading-order potentials in the incoming and outgoing states are generated by Higgs-doublet and ${\rm SU}_L(2)\times {\rm U}_Y(1)$ gauge-boson exchanges, and are determined by the Yukawa couplings and the gauge representations of the interacting particles; they are in general different for the scattering and bound states in a transition via $H$ emission.
If the DM chemical decoupling occurs partially or entirely below the electroweak phase transition, the BSF processes via Higgs doublet emission are encompassed by BSF via $W^\pm$ or $Z$ emission that must, in the appropriate limit, exhibit a large enhancement due to the Goldstone boson equivalence theorem~\cite{Ko:2019wxq}.

\subsection{Bound-state formation processes \label{sec:Models_BSF}}

Having seen examples of Abelian and non-Abelian theories in which BSF via monopole transitions may occur, we now review their amplitudes and cross-sections. Analytic formulae can be obtained if the scattering-state and bound-state wavefunctions are Coulombic.\footnote{
The wavefunctions of a Yukawa potential can be well approximated by a Coulomb potential if the mediator mass, $m_{\rm med}$, is sufficiently small, $m_{\rm med} \ll \mu \aB/n$ for bound states and $m_{\rm med} \lesssim~\mu \vrel$ for scattering states, where $\mu$ and $\vrel$ are the reduced mass and relative velocity of the interacting particles, and $\aB$ is the coupling strength of the bound-state potential~\cite{Petraki:2016cnz} and $n$ is the bound-state principal quantum number. For combined Yukawa plus Coulomb potentials, these conditions may be significantly relaxed~\cite{Harz:2017dlj,Harz:2019rro}. The Coulomb approximation is often satisfactory, in the relevant parameter space, during DM freeze-out, but to a more limited extent for the indirect-detection signals produced in galaxies today or during CMB~\cite{Cirelli:2016rnw}. See \cref{sec:UnitarityRestore_BeyondCoulomb} for considerations beyond the Coulomb approximation.
}

\subsubsection{Setup \label{sec:Models_BSF_Setup}}

We consider pairs of particles of masses $m_1$ and $m_2$, and define their total and reduced masses, 
\begin{align}
\mT \equiv m_1 + m_2 
\quad \text{and} \quad
\mu \equiv \dfrac{m_1 m_2}{m_1+m_2} ,
\label{eq:masses}
\end{align}
We will be interested in transitions between scattering and bound states where the interactions can be well approximated by Coulomb potentials, that we shall parametrize respectively as\footnote{It is possible that the particles in the scattering and bound states are different (e.g.~BSF via $H$ emission described above), therefore the masses are in general different. While our computations can be generalized to encompass this possibility, for simplicity, we will take the masses of the incoming and outgoing particles to be the same.}
\begin{align}
\VS = -\aS/r 
\quad\text{and}\quad
\VB = -\aB/r.
\label{eq:Potentials_SSandBS}
\end{align}
Note that $\aB>0$, while $\aS$ may be positive, negative or zero. The scattering-state and bound-state spectra are characterized by the momentum in the center-of-momentum (CM) frame, $\vb{k}$, and the discrete principal and angular-momentum quantum numbers, $n\ellB\mB$, respectively. For convenience, here and in the following, we define the Bohr momenta, and their ratios to the scattering-state momentum,
\begin{align}
&\kappaS \equiv \mu \aS ,&  
\quad
&\kappaB \equiv \mu \aB ,&  
\quad
&\kappan \equiv \mu \aB/n ,& 
\label{eq:kappas_def}
\\
&\zetaS \equiv \kappaS/k = \aS/\vrel ,&  
\quad
&\zetaB \equiv \kappaB/k = \aB/\vrel ,&
\quad
&\zetan \equiv \kappan/k = \zetaB/n .&
\label{eq:zetas_def}
\end{align}
The energies in the CM frame are 
$\ES = \mT+{\cal E}_{\vb{k}}$ and
$E_{n} = \mT+{\cal E}_{n}$, 
with
\begin{align}
{\cal E}_{\vb{k}} 
=\dfrac{\vb{k}^2}{2\mu} 
=\dfrac{\mu\vrel^2}{2}
\quad\text{and}\quad
{\cal E}_{n} 
=-\dfrac{\kappaB^2}{2\mu n^2}
=-\dfrac{\mu \aB^2}{2n^2} .
\label{eq:Energies}
\end{align}
The scattering and bound states are described by wavefunctions 
$\psi_{\vb{k}}(\bf r;~\kappaS)$, 
$\psi_{n\ellB\mB}(\vb{r};~\kappaB)$ 
and 
$\tilde\psi_{\vb{k}}(\bf p;~\kappaS)$, 
$\tilde\psi_{n\ellB\mB}({\bf p};~\kappaB)$ 
in position and momentum space, respectively. Since we will be dealing with transitions between states governed by different potentials, we denote the Bohr momenta of the corresponding potential explicitly, as an argument of the wavefunction. When describing identical particles, these wavefunctions are properly (anti-)symmetrized. For clarity, we will denote the wavefunctions for distinguishable particles with capital $\Psi$ and the corresponding quantum numbers.
We collect our conventions for the Fourier transform and partial-wave analysis of the wavefunctions in \cref{app:OverlapIntegrals}.

Bound states may form from scattering states with emission of a particle that carries away energy $\omega_{n}$ and momentum $\prad_{n}$; it will be convenient to regard them as functions of the energy in the CM frame, parametrized in terms of the first Mandelstam variable $s$. For on-shell scattering-state momentum $k$~\cite{Petraki:2015hla}, 
\begin{subequations}
\label{eq:BSFenergetics}
\label[pluralequation]{eqs:BSFenergetics}
\begin{align}
\sqrt{s} &= \mT + {\cal E}_{\vb{k}},
\label{eq:ScatteringStates_OnShellRelation}
\\
\omega_{n}(s) 
&= {\cal E}_{\vb{k}} - {\cal E}_{n} 
= \dfrac{k^2}{2\mu}(1+\zetan^2) ,
\label{eq:omega}
\\
\prad_{n} (s) &= \omega_{n} (s) \, \Phi_{n} (s) ,
\label{eq:prad}
\\
\Phi_{n} (s) 
&\equiv 
\left(1-\dfrac{\mrad^2}{\omega_{n}^2(s)}\right)^{1/2}    
=\left(1-
\dfrac
{4\mu^2 \mrad^2}
{(\kappan^2+k^2)^2}
\right)^{1/2},
\label{eq:Phi_n}
\end{align}
\end{subequations}
where $\Phi_{n} (s)$ is the phase-space suppression factor due to the mass $\mrad$ of the radiated particle. We reiterate that \cref{eqs:BSFenergetics} hold for on-shell momenta $k$. In \cref{sec:UnitarityRestore}, we will also consider off-shell scattering states and amplitudes, for which \cref{eqs:BSFenergetics} do not hold, with $k$ being independent of $s$.

We also define the partial-wave unitarity cross-section as in~\cite{Flores:2024sfy},
\begin{align}
\sigma_\ell^U (k) \equiv \sym_\ell \dfrac{4\pi (2\ell+1) }{\vb{k}^2} ,
\label{eq:sigmaU}
\end{align}
where $\sym_\ell$ is a symmetry factor for the incoming state: 
$\sym_\ell=1$ if the incoming particles are distinguishable,  while
$\sym_\ell=2$~or~0 if they are identical and $\ell$ is a surviving or vanishing mode, respectively, due to the (anti)symmetrization of the wavefunction. Although the symmetry factor of a state does not depend on the partial wave, this effective description arises after partial-wave analysis, and it will prove convenient to define this $\ell$-dependent symmetry factor.

\subsubsection{Amplitudes \label{sec:Models_BSF_Amplitudes}}

We distinguish between the irreducible and full BSF amplitudes, that we denote as 
${\cal A}_{\vb{k} \to n\ellB\mB}^{\BSF}$ and
${\cal M}_{\vb{k} \to n\ellB\mB}^{\BSF}$, respectively. 
The irreducible amplitudes exclude scattering-state self-energy insertions, while the full amplitudes include the self-energy associated with the Coulomb interaction in the scattering state, \cref{eq:Potentials_SSandBS},\footnote{
In \cref{sec:UnitarityRestore}, we will refer to the full amplitudes and cross-sections presented here as `unregulated', to distinguish them from the `regulated' ones obtained by resumming the BSF contributions in the scattering-state self-energy in addition to the Coulomb exchanges.
} 
%
and are related as follows (see e.g.~\cite{Petraki:2015hla})
\begin{align}
{\cal M}_{\vb{k} \to n\ellB\mB}^{\BSF} = 
\int \dfrac{\dd^3 \vb{p}}{(2\pi)^3}
\, \tilde{\psi}_{\vb{k}}^{} (\vb{p};~\kappaS)
\, {\cal A}_{{\bf p} \to n\ellB\mB}^{\BSF} .
\label{eq:BSF_McalvsAcal}
\end{align}
For monopole transitions, they are~\cite{Oncala:2019yvj}
\begin{subequations}
\label{eq:BSF_Amplitudes}
\label[pluralequation]{eqs:BSF_Amplitudes}
\begin{align}
\im{\cal A}_{\vb{k} \to n\ellB\mB}^{\BSF} 
&= - \im 
\yrad \, \mT
\sqrt{2 \mu}
\ \tilde{\Psi}_{n\ellB\mB}^{*} (\vb{k};~\kappaB),
\label{eq:BSF_Acal}
\\[1em]
\label{eq:BSF_Mcal}
\im {\cal M}_{\vb{k} \to n\ellB\mB}^{\BSF} 
&= - \im \yrad \dfrac{\mT \sqrt{2\mu}}{\kappaB^{3/2}} 
\ \times
\\
\nn
&\times
\begin{cases}
{\cal R}_{\vb{k},n\ellB\mB} 
(\kappaS,\kappaB) ,
& \text{distinguishable particles},
\\[1ex]
\dfrac{
{\cal R}_{\vb{k}, n\ellB\mB} 
(\kappaS,\kappaB)
\pm
{\cal R}_{-\vb{k}, n\ellB\mB} 
(\kappaS,\kappaB)
}{\sqrt{2}},
& \text{identical bosons / fermions}. 
\end{cases}
\end{align}
\end{subequations}
We have introduced the dimensionless scattering-bound overlap integral ${\cal R}_{\vb{k}, n\ellB\mB}$, and later on we will also need the bound-bound counterpart, ${\cal R}_{n\ellB\mB;~n'\ellB'\mB'}$,
\begin{subequations}
\label{eq:Rcal_def}
\label[pluralequation]{eqs:Rcal_def}
\begin{align}
{\cal R}_{\vb{k}, n\ellB\mB}^{} 
(\kappaS,\kappaB)
&\equiv 
\kappaB^{3/2} \, 
\int \dd^3 r 
\ \Psi_{n\ellB\mB}^* (\vb{r};~\kappaB) 
\ \Psi_{\vb{k}}^{} (\vb{r};~\kappaS),
\label{eq:Rcal_ScattBound_def}
\\
{\cal R}_{n\ellB\mB;~n'\ellB'\mB'}^{} 
(\kappaB,\kappaB')
&\equiv 
\int \dd^3 r 
\ \Psi_{n'\ellB'\mB'}^* (\vb{r};~\kappaB') 
\ \Psi_{n\ellB\mB}^{} (\vb{r};~\kappaB).
\label{eq:Rcal_BoundBound_def}
\end{align}    
\end{subequations}
We emphasize that the wavefunctions in \cref{eq:BSF_Acal,eqs:Rcal_def} refer to distinguishable particles; the (anti)-symmetrization for identical particles is directly shown in \cref{eq:BSF_Mcal}.  
We compute ${\cal R}_{\vb{k}, n\ell m}$ and ${\cal R}_{n\ellB\mB;~n'\ellB'\mB'}$ in \cref{app:OverlapIntegrals}, following Refs.~\cite{Oncala:2019yvj,Oncala:2021tkz}, but using somewhat different notation to better exhibit their analytic properties. 

In \cref{eqs:BSF_Amplitudes}, $\yrad$ is related to the coupling in the Lagrangian that allows for the emission of the charged boson, and must be determined by considering other model-dependent factors, e.g.~those that arise from the interference of Feynman diagrams contributing to the process.\footnote{%
For the transition \eqref{eq:BSF_monopole_XXtoXXdagger} of the model \eqref{eq:L_Abelian}, $\yrad = y_{X\varrho}^{} 2\delta_{\ell,\rm even}$, and the symmetry factor is $\sym_\ell = 2\delta_{\ell,\rm even}$. Note that the factor $2\delta_{\ell,\rm even}$ in $\yrad$ arises from the interference of two Feynman diagrams at leading order, while in $\sym_\ell$ it arises from the symmetrization of the incoming $XX$ wavefunction~\cite{Oncala:2019yvj}.}  

\subsubsection{Partial-wave analysis \label{sec:Models_BSF_PWA}}

Next, we analyze the amplitudes in partial waves, as follows\footnote{
We assume the orthonormality condition 
$\int \dd\Omega 
\, Y_{\ell m}^{} (\Omega) 
\, Y_{\ell' m'}^*(\Omega) 
= \delta_{\ell\ell'}\delta_{mm'}$ 
for the spherical harmonics, and recall the useful decomposition 
$P_\ell(\hat{\vb{a}}\cdot\hat{\vb{b}}) 
= [(4\pi)/(2\ell+1)]
\sum_{m=-\ell}^\ell 
Y_{\ell m}^{} (\Omega_{\vb{a}}) \,
Y_{\ell m}^*(\Omega_{\vb{b}})$ 
for the Legendre polynomials.}
%
\begin{subequations}
\label{eq:BSF_Amplitudes_PW}
\label[pluralequation]{eqs:BSF_Amplitudes_PW}
\begin{align}
{\cal A}_{\vb{k}\to n\ellB\mB}^{\BSF} &= 
16\pi \sqrt{4\pi} 
\sum_{\ell=0}^{\infty}
\sum_{m=-\ell}^{\ell} 
Y_{\ell m}^* (\Omega_{\vb{k}}) 
\, {\cal A}_{\ell m}^{\BSF} (k;n\ellB\mB),
\label{eq:BSF_Acal_PW}
\\
{\cal M}_{\vb{k}\to n\ellB\mB}^{\BSF} &= 
16\pi \sqrt{4\pi} 
\sum_{\ell=0}^{\infty}
\sum_{m=-\ell}^{\ell} 
Y_{\ell m}^* (\Omega_{\vb{k}}) 
\, {\cal M}_{\ell m}^{\BSF} (k;n\ellB\mB),
\label{eq:BSF_Mcal_PW}
\end{align}
\end{subequations}
with their inverse being
\begin{subequations}
\label{eq:BSF_Amplitudes_PW_inv}
\label[pluralequation]{eqs:BSF_Amplitudes_PW_inv}
\begin{align}
{\cal A}_{\ell m}^{\BSF} (k;n\ellB\mB) 
&= \dfrac{1}{16\pi \sqrt{4\pi}} 
\int \dd\Omega_{\vb{k}} 
Y_{\ell m}^{} (\Omega_{\vb{k}}) 
\, {\cal A}_{\vb{k}\to n\ellB\mB}^{\BSF} ,
\label{eq:BSF_Acal_PW_inv}
\\
{\cal M}_{\ell m}^{\BSF} (k;n\ellB\mB) 
&= \dfrac{1}{16\pi \sqrt{4\pi}} 
\int \dd\Omega_{\vb{k}} 
Y_{\ell m}^{} (\Omega_{\vb{k}}) 
\, {\cal M}_{\vb{k}\to n\ellB\mB}^{\BSF} ,
\label{eq:BSF_Mcal_PW_inv}
\end{align}
and, by virtue of \cref{eq:BSF_McalvsAcal}, 
\begin{align}
{\cal M}_{\ell m}^{\BSF} (k;n\ellB\mB)
&= \dfrac{1}{2\pi^2} 
\int_0^\infty dp \ p^2 
\ \tilde\psi_{k,\ell}^{} (p;~\kappaS)
\ {\cal A}_{\ell m}^{\BSF} (p;n\ellB\mB),
\label{eq:BSF_McalvsAcal_PW}
\end{align}
\end{subequations}
where, upon partial-wave analysis, 
\begin{align}
\tilde\psi_{k,\ell}^{} (p;~\kappaS) 
= \sqrt{\sym_\ell} 
\ \tilde\Psi_{k,\ell}^{} (p;~\kappaS) ,   
\label{eq:Wavefunctions_SymmetryFactor}
\end{align}
and similarly for the position-space wavefunctions.  The symmetry factor $\sym_\ell$ has been defined below \cref{eq:sigmaU}. Considering the amplitudes \eqref{eqs:BSF_Amplitudes}, and that 
$\Psi_{n\ellB\mB}^*$ and ${\cal R}_{\vb{k}, n\ellB\mB}$ are proportional to $Y_{\ellB \mB}^* (\Omega_{\vb{k}})$,
[cf. \cref{eq:PW_Wavefun,eq:Rcal_final_app}], we find
\begin{subequations}
\label{eq:BSF_lm-nl}
\label[pluralequation]{eqs:BSF_lm-nl}
\begin{align}
{\cal A}_{\ell m}^{\BSF} (k; n\ellB\mB) 
&=
\delta_{\ell \ellB}^{} 
\delta_{m\mB}^{}
{\cal A}_{n\ell}^{\BSF} (k) ,
\label{eq:BSF_lm-nl_Acal}
\\
{\cal M}_{\ell m}^{\BSF} (k; n\ellB\mB) 
&=
\delta_{\ell \ellB}^{} 
\delta_{m\mB}^{}
{\cal M}_{n\ell}^{\BSF} (k) ,
\label{eq:BSF_lm-nl_Mcal}
\end{align}
\end{subequations}
with
\begin{subequations}
\label{eq:BSF_nl}
\label[pluralequation]{eqs:BSF_nl}
\begin{align}
{\cal A}_{n\ell}^{\BSF} (k) 
&\equiv
-\dfrac{\yrad}{16\pi\sqrt{4\pi}}
\ \mT\sqrt{2\mu} 
\ \tilde\Psi_{n\ell}^* (k;~\kappaB) 
=
\dfrac{\yrad}{16\sqrt{\pi}}
\dfrac{\mT\sqrt{2\mu}}{\kappan^{3/2}}
\times
\rM_{n\ell} (0,\zetaB) ,
\label{eq:BSF_nl_Acal}
\\
{\cal M}_{n\ell}^{\BSF} (k) 
&\equiv
\sqrt{\sym_\ell}
\dfrac{\yrad}{16\sqrt{\pi}}
\dfrac{\mT\sqrt{2\mu}}{\kappan^{3/2}}
\left(1-\dfrac{\kappaS}{\kappaB}\right)
\times
\rM_{n\ell} (\zetaS,\zetaB) ,
\label{eq:BSF_nl_Mcal}
\end{align}
and, as in \cref{eq:BSF_McalvsAcal_PW}, 
\begin{align}
{\cal M}_{n\ell}^{\BSF} (k)
&= \dfrac{1}{2\pi^2} 
\int_0^\infty \dd p \ p^2 
\ \tilde{\psi}_{k,\ell}^{} (p;~\kappaS)
\ {\cal A}_{n\ell}^{\BSF} (p)
,
\label{eq:BSF_nl_McalvsAcal}
\end{align}
\end{subequations}
where $\rM_{n\ell} (\zetaS,\zetaB)$ encapsulates the dependence of the amplitude on the incoming momentum $k$, 
\begin{empheq}[box=\myshadebox]{align}
&\rM_{n\ell} (\zetaS,\zetaB) 
\equiv 
-\dfrac{1}{2\pi \, n^{3/2} \, (1-\kappaS/\kappaB)}
\int \dd\Omega_{\vb{k}} 
\, Y_{\ell m} (\Omega_{\vb{k}}) 
\, {\cal R}_{\vb{k},n\ell m} (\kappaS,\kappaB)
\nn 
\\
&=
\im^\ell (-1)^{n-\ell} 
\times
\frac{2^{2\ell+3}}{(2\ell+1)!} 
\left[\frac{n (n+\ell)!}{(n-\ell-1)!}\right]^{1/2} 
\ e^{\pi\zetaS/2} 
\ \Gamma(1+\ell-\im\zetaS)
\nn
\\
&\times 
\dfrac
{\zetan^{\ell+4}}
{(\zetan\mp\im)^{2\ell+4}}
\left(
\dfrac
{\zetan\mp\im}
{\zetan\pm\im}
\right)^{n+1 \mp \im \zetaS}
{}_2F_1 \left( 
1+\ell-n, ~ 
1+\ell \pm \im \zetaS; ~
2\ell+2; ~
\mp\frac{4 \im \zetan}{(\zetan \mp \im)^2}
\right) ,
\label{eq:BSF_rMnl}
\end{empheq}
with
$\rM_{n\ell} (0,\zetaB) = 
-[\kappan^{3/2} / (2\pi) ]
\ \tilde{\Psi}_{n\ell}^*(k;~\kappaB)$.
The two signs in \cref{eq:BSF_rMnl} arise from a Pfaff transformation for the Hypergeometric function. 

The analytic properties of $\rM_{n\ell}$ in the complex momentum plane will be considered in detail in \cref{sec:UnitarityRestore}, as they determine the result of the resummation of the BSF processes in the self-energy of the incoming state and the unitarization of the BSF cross-sections.

\subsubsection{Cross-sections \label{sec:Models_BSF_CrossSections}}

The differential BSF cross-sections are
\begin{align}
\dfrac
{d\sigma_{\vb{k}\to n\ellB\mB}^{\BSF} }
{d\Omega^{\rm rad}} 
&= 
\dfrac{1}{64\pi^2 s} 
\dfrac{\prad_n(s)}{k} 
|{\cal M}_{\vb{k}\to n\ellB\mB}^{\BSF}|^2 ,
\label{eq:BSF_sigma_differential}
\end{align}
where $\Omega^{\rm rad}$ is the solid angle of the radiated boson. Considering \cref{eq:BSF_Mcal_PW,eq:BSF_lm-nl_Mcal}, and integrating over $\dd\Omega^{\rm rad}$,
\begin{align}
\sigma_{\vb{k}\to n\ellB\mB}^{\BSF} &=
64\pi^2
\, \dfrac{\prad_n(s)}{s \, k}
\ |Y_{\ellB \mB} (\Omega_{\vb{k}})|^2
\ |{\cal M}_{n\ellB}^{\BSF} (k)|^2 .
\label{eq:BSF_sigma}
\end{align}
The partial-wave BSF cross-section for capture to the $n$ bound level is
\begin{align}
\sigma_{n\ell}^\BSF (k) 
=
\sum_{m=-\ell}^\ell
\sigma_{\vb{k}\to n\ell m}^{\BSF}
=
(2\ell+1)
\, \dfrac{16\pi\prad_n(s)}{s \, k}
\ |{\cal M}_{n\ell}^{\BSF} (k)|^2.
\label{eq:BSF_sigma_PW}
\end{align}
Considering \cref{eq:BSFenergetics,eq:BSF_nl_Mcal}, and setting $s\to \mT^2$, consistently with the non-relativistic approximations employed in computing the amplitudes of \cref{sec:Models_BSF_Amplitudes,sec:Models_BSF_PWA}, we arrive at
\begin{align}
\sigma_{n\ell}^{\BSF} (k) =
\dfrac{1}{4} \, \sigma_\ell^U (k)
\times
\Phi_{n} (s) 
\ \arad
\ \left(1-\frac{\kappaS}{\kappaB}\right)^2
\times r_{n\ell} (\zetaS,\zetaB),
\qquad
n \geqslant 1+\ell,
\label{eq:BSF_sigma_PW_final}
\end{align}
where $\sigma_\ell^U$ is the partial-wave unitarity cross-section given in \cref{eq:sigmaU}, we defined
\begin{align}
\arad \equiv \dfrac{\yrad^2}{16\pi}, 
\label{eq:alpha_rad_def}
\end{align}
and 
\begin{empheq}[box=\myshadebox]{align}
r_{n\ell}  (\zetaS,\zetaB)   
&\equiv
\dfrac{1+\zetan^2}{\zetan^3}
\ |\rM_{n\ell} (\zetaS,\zetaB) |^2   
\nn \\
&=
\left[\frac{2^{2\ell+3}\ell!}{(2\ell+1)!}\right]^2
\frac{n (n+\ell)!}{(n-\ell-1)!} 
\times
S_\ell (\zetaS)
\, e^{-4\zetaS \, \arccot (\zetan)}
\left[ \frac{\zetan^{2\ell+5}}{(1+\zetan^2)^{2\ell+3}} \right]
\nn \\  
&\times
\left|
{}_2F_1 \left( 
1+\ell-n, ~ 
1+\ell \pm \im \zetaS; ~
2\ell+2; ~
\mp\dfrac{4 \im \zetan}{(\zetan\mp\im)^2}
\right)
\right|^2 , 
\label{eq:BSF_rnl_def}
\end{empheq}
with $S_\ell (\zetaS)$ being the Coulomb Sommerfeld factor for $\ell$-wave processes~\cite{Cassel:2009wt}, 
\begin{align}
S_\ell(\zetaS) 
\equiv 
\left|\dfrac{1}{\ell!} \, e^{\pi \zetaS/2} \, \Gamma(1+\ell -\im\zetaS) \right|^2
=\frac{2\pi\zetaS}{1-e^{-2\pi\zetaS}}~
\prod_{j=1}^{\ell} \left(1+\frac{\zetaS^2}{j^2}\right) .
\label{eq:S_ell}
\end{align}
The $r_{n\ell}  (\zetaS,\zetaB)$ factors of \cref{eq:BSF_rnl_def} will be the focus of the next section.
 
\clearpage
\section{Unitarity violation \label{sec:UnitarityViol}}

The unitarity of the $\mathbb{S}$ matrix implies upper bounds on partial-wave elastic and inelastic cross-sections. 
For inelastic cross-sections, this is 
\begin{align}
\sigma_\ell^{\inel} (k) \leqslant \sigma_\ell^U(k) /4,
\label{eq:UnitarityLimit_Inelastic}
\end{align}
where $\sigma_\ell^U$ has already been defined in \cref{eq:sigmaU}. A more stringent upper bound can be derived as a function of the elastic cross-section~\cite{Hui:2001wy,Flores:2024sfy,Flores:2025uoh}. 

To investigate the BSF cross-sections in relation to the unitarity bounds, we must consider the \emph{inclusive} BSF cross-section that sums over all bound levels formed from the same scattering-state partial wave. Neglecting any phase-space suppression due to a possible non-vanishing mediator mass, $\Phi_n (s)\to 1$, \cref{eq:BSF_sigma_PW_final} yields
\begin{empheq}[box=\myshadebox]{align}
\sigma_\ell^{\BSF} (k) 
\equiv \sum_{n=1+\ell}^\infty 
\sigma_{n\ell}^\BSF (k)
= \dfrac{1}{4} \sigma_\ell^U (k)
\ \arad
\ \left(1-\dfrac{\kappaS}{\kappaB}\right)^2
\times r_\ell (\zetaS,\zetaB) ,
\label{eq:BSF_sigma_lS_total}
\end{empheq}
where
\begin{align}
r_\ell (\zetaS,\zetaB)
\equiv \sum_{n=1+\ell}^\infty r_{n\ell}  (\zetaS,\zetaB) 
\approx \int_{1+\ell}^\infty \dd n \ r_{n\ell}  (\zetaS,\zetaB) ,   
\label{eq:rl_def}
\end{align}
with $r_{n\ell}  (\zetaS,\zetaB)$ given by \cref{eq:BSF_rnl_def}. 
The BSF cross-sections computed according to the above violate the unitarity bound \eqref{eq:UnitarityLimit_Inelastic} if 
$\arad(1-\kappaS/\kappaB)^2 r_\ell (\zetaS,\zetaB)>1$. 
In the following, we aim to estimate $r_\ell (\zetaS,\zetaB)$ using analytical approximations. We will find that for $\kappaS / \kappaB \leqslant 1$, $r_\ell$ grows at low velocities, thereby indicating violation of unitarity at arbitrarily low couplings, $\arad$. 
We emphasize that this corresponds to the limit of massless radiation; for $\mrad \neq 0$, the sum in \cref{eq:BSF_sigma_lS_total} is truncated at some finite level $n$ by phase space, which can soften or curtail the growth.

Before we begin, we note that \cref{app:IdentitiesApprox} collects mathematical identities and approximations that will be useful in the following. 

\subsection
[Kramers-like formulae for individual partial waves: $\alpha_{\scriptscriptstyle{\cal S}}=0$]
{Kramers-like formulae for individual partial waves: $\bm{\aS=0}$  \label{sec:UnitarityViol_alphaS=0}}

\subsubsection{Approximating the cross-sections \label{sec:UnitarityViol_alphaS=0_rnl}}

For $\aS=0$, the quadratic transformation of \cref{eq:2F1_QuadraticTransf} 
allows us to rewrite \cref{eq:BSF_rnl_def} as follows
\begin{align}
\label{eq:rnl_alphaS=0}
&r_{n\ell}  (0,\zetaB) =
\\
&=
\left[ \dfrac{2^{2\ell+3} \ell!}{(2\ell+1)!} \right]^2
\dfrac{n (n+\ell)!}{(n-\ell-1)!}
\left[\dfrac{\zetan^{2\ell+5}}{(1+\zetan^2)^{2\ell+3}}\right]
\times
\left|{}_2F_1 \left[
\dfrac{-n+\ell+1}{2},
\dfrac{n+\ell+1}{2};~
\ell+\dfrac{3}{2};~
\left(
\dfrac{2\zetan}{1+ \zetan^2}
\right)^2
\right] \right|^2  .
\nn
\end{align}
As we shall see, for $\aS=0$ (and in fact for any $\aS <\aB$, cf.~\cref{sec:UnitarityViol_alphaSnot0}), the sum over bound levels, $r_\ell$, is dominated by capture to excited levels, $n \gg \ell$. For large $n$, we set $(n+\ell)! / (n-\ell-1)! \approx n^{2\ell+1}$ as per \cref{eq:Stirling_RatioOfGammas}. Moreover, the Hypergeometric function in \cref{eq:rnl_alphaS=0} can be expanded according to \cref{eq:HypergeometricSeries}, as a power series in its last argument, which is bounded from above, 
$2\zetan / (1+ \zetan^2) \leqslant 1$, 
with its maximum value obtained for $\zetan=1$, while
$2\zetan / (1+ \zetan^2) \ll 1$ 
for both $\zetan \ll 1$ and $\zetan \gg 1$. 
This implies that, apart from a narrow range of $\zetan$ around 1, the power series \eqref{eq:HypergeometricSeries} is dominated by $j \ll [n\pm(\ell+1)]/2$. Using again \cref{eq:Stirling_RatioOfGammas},
we set 
$([\mp n + \ell+1]/2)_j \approx (\mp n/2)^j$, which allows us to approximate ${}_2F_1$ with an ${}_0F_1$ in \cref{eq:rnl_alphaS=0}.\footnote{
We recall here the related limits,  
$\displaystyle \lim_{a\to\infty} 
{}_2F_1 (a,b;c;z/a)={}_1F_1 (b;c;z)$ 
and 
$\displaystyle \lim_{a\to\infty} 
{}_1F_1 (a;c;z/a)={}_0F_1 (c;z)$.
}
Collecting the above, we find that, for $n \gg \ell+1$,
\begin{align}
r_{n\ell}  (0,\zetaB) 
&\approx
\left[ \dfrac{2^{2\ell+3} \ell!}{(2\ell+1)!} \right]^2
n^{2\ell+2}
\left[\dfrac{\zetan^{2\ell+5}}{(1+\zetan^2)^{2\ell+3}}\right]
\times \left|{}_0F_1 \left[
\ell+\dfrac{3}{2};~
-\dfrac{n^2}{4}\left(
\dfrac{2\zetan}{1+ \zetan^2}
\right)^2
\right] \right|^2
\nn \\
&= 
2^6 \zetaB^2
\left[\dfrac{\zetaB^3/n^3}{(1+\zetaB^2/n^2)^{3}}\right]
\times \left|j_{\ell} \left(\dfrac{2\zetaB}{1+\zetaB^2/n^2}\right)
\right|^2 ,
\label{eq:rnl_alphaS=0_nggl}
\end{align}
where $j_\ell$ is the spherical Bessel function, and in the second step, we used the identity~\cite[pg. 38]{Petkovsek:1996aeb}
\begin{align}
{}_0F_1 \left(\ell+\dfrac{3}{2};-\dfrac{x^2}{4} \right)
&= 
\dfrac{(2\ell+1)!}{2^{\ell} \, \ell!}
\, \dfrac{j_\ell (x)}{x^\ell} .
\label{eq:Identity_0F1}
\end{align}
Note that the $\ell$-dependence of \cref{eq:rnl_alphaS=0_nggl} arises only from  $j_\ell$.

\begin{figure}[p!]
\centering
\includegraphics[width=0.45\linewidth]{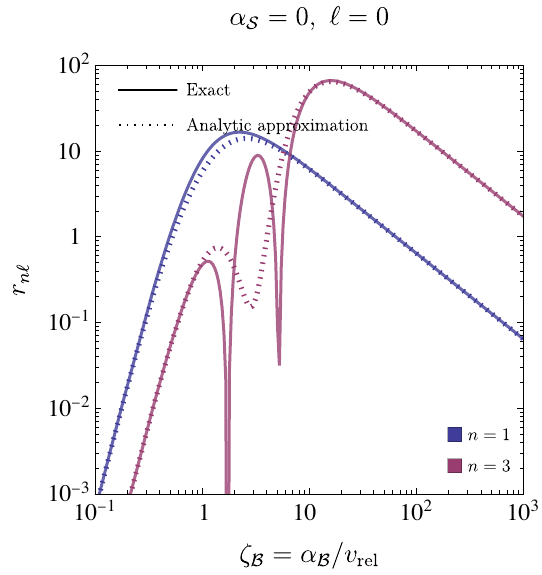}
~~
\includegraphics[width=0.45\linewidth]{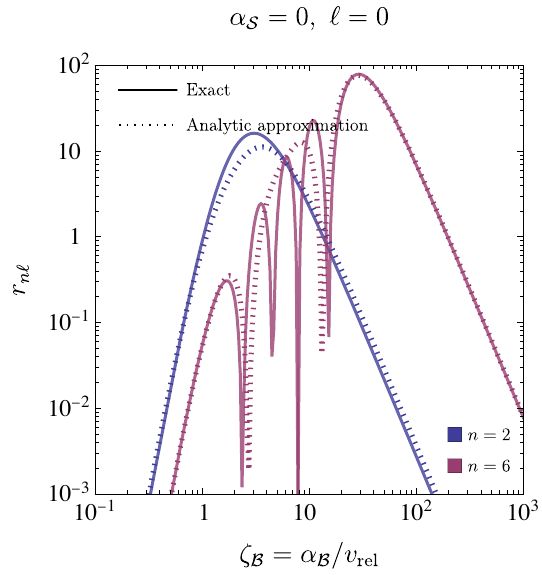}
\caption{\label{fig:rnl_Levels_lambda=0} 
The factors 
$r_{n\ell} \propto \sigma_\ell (k;n) / \sigma_\ell^U (k)$ vs $\zetaB$, for $\aS=0$ and different values of $n$, $\ell$, as denoted in the labels. We compare the original expression \eqref{eq:rnl_alphaS=0} with the analytic approximation \eqref{eq:rnl_alphaS=0_nggl}. While the phase of the oscillations is different, both curves are enveloped by the same function. Larger $n$ implies more oscillations, higher maximum value of $r_{n\ell}$ and larger $\zetaB$ value at which this maximum occurs.
}

\vspace{1em}

\includegraphics[width=0.45\linewidth]{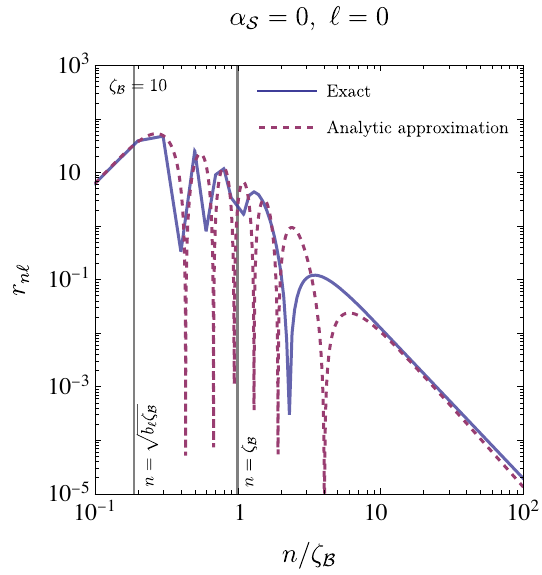}
~~
\includegraphics[width=0.45\linewidth]{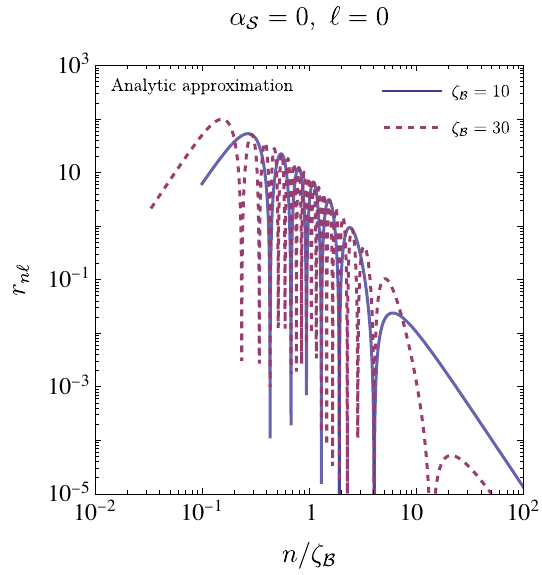}
\caption{\label{fig:rnl_Spectrum_lambda=0}
$r_{n\ell}$ vs $n/\zetaB$, for $\aS=0$ and $\ell=0$.
\emph{Left:}~We compare the original expression \eqref{eq:rnl_alphaS=0}, with the analytic approximation \eqref{eq:rnl_alphaS=0_nggl}. (The curve corresponding to \cref{eq:rnl_alphaS=0} is not smooth as it connects the discrete, physical values of $n$.)  In the range $\sqrt{b_\ell \zetaB}<n<\zetaB$ (encompassed within the vertical gray lines), $r_{n\ell} \propto 1/n$, while for $n>\zetaB$, $r_{n\ell} \propto 1/n^3$ [cf.~\cref{eq:rnl_alphaS=0_nSpectrum}].
\emph{Right:}~The oscillations increase with $\zetaB$. 
}
\end{figure}

Both \cref{eq:rnl_alphaS=0_nggl,eq:rnl_alphaS=0} exhibit oscillations in $n$ and $\zetan$. Although the phase of the oscillations is different, meaning that the discrepancy between the two can be significant for a given $n,\ell$ and $\zetaB$, we consider \eqref{eq:rnl_alphaS=0_nggl} to be a good approximation of \eqref{eq:rnl_alphaS=0}, in the sense that both expressions are enveloped by the same function with respect to $\zetaB$. In \cref{fig:rnl_Levels_lambda=0,fig:rnl_Spectrum_lambda=0}, we compare the two expressions, as well as the spectrum with respect to $n$, for different values of $\zetaB$.

\subsubsection*{Spectrum as a function of  \texorpdfstring{$\bm{n}$}{n}}

We may further simplify \cref{eq:rnl_alphaS=0_nggl}, using the asymptotic forms of $j_\ell$. We shall set
\begin{align}
j_\ell (x) \sim 
\begin{cases}
\dfrac{2^\ell \ell!}{(2\ell+1)!} \, x^\ell  ,   
&  0<x \lesssim 2 b_\ell ,
\\[1em]
\dfrac{1}{x} \, \sin(x-\ell\pi/2) ,    
& 2 b_\ell < x ,
\end{cases}
\label{eq:Bessel_Asymptotics}
\end{align}
with $b_\ell$ chosen to ensure continuity when setting the oscillatory factor to $1/\sqrt{2}$,
\begin{align}
b_\ell \equiv \dfrac{1}{2}  
\left[\dfrac{(2\ell+1)!}{2^{\ell+1/2}\ell!} \right]^{1/(\ell+1)} 
\simeq \dfrac{1+\ell}{e} .
\label{eq:bl_def}
\end{align}
The last approximation is obtained using Stirling's formula. Setting $x=2\zetaB/(1+\zetaB^2/n^2)$, as in \cref{eq:rnl_alphaS=0_nggl}, the intervals of \cref{eq:Bessel_Asymptotics} correspond approximately to 
\begin{subequations}
\label{eq:Bessel_Intervals}
\label[pluralequation]{eqs:Bessel_Intervals}
\begin{align}
&x \in [0,2b_\ell]:&
&\zetaB \lesssim b_\ell 
\quad\text{or}\quad 
\zetaB \gtrsim n^2/b_\ell >n,
\label{eq:Bessel_Intervals_z-small}
\\
&x \in (2b_\ell,\infty):&
&b_\ell < \zetaB < n^2/b_\ell .
\label{eq:Bessel_Intervals_z-large}
\end{align}
\end{subequations}
Combining \cref{eq:rnl_alphaS=0_nggl,eq:Bessel_Asymptotics,eq:Bessel_Intervals}, we determine the spectrum as a function of $n$, in different velocity ranges:
\begin{description}
\item[Large velocities, $\bm{\zetaB < b_\ell}$~:] 
\begin{subequations}
\label{eq:rnl_alphaS=0_nSpectrum}
\label[pluralequation]{eqs:rnl_alphaS=0_nSpectrum}
\begin{align}
r_{n\ell} \approx
8\left(\dfrac{\zetaB}{b_\ell}\right)^{2\ell+2}
\dfrac{(\zetaB/n)^3}{(1+\zetaB^2/n^2)^{2\ell+3}}  
\approx
\dfrac{8\zetaB^{2\ell+5}}{b_\ell^{2\ell+2}}  
\dfrac{1}{n^3} .
\label{eq:rnl_alphaS=0_nSpectrum_VerySmallZetaB}
\end{align}

\item[Intermediate velocities, $\bm{b_\ell < \zetaB < (1+\ell)^2/b_\ell}$~:]  
\begin{align}
r_{n\ell} 
\approx 2^4 \ \dfrac{(\zetaB/n)^3}{1+\zetaB^2/n^2}
\sin^2 \left(\dfrac{2\zetaB}{1+\zetaB^2/n^2} -\dfrac{\ell \pi}{2}\right)
\sim \dfrac{2^3 \zetaB^3}{n^3} ,
\label{eq:rnl_alphaS=0_nSpectrum_SmallZetaB}
\end{align}
where in the last step, we replaced $\sin^2(\cdot) \to 1/2$.

\item[Low velocities, $\bm{(1+\ell)^2/b_\ell < \zetaB}$~:] 
\begin{align}
r_{n\ell} \approx 
\begin{cases}
8\left(\dfrac{\zetaB}{b_\ell}\right)^{2\ell+2}
\dfrac{(\zetaB/n)^3}{(1+\zetaB^2/n^2)^{2\ell+3}}
\simeq
\dfrac{8 \, n^{4\ell+3}}{b_\ell^{2\ell+2} \zetaB^{2\ell+1}},
~~~~& 1+\ell \leqslant n < \sqrt{b_\ell \zetaB},
\\[1em]
2^4 \ \dfrac{(\zetaB/n)^3}{1+\zetaB^2/n^2}
\ \sin^2 \left(\dfrac{2\zetaB}{1+\zetaB^2/n^2} -\dfrac{\ell \pi}{2}\right)
\sim \dfrac{2^3 \zetaB}{n},
~~~~& \sqrt{b_\ell \zetaB} < n < \zetaB, 
\\[1em]
2^4 \ \dfrac{(\zetaB/n)^3}{1+\zetaB^2/n^2}
\ \sin^2 \left(\dfrac{2\zetaB}{1+\zetaB^2/n^2} -\dfrac{\ell \pi}{2}\right)
\sim \dfrac{2^3 \zetaB^3}{n^3},
~~~~& \zetaB < n .
\end{cases}
\label{eq:rnl_alphaS=0_nSpectrum_LargeZetaB}
\end{align}
where as before, we replaced $\sin^2(\cdot) \to 1/2$.

\end{subequations}
\end{description}

\subsubsection{Summing over bound levels \label{sec:UnitarityViol_alphaS=0_Kramers}}

From \cref{eq:rl_def,eq:rnl_alphaS=0_nggl}, setting \begin{align}
x=\dfrac{2\zetaB}{1+\zetaB^2/n^2} 
\quad \Leftrightarrow \quad
n=\sqrt{\dfrac{x}{2\zetaB-x}} \, \zetaB,
\label{eq:xvar_def}
\end{align} 
we find
\begin{align}
r_\ell (0,\zetaB) \approx  
8\zetaB
\int_{2\zetaB/[1+\zetaB^2/(1+\ell)^2]}^{2\zetaB} 
\dd x \ x \ j_\ell^2 (x) .
\label{eq:rl_alphaS=0_integral}
\end{align}
The integral can be computed analytically and expressed in terms of Hypergeometric functions, 
\begin{align}
&r_\ell (0,\zetaB) 
\approx
\dfrac{4\zetaB}{1+\ell}
\left(\dfrac{2^\ell \ell!}{(2\ell+1)!}\right)^2
\times
\Bigg\{
(2\zetaB)^{2+2\ell}
{}_2F_3 \left[
1+\ell,1+\ell;
~~\dfrac{3}{2}+\ell,2+\ell,2+2\ell;
~~-(2\zetaB)^2 
\right]
\nn \\
&
-\left(\dfrac{2\zetaB}{1+\zetaB^2/(1+\ell)^2}\right)^{2+2\ell}
{}_2F_3 \left[
1+\ell,1+\ell;
~~\dfrac{3}{2}+\ell,2+\ell,2+2\ell;
~~-\left(\dfrac{2\zetaB}{1+\zetaB^2/(1+\ell)^2}\right)^2 
\right]
\Bigg\}.
\end{align}
It is, however, more instructive to consider the asymptotic forms of $j_\ell$ at small and large $x$, given by  \cref{eq:Bessel_Asymptotics}, to obtain simpler approximate expressions. Combining \cref{eq:rl_alphaS=0_integral,eq:Bessel_Asymptotics,eq:Bessel_Intervals}, we discern the following cases:
\begin{subequations}
\label{eq:rl_alphaS=0}
\label[pluralequation]{eqs:rl_alphaS=0}
\begin{description}
\item[Large velocities, 
$\bm{\zetaB \lesssim b_\ell}$~:] 
\begin{align}
r_\ell(0,\zetaB) 
\approx 
8\zetaB 
\ \dfrac{1}{2(2b_\ell)^{2+2\ell}}
\int_{2\zetaB/[1+\zetaB^2/(1+\ell)^2]}^{2\zetaB} 
\dd x \ x^{1+2\ell}
\approx
\dfrac{4\zetaB^{5+2\ell}}{(1+\ell)^2 b_\ell^{2+2\ell}},
\label{eq:rl_alphaS=0_zetaBsmall}
\end{align}
where we kept the lowest order term in $\zetaB/(1+\ell)$. 
\item[Intermediate velocities, $\bm{b_\ell < \zetaB < (1+\ell)^2/b_\ell \simeq (1+\ell) e}$~:]
\begin{align}
r_\ell(0,\zetaB) 
\approx 
8\zetaB 
\int_{2\zetaB/[1+\zetaB^2/(1+\ell)^2]}^{2\zetaB} 
\dd x \ \dfrac{\sin^2(x-\ell\pi/2)}{x}
\approx 
4\zetaB \ln \left[1+ \dfrac{\zetaB^2}{(1+\ell)^2}\right] .
\label{eq:rl_alphaS=0_zetaBmed}
\end{align}
where we replaced the oscillatory factor $\sin^2(\cdot) \to 1/2$.
\item[Low velocities, $\bm{(1+\ell)^2/b_\ell \lesssim \zetaB}$~:]
\begin{align}
r_\ell(0,\zetaB) 
&\approx 
8\zetaB \left[
\dfrac{1}{2(2b_\ell)^{2+2\ell}}
\int_{2\zetaB/[1+\zetaB^2/(1+\ell)^2]}^{2b_\ell} 
\dd x \ x^{1+2\ell}
+
\int_{2b_\ell}^{2\zetaB} 
\dd x \ \dfrac{\sin^2(x-\ell\pi/2)}{x}
\right]
\nn \\
&\approx
4\zetaB \left[
\dfrac{1}{2(1+\ell)} \left[
1-\left(\dfrac{(1+\ell)^2/b_\ell}{\zetaB}\right)^{2\ell + 2}
\right]
+\ln \left(\dfrac{\zetaB}{b_\ell}\right)
\right],
\label{eq:rl_alphaS=0_zetaBlarge}
\end{align}
where, as above, we replaced $\sin^2(\cdot)\to1/2$, and kept the lowest order term in $(1+\ell)/\zetaB$. 

The two terms in \cref{eq:rl_alphaS=0_zetaBlarge} arise from 
$n\in \big[1+\ell, \sqrt{b_\ell \zetaB}\,\big)$ and 
$n \in \big[\!\sqrt{b_\ell \zetaB},\infty\big)$, respectively. The latter increases faster with $\zetaB$, exhibits only a mild dependence on $\ell$, and dominates in the entire relevant $\zetaB$ range. 

The sub-intervals $\sqrt{b_\ell \zetaB} < n < \zetaB$ and $\zetaB < n$ 
(corresponding to 
$\frac{2\zetaB}{1+\zetaB/b_\ell} < x < \zetaB$
and
$\zetaB < x < 2\zetaB$)
--- which are characterized by different spectral scalings, $r_{n\ell} \propto 1/n$ and $1/n^3$ respectively, as shown in~\cref{eq:rnl_alphaS=0_nSpectrum} --- contribute with ratio $\ln[\zetaB/(2b_\ell)]:\ln 2$. That is, the interval $\sqrt{b_\ell \zetaB} < n < \zetaB$ dominates at sufficiently low velocities, $\zetaB \gtrsim 4b_\ell$. 
\end{description}
\end{subequations}

\begin{figure}[ht!]
\centering
\includegraphics[width=0.6\linewidth]{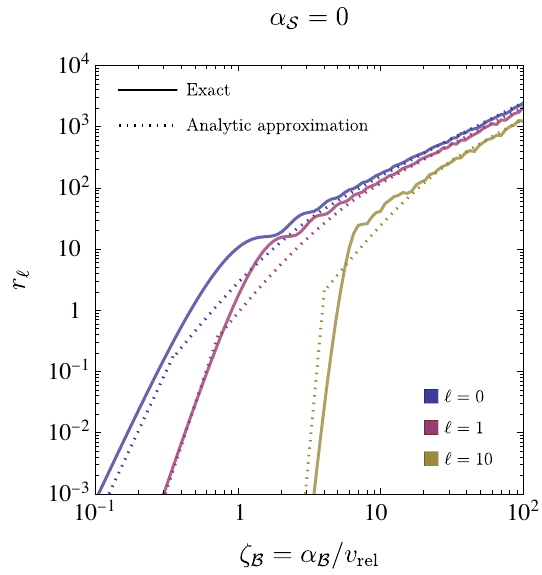}
\caption{$r_\ell = \sum_{n=1+\ell}^\infty r_{n\ell}$  vs $\zetaB$, for different values of $\ell$. At $\zetaB \gg 1+\ell$, $r_\ell$ depends only mildly on $\ell$. 
We compare $r_\ell$ calculated numerically using the full expression \eqref{eq:rnl_alphaS=0} for $r_{n\ell}$, with the analytic approximation \eqref{eq:rl_alphaS=0}.
The growth of $r_\ell$ with $\zetaB$ implies that unitarity is violated at sufficiently low velocities, for arbitrarily low couplings, $\arad$ [cf.~\cref{eq:BSF_sigma_lS_total}].
\label{fig:rl_lambda=0}
} 
\end{figure}

Summarizing the above,
\begin{empheq}[box=\myshadebox]{align}
r_\ell (0,\zetaB) \approx 
\begin{cases}
\dfrac{4\zetaB^{2\ell+5}}{(1+\ell)^2 b_\ell^{2\ell+2}},
& \zetaB \lesssim b_\ell  ,
\\[1em]
4 \zetaB \ln \left[1+\dfrac{\zetaB^2}{(1+\ell)^2}\right],
& b_\ell < \zetaB < \dfrac{(1+\ell)^2}{b_\ell} ,      
\\[1em]
4\zetaB \left\{
\dfrac{1}{2(1+\ell)}
\left[1-\left(\dfrac{(1+\ell)^2/b_\ell}{\zetaB}\right)^{2\ell+2}\right]
+ \ln\left(\dfrac{\zetaB}{b_\ell}\right)
\right\},
& \dfrac{(1+\ell)^2}{b_\ell} \lesssim \zetaB  .
\end{cases}
\label{eq:rl_alphaS=0_together}
\end{empheq}
We sketch \cref{eq:rl_alphaS=0_together} in \cref{fig:rl_lambda=0}, where we compare it with the result obtained using the exact formula \eqref{eq:rnl_alphaS=0}. Evidently, the factors $r_\ell$ increase with $\zetaB$ for all partial waves. Combined with \cref{eq:BSF_sigma_lS_total}, this implies that unitarity is violated even for arbitrarily small couplings, $\arad$, at sufficiently low velocities,
\begin{align}
4 \arad \, \zetaB \ln(\zetaB/b_\ell) \gtrsim  1.   
\label{eq:UniViolation_Regime_alphaS=0}
\end{align}

\Cref{eq:rl_alphaS=0_together} is the analogue of Kramers formula in QED~\cite{Kramers1923XCIIIOT}, but \emph{for individual $\ell$ values}. In fact, the dependence of $r_{n\ell}$ on $n$ for $\sqrt{b_\ell \zetaB} < n <\zetaB$ is the same as in QED, $r_{n\ell} \propto 1/n$ [cf.~\cref{eq:rnl_alphaS=0_nSpectrum}], and the corresponding contribution to $r_\ell$ is the analogue of Kramers logarithm. However, there is an extra overall $\zetaB$ factor with respect to QED, leading to significant growth of $r_\ell$ with $\zetaB$. Further summing \eqref{eq:rl_alphaS=0_together} over $\ell$, we find
\begin{align}
\sum_{\ell} 
\dfrac
{r_\ell \, \sigma_\ell^U}
{\sigma_{\ell=0}^U} 
&\approx 
\int_0^\infty \dd\ell \, (2\ell+1) \, r_\ell 
\approx 4\zetaB \times (0.4+2.5)\zetaB^2
\approx 12 \, \zetaB^3,  
\label{eq:SumOverPW_alphaS=0}
\end{align}
where we kept the leading contributions at large $\zetaB$,  arising from the lower and intermediate intervals in \cref{eq:rl_alphaS=0_together}, corresponding to 
$0\leqslant 1+\ell < \zetaB/e$ and $\zetaB/e\lesssim 1+\ell \lesssim e\zetaB$, respectively, and used numerical fits for the final estimate. \Cref{eq:SumOverPW_alphaS=0} shows very significant growth with decreasing velocity compared to QED.

\medskip

To estimate the range of $n$ that must be considered to achieve precision $\mathtt{p}$ in $r_\ell$, we require
\begin{align} 
\int_{n_{\max}(\mathtt{p})}^\infty \dd n \ r_{n\ell}  (0,\zetaB)
=8\zetaB \int_{x_{\max}(\mathtt{p})}^{2\zetaB} 
\dd x \ x \ j_\ell^2 (x)
=\mathtt{p} \, r_\ell  (0,\zetaB),
\end{align}
where $x_{\max}(\mathtt{p})$ and $n_{\max}(\mathtt{p})$ are the maximum $x$ and $n$ that we need to consider, and are related via \cref{eq:xvar_def}. For $\zetaB > b_\ell$, we find 
$x_{\rm max}(\mathtt{p}) =2\zetaB/(\zetaB/b_\ell)^{\mathtt{p}}$, or
\begin{align}
n_{\max}(\mathtt{p}) 
\approx \dfrac{\zetaB}{\sqrt{(\zetaB/b_\ell)^\mathtt{p}-1}}
\approx \dfrac{\zetaB}{\sqrt{\mathtt{p}\ln(\zetaB/b_\ell)}}, 
\label{eq:Nmax}
\end{align}
where we expanded in $\mathtt{p} \ll 1$, and kept the dominant term. For example, for precision $\mathtt{p} = 1\%$, we must consider $n\leqslant n_{\max} \approx 10 \zetaB/\sqrt{\ln (\zetaB/b_\ell)}$. This is similar to the estimate of  Ref.~\cite[footnote~1]{Binder:2023ckj}.


\clearpage
\subsection
[Kramers-like formulae for individual partial waves: $\alpha_{\scriptscriptstyle{\cal S}}\neq0$]
{Kramers-like formulae for individual partial waves: $\bm{\aS\neq0}$  \label{sec:UnitarityViol_alphaSnot0}}

We define the ratio of the scattering-state to bound-state Bohr momenta,
\begin{align}
\lambda 
\equiv \kappaS / \kappaB 
= \aS/\aB 
=\zetaS / \zetaB.
\label{eq:lambda_def}
\end{align}
In the following, we consider $r_{n\ell}  (\zetaS,\zetaB)$ and $r_\ell(\zetaS,\zetaB)$ at constant $\lambda$, noting that in certain theories such as QCD, the running of the couplings renders $\lambda$ (mildly) velocity dependent (see e.g.~\cite{Harz:2018csl,Harz:2019rro}).

\Cref{fig:rnl_Levels_l=0} illustrates the dependence of $r_{n\ell}$ on $\vrel$, for different values of $\lambda$, and various levels $n$. For $\lambda <0$, the behavior is determined primarily by the interplay between the Sommerfeld suppression factor $S_0 (\zetaS)$ and the exponential factor $e^{-4\zetaS\arccot(\zetaB/n)}$ in \cref{eq:BSF_rnl_def}. The latter causes $r_{n\ell}$ to increase as $\vrel$ decreases, until this factor saturates to $e^{-4\zetaS(n/\zetaB)} = e^{+4n|\lambda|}$ at $\zetaB/n \gg 1$ and the growth of $r_{n\ell}$ is overtaken by the exponential Sommerfeld suppression, $S_0 (\zetaS) \propto e^{-2\pi |\zetaS|}$. As a result, higher $n$ levels contribute larger peaks occurring at larger $\zetaB$. On the other hand, for $\lambda >0$, the $r_{n\ell}$ factors saturate to a constant value at low $\vrel$, owing to the Sommerfeld enhancement due to the attractive interaction in the scattering state. Higher $n$ are important for $\lambda \leqslant 1$.\footnote{\label{foot:lambda=1}
For $\lambda=1$, the monopole modes vanish due to the orthogonality of the wavefunctions. Having factored the $(1-\lambda)^2$ suppression out of $r_{n\ell}$ in \cref{eq:BSF_rnl_def}, we are still interested in understanding the behavior of $r_{n\ell}$ around $\lambda \sim 1$, as this allows for more thorough insight into the properties of $r_{n\ell}$ and $r_\ell$, and is also relevant for higher multipole transitions, which do not vanish at $\lambda=1$, and exhibit similar qualitative features.}

\subsubsection{Asymptotic scaling at low velocities, \texorpdfstring{$\zetaB, |\zetaS| \gg n$}{zetaB, zetaS >> n} \label{sec:UnitarityViol_alphaSnot0_Asymptote_LowV}}

\begin{figure}[h!]
\centering
\includegraphics[width=0.32\linewidth]{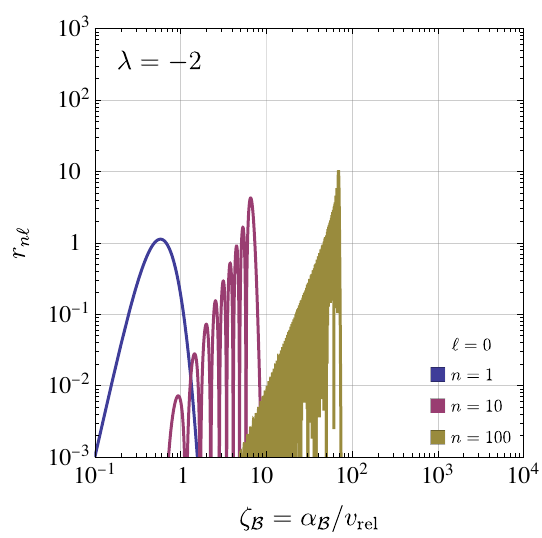}
\includegraphics[width=0.32\linewidth]{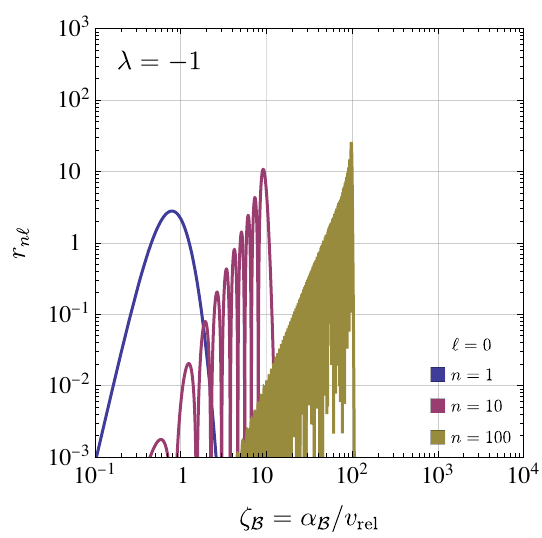}
\includegraphics[width=0.32\linewidth]{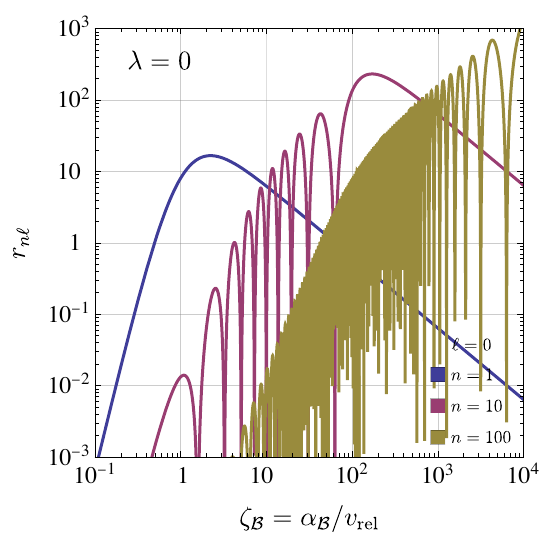}
\\[0.5em]
\includegraphics[width=0.32\linewidth]{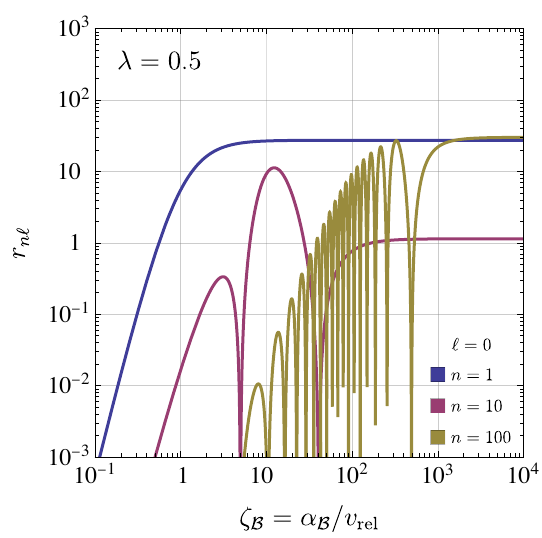}
\includegraphics[width=0.32\linewidth]{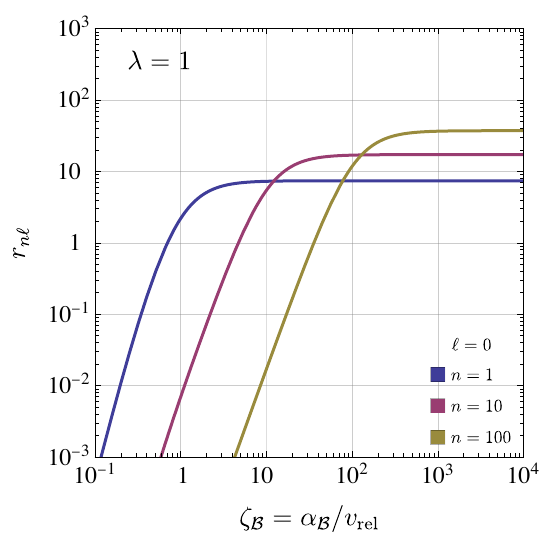}
\includegraphics[width=0.32\linewidth]{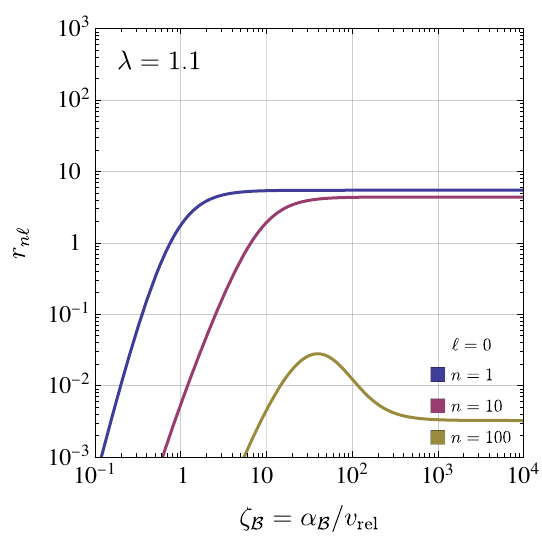}
\caption{\label{fig:rnl_Levels_l=0}
$r_{n\ell}$ vs $\zetaB$, for $\ell=0$, and different values of  $\lambda=\aS/\aB$ and $n$, as denoted in the labels. The asymptotic behavior at low velocities and large $n$, $\zetaB , |\zetaS| \gg n \gg \ell+1$ does not depend on $\ell$ [cf.~\cref{eq:rnl_LowVelocities_lambdaCases}]. (This does not hold for $\lambda=0$.)  
\emph{Upper row:} $\lambda \leqslant 0$; the excited levels contribute at lower velocities and at increasing strength. The Sommerfeld suppression for $\lambda<0$ gives rise to a sharp cutoff at low velocities. 
\emph{Lower row:} $\lambda > 0$. Capture into excited levels is significant for $\lambda \lesssim 1$, but becomes exponentially suppressed for $\lambda\gg1$. 
(For $\lambda=1$, see \cref{foot:lambda=1}.) 
The Sommerfeld enhancement implies that $r_{n\ell}$ asymptotes to a constant value at low velocities, for $\lambda >0$ and any $\ell$. However, $\lambda\neq 1$ gives rise to non-monotonic dependence on $\vrel$. 
}
\end{figure}

We expand the hypergeometric function in \cref{eq:BSF_rnl_def} as a finite polynomial using the series representation \eqref{eq:HypergeometricSeries}. Given that the summation index $j$ spans the range $1+\ell+j \in [1+\ell, n]$, the ratio of $\Gamma$ functions originating from the second argument of ${}_2F_1$ can be approximated using the Stirling formula of \cref{eq:Stirling_Gamma_Im>Re_Ratio}, to be $\Gamma(1+\ell+j+\im\zetaS) / \Gamma(1+\ell+\im\zetaS) \approx (\im \zetaS)^j$. 
This approximation is justified because the imaginary part of the Gamma function arguments dominates the real counterparts, i.e., $\zetaS \gg n \geqslant 1+\ell+j \geqslant 1+\ell$.
This allows us to transform the ${}_2F_1$ expansion into an ${}_1F_1$ function, 
\begin{align}
{}_2F_1 \left(
-n+\ell+1,
1+\ell +\im \zetaS;~
2\ell+2;~
-\dfrac{4\im \zetan}{(\zetan-\im)^2}
\right) 
\approx
{}_1F_1 \left( 
-n+1+\ell; ~ 
2\ell+2; ~
4 \lambda n
\right).
\label{eq:2F1_Asymptotic_LowVrel}
\end{align}
Setting also $\arccot (\zetan) \approx 1/\zetan$ for $\zetan \gtrsim 1$, and 
\begin{align}
S_\ell(\zetaS) \approx 
\dfrac{2\pi|\zetaS|^{2\ell+1} }{(\ell!)^2} \, e^{2\pi \min(\zetaS,0)} ,
\qquad \text{for} \quad |\zetaS| \gtrsim 1+\ell,
\label{eq:SommerfeldFactor_LargeZetaS}
\end{align}
\cref{eq:BSF_rnl_def} becomes
\begin{align}
r_{n\ell} 
&\approx       
(2\pi)\left[\frac{2^{2\ell+3}}{(2\ell+1)!}\right]^2
\frac{n^{2\ell+2} (n+\ell)!}{(n-\ell-1)!} 
|\lambda|^{2\ell+1}
e^{-4\lambda n}
\left|
{}_1F_1 \left( 
-n+\ell+1; ~ 
2\ell+2; ~
4 \lambda n
\right)
\right|^2
\nn \\
&\times
e^{2\pi \min (\zetaS,0)} .
\label{eq:rnl_LowVelocities}
\end{align}
The last factor in the above encapsulates the velocity scaling in this regime: for $\aS>0$, $r_{n\ell}$ asymptote to a constant at low $\vrel$, while for $\aS<0$ they decay exponentially with decreasing $\vrel$.  The confluent Hypergeometric function ${}_1F_1$ in \cref{eq:rnl_LowVelocities} is related to the Laguerre polynomials~\cite[13.6.19]{NIST:DLMF}
\begin{equation}
{}_1F_1  \left( 
1+\ell-n; ~ 
2\ell+2; ~
4 \lambda n
\right) = 
\dfrac{(n-\ell-1)! \, (2\ell+1)!}{(n+\ell)!}
\ L_{n-\ell-1}^{(2\ell+1)} (4\lambda n),
\label{eq:Hypergeometric_Laguerre}
\end{equation}
thus, at $n\gg 1+\ell$, \cref{eq:rnl_LowVelocities} becomes
\begin{align}
r_{n\ell} 
&\approx     
2^5\pi 
\, n
\, \left[
(4|\lambda|)^{\ell+1/2}
\, e^{-2\lambda n}
\, L_{n-\ell-1}^{(2\ell+1)}(4 \lambda n)
\right]^2
\times
e^{2\pi \min (\zetaS,0)} .
\label{eq:rnl_LowVelocities_Laguerre}
\end{align}
The behavior of the Laguerre polynomials, $L_{n-\ell-1}^{(2\ell+1)} (4\lambda n)$, varies qualitatively as follows~\cite{Temme:1990}
\begin{align*}
\begin{array}{lll}
(i)   & \lambda <0 & \text{monotonic region,}  \\
(ii)  & 0<\lambda < 1 &\text{oscillatory region,}  \\
(iii) & \lambda =1 & \text{turning point,} \\
(iv)  & 1<\lambda & \text{monotonic region.}
\end{array}
\end{align*}
For $n \gg 1+\ell$, the asymptotic forms of $L_{n-\ell-1}^{(2\ell+1)} (4\lambda n)$ are reviewed in \cref{app:Laguerre}. Considering \cref{eq:Laguerre_Asymptotes_Bessel,eq:Laguerre_Asymptotes_Airy}, and replacing the oscillatory factors $\sin^2(\cdot) \to 1/2$, \cref{eq:rnl_LowVelocities_Laguerre} becomes
\begin{align}
r_{n\ell}  (\zetaS,\zetaB) \sim 
\begin{cases}
\dfrac{4}{\sqrt{|\lambda| (1+|\lambda|)}} 
\ e^{4n \qty[\!\sqrt{|\lambda|(1+|\lambda|)}+\arcsinh \sqrt{|\lambda|} \, ]} 
\ e^{2\pi \zetaS}, 
&  \lambda <0,
\\[1em]
\dfrac{8}{\sqrt{\lambda (1-\lambda)}},
&  0 < \lambda < 1,
\\[1em]
\left(\dfrac{2^5\pi}{18^{2/3} [\Gamma(2/3)]^2}\right) n^{1/3} \simeq 8n^{1/3},
&  \lambda = 1,
\\[1em]
\dfrac{4}{\sqrt{\lambda (\lambda-1)}}  
\, e^{-4n\qty[\!\sqrt{\lambda(\lambda-1)}-\arccosh\sqrt{\lambda}]} , 
&  1 < \lambda.
\end{cases}
\label{eq:rnl_LowVelocities_lambdaCases}
\end{align}
Note that \cref{eq:rnl_LowVelocities_lambdaCases} has no $\ell$ dependence except for its range of validity, $\zetaB, |\zetaS| \gg n \gg \ell+1$. Moreover, for $\lambda <0$, it suggests a sharp cutoff at
\begin{align}
\zetaB \gtrsim 
\dfrac{n}{\sqrt{|\lambda|}} \times 
\dfrac{2}{\pi}
\qty(
\!\sqrt{1+|\lambda|}+
\dfrac{\arcsinh \sqrt{|\lambda|} }{\sqrt{|\lambda|}}
) 
= \dfrac{n}{\sqrt{|\lambda|}} \times 
\dfrac{4}{\pi}
\qty(1+\dfrac{|\lambda|}{6}+{\cal O}(\lambda^2)),   
\label{eq:lambda<0_zetaB_cutoff}
\end{align}
as can be observed in \cref{fig:rnl_Levels_l=0}.

\subsubsection{Asymptotic scaling at highly excited states, \texorpdfstring{$n \gg \zetaB$}{n>>zetaB} \label{sec:UnitarityViol_alphaSnot0_Asymptote_nLarge}}

Highly excited states, $n \gg \zetaB$, have not reached their asymptotic value computed in \cref{sec:UnitarityViol_alphaSnot0_Asymptote_LowV}. In this regime, for $n\to \infty$, the hypergeometric function of \cref{eq:BSF_rnl_def} becomes insensitive to $n$,
\begin{align}
{}_2F_1 \left(
-n+\ell+1,
1+\ell +\im \zetaS;~
2\ell+2;~
-\dfrac{4\im \zetan}{(\zetan-\im)^2}
\right) 
~\approx~~
&{}_2F_1 \left(
-n+\ell+1,
1+\ell +\im \zetaS;~
2\ell+2;~
4\im \zetan
\right)
\nn \\
\overset{n\to \infty}{\longrightarrow}~
&{}_1F_1 \left( 
1+\ell+\im \zetaS; ~ 
2\ell+2; ~
-\im 4\zetaB
\right).
\label{eq:2F1_Asymptotic_nLarge}
\end{align}
Setting now $\arccot (\zetan) \simeq \pi/2$ in \cref{eq:BSF_rnl_def}, the factor $e^{-4\zetaS \arccot(\zetan)}$ exactly counteracts the exponential Sommerfeld suppression of \cref{eq:SommerfeldFactor_LargeZetaS} for $\lambda <0$, while it introduces an exponential suppression for $\lambda >0$. Making other evident relevant approximations, \cref{eq:BSF_rnl_def} for $n \gg \zetaB$ and $|\zetaS| \gtrsim 1+\ell$ becomes
\begin{align}
r_{n\ell}  (\zetaS,\zetaB) 
&\sim 
\left[\dfrac{2^{2\ell+3} \ell!}{(2\ell+1)!} \right]^2
\zetaB^{2\ell+2} 
\left(\dfrac{\zetaB}{n}\right)^3
\times
e^{-2\pi\zetaS}
S_\ell(\zetaS)
\times 
|{}_1F_1 (
1+\ell+\im \zetaS;
~2\ell+2;
~ -\im 4\zetaB)|^2 .
\label{eq:rnl_nLarge_inter}
\end{align}
To further simplify the velocity dependence of \cref{eq:rnl_nLarge_inter}, we employ the integral representation of ${}_1F_1$ and apply the method of steepest descent. The details are presented in \cref{app:1F1}.  Considering \cref{eq:1F1_summary_largeZetaS}, the above becomes
\begin{align}
r_{n\ell}  (\zetaS,\zetaB) 
&\sim 
8\left(\dfrac{\zetaB}{n}\right)^3
\times 
\begin{cases}
\dfrac{1}{\sqrt{1+|\lambda|}},
& \lambda < 0,
\\[1em]
\dfrac{1}{\sqrt{1-\lambda}},
& 0< \lambda < 1,
\\[1em]
\zetaB^{1/3},
& \lambda = 1,
\\[1em]
\dfrac{
\exp\qty[-4\zetaB 
\qty(\lambda \arccos(1/\sqrt{\lambda})-\sqrt{\lambda-1})]}
{2\sqrt{\lambda-1}},
& \lambda > 1.
\end{cases}
\label{eq:rnl_nLarge}
\end{align}
While \cref{eq:rnl_nLarge} holds asymptotically in the limit $n\to \infty$, we shall use it to estimate $r_{n\ell}$ in the range $n \gtrsim \zetaB$ imposed by the first approximation in \cref{eq:2F1_Asymptotic_nLarge}. For $\lambda <0$, however, the Sommerfeld suppression introduces a sharp cutoff roughly at $\zetaB \gtrsim n / \sqrt{|\lambda|}$ (cf.~\cref{eq:lambda<0_zetaB_cutoff}); we thus take $n > \zetaB \max(1,\sqrt{|\lambda|})$.
Considering this, and upon summation over bound levels, the scaling of \cref{eq:rnl_nLarge} with $n$ implies a growth factor at low velocities,
$\sim 8\int_{\sqrt{|\lambda|}\zetaB}^\infty \dd n\ (\zetaB/n)^3 = 4\zetaB/|\lambda|$ for $\lambda <-1$ and
$\sim 8\int_{\zetaB}^\infty\dd n\ (\zetaB/n)^3 = 4\zetaB$ for $\lambda \geqslant-1$.

\subsubsection{Summing over bound levels \label{sec:UnitarityViol_alphaSnot0_Kramers}}

Using the analytical approximations \cref{eq:rnl_LowVelocities_lambdaCases,eq:rnl_nLarge}, we shall now estimate the contribution to $r_\ell (\zetaS,\zetaB)$ from $n\in [1+\ell,\min (\zetaB,|\zetaS|) )$ and $n\in (\zetaB,\infty)$, respectively, assuming always $\zetaB \gg 1+\ell$. 

\begin{subequations}
\label{eq:rl_zetaSneq0}
\label[pluralequation]{eqs:rl_zetaSneq0}
\setlist[description]{}
\begin{description}
\item[$1\ll\lambda:$] 
Given the exponential suppression with $n$, found in \cref{eq:rnl_LowVelocities_lambdaCases}, which becomes steeper for larger $\lambda$, it suffices to consider only the lowest-lying level, $n=1+\ell$, for a given $\ell$. From \cref{eq:BSF_rnl_def}, we find
\begin{align}
r_\ell(\zetaS,\zetaB) \sim 
2^{2\ell+5} \sqrt{\pi (1+\ell)}
\, \lambda^{2\ell+1} 
\, e^{-2(1+\ell)(2\lambda-1)},
\label{eq:rl_lambda_gg_1}
\end{align}
where we took into account that $\zetaB,\zetaS \gg 1+\ell$, and used the Stirling approximation on the $\ell$-dependent factors to simplify the final expression. By continuity, highly-excited states, $n> \zetaB$, are entirely negligible, as can be confirmed from \cref{eq:rnl_nLarge}.

\item[$1 \lesssim \lambda:$] If $\lambda$ is not much larger than 1, and $\ell$ is not too large, then the exponential suppression with $n$ of \cref{eq:rnl_LowVelocities_lambdaCases} is mild, and several levels contribute significantly to $r_\ell$. Integrating over $n$, we find
\begin{align}
r_\ell(\zetaS,\zetaB) \sim 
\int_{1+\ell}^{\zetaB} \dd n \, r_{n\ell}  (\zetaS,\zetaB)
=
\dfrac
{e^{
-4(1+\ell) 
\qty[\!\sqrt{\lambda(\lambda-1)}-\arccosh \sqrt{\lambda}]
}}
{
\sqrt{\lambda(\lambda-1)} \,
\qty[\!\sqrt{\lambda(\lambda-1)}-\arccosh \sqrt{\lambda}]
},
\label{eq:rl_lambda_simeq_1}
\end{align}
where we dropped the term decaying with $\zetaB$. The integration performed in \cref{eq:rl_lambda_simeq_1} is justified if 
$4(1+\ell) 
\qty[\!\sqrt{\lambda(\lambda-1)}-\arccosh \sqrt{\lambda}] \lesssim 1$. 
For larger $\ell$ and/or $\lambda$, \cref{eq:rl_lambda_gg_1} provides a better estimate. As in the case of $1\ll\lambda$, the contribution of highly excited states, $n>\zetaB$, is entirely negligible.
 
\item[$\lambda = 1:$] Integrating over $n$ yields a contribution growing with $\zetaB$,
\begin{align}
r_\ell(\zetaS,\zetaB) 
&\sim 
\int_{1+\ell}^{\zetaB} \dd n \, 
[r_{n\ell}  (\zetaS,\zetaB)]_{\rm \cref{eq:rnl_LowVelocities_lambdaCases}}
+
\int_{\zetaB}^\infty \dd n \, 
[r_{n\ell}  (\zetaS,\zetaB)]_{\rm \cref{eq:rnl_nLarge}}
\nn \\
&\approx (6+4) \zetaB^{4/3}
= 10\zetaB^{4/3}.
\label{eq:rl_lambda=1}
\end{align}
The two regimes, $n< \zetaB$ and $n>\zetaB$, yield contributions that grow at the same rate at low velocities, with the former being somewhat more dominant. 

\item[$0 < \lambda < 1$:] Similarly to above, we find
\begin{align}
r_\ell(\zetaS,\zetaB) 
&\sim 
\int_{1+\ell}^{\lambda\zetaB} \dd n 
\, [r_{n\ell}  (\zetaS,\zetaB)]_{\rm \cref{eq:rnl_LowVelocities_lambdaCases}}
+
\int_{\zetaB}^\infty \dd n \, 
[r_{n\ell}  (\zetaS,\zetaB)]_{\rm \cref{eq:rnl_nLarge}}
\nn \\
&\sim 
4\zetaB 
\ \dfrac{2\sqrt{\lambda}+1}{\sqrt{1-\lambda}},
\label{eq:rl_0<lambda<1}
\end{align}
where we kept only the modes growing with $\zetaB$.

\item[$\lambda < 0:$] The exponential growth with $n$, indicated in \cref{eq:rnl_LowVelocities_lambdaCases}, is counteracted by the exponential decay due to $\zetaS<0$. The balance is sensitive to the range of $n$ considered. 
For the validity of the approximation leading to \cref{eq:rnl_LowVelocities_lambdaCases}, we now require $\zetaB/n$ and $|\zetaS|/n \gtrsim f$ with $f\gtrsim 1$. Integrating over $n \in [1+\ell,f^{-1} \zetaB \min (1,|\lambda|)]$, we find
\begin{align}
&[r_\ell(\zetaS,\zetaB)]_{\rm \cref{eq:rnl_LowVelocities_lambdaCases}} 
\sim 
\int_{1+\ell}^{f^{-1} \zetaB\min(1,|\lambda|)} 
\dd n \, r_{n\ell}  (\zetaS,\zetaB)
\nn \\
&=
\dfrac
{\exp\qty{\zetaB \qty(
4 f^{-1} \min(1,|\lambda|)
\qty[\!\sqrt{|\lambda| (1+|\lambda|)}+\arcsinh\sqrt{|\lambda|}]
-2\pi|\lambda|)}
}
{
\sqrt{|\lambda| (1+|\lambda|)} \,
\qty[\!\sqrt{|\lambda| (1+|\lambda|)}+\arcsinh\sqrt{|\lambda|}]
}.
\label{eq:rl_lambda<0_nLow}
\end{align}
The above is decaying with $\zetaB$, for any $\lambda$, if we set $f \gtrsim 1.45$.\footnote{For $f <1.45$, there is a finite range of $\lambda$ values for which \cref{eq:rl_lambda<0_nLow} grows exponentially with $\zetaB$. This is spurious growth, borne by the fact that the validity of the approximation leading to \cref{eq:rnl_LowVelocities_lambdaCases} may only be marginal for $f$ very close to 1. Indeed, we find no exponential growth in the numerical evaluation of $r_\ell$ (cf.~\cref{fig:rl}).}

On the other hand, highly excited levels, $n \gtrsim \zetaB\max(1,\sqrt{|\lambda|})$, yield a clearly growing component at low velocities,
\begin{align}
r_\ell(\zetaS,\zetaB) \sim
\int_{\zetaB\max(1,\sqrt{|\lambda|})}^\infty \dd n \, 
[r_{n\ell}  (\zetaS,\zetaB)]_{\rm \cref{eq:rnl_nLarge}}
\sim
\dfrac{4\zetaB}{\max(1,|\lambda|)\sqrt{1+|\lambda|}}.
\label{eq:rl_lambda<0}
\end{align}

\end{description}
\end{subequations}
The low-lying bound levels, $n \lesssim \min(\zetaB,|\zetaS|)$, account for nearly all of $r_\ell$ for $\lambda>0$. On the other hand, the growth of $r_\ell$ at low velocities for $\lambda <0$ arises from higher levels, $n \gtrsim \zetaB$.

Collecting the above approximations, we find for $\zetaB \gtrsim 1+\ell$, 
\begin{empheq}[box=\myshadebox]{align}
r_\ell(\zetaS,\zetaB) \sim
\begin{cases}
4\zetaB 
\, \dfrac{1}{\max(1,|\lambda|) \sqrt{1+|\lambda|}} ,
& \lambda <0,
\\[1.5em]
4\zetaB 
\, \dfrac{1+2\sqrt{\lambda}}{\sqrt{1-\lambda}},
& 0 < \lambda < 1,
\\[1em]
10 \, \zetaB^{4/3},
& \lambda = 1, 
\\[0.5em]
\dfrac
{e^{-4(1+\ell) 
\qty[\!\sqrt{\lambda(\lambda-1)}-\arccosh \sqrt{\lambda}]
}}{
\sqrt{\lambda(\lambda-1)} \,
\qty[\!\sqrt{\lambda(\lambda-1)}-\arccosh \sqrt{\lambda}]
},
& 1 < \lambda \lesssim {\cal O}(1),
\\[2em]
2^{2\ell+5} \sqrt{\pi (1+\ell)}
\, \lambda^{2\ell+1} 
\, e^{-2(1+\ell)(2\lambda-1)},
& 1\ll \lambda.
\end{cases} 
\label{eq:rl_zetaSneq0_together}
\end{empheq}
In \cref{fig:rl}, we show $r_{\ell}$ as a function of $\vrel$ for representative values of $\lambda$, and different partial waves. We compare the numerical evaluation using \cref{eq:BSF_rnl_def}, to the analytical approximations of \cref{eq:rl_zetaSneq0_together}. 
We find numerically that \cref{eq:rl_zetaSneq0_together} reproduces the correct low-velocity growth, while exhibiting ${\cal O}(1)$ mismatches in the overall normalization in some cases. Note that the limits $\lambda\to0$ and $\lambda\to1$ are non-uniform: compared to the $0<\lambda<1$ behavior at large $\zetaB$, the $\lambda\to0$ case exhibits an extra $\ln\zetaB$ enhancement (cf.~\cref{eq:rl_alphaS=0_together}), while $\lambda\to1$ is enhanced by a factor $\sim \zetaB^{1/3}$.

Considering \cref{eq:BSF_sigma_lS_total,eq:rl_zetaSneq0_together}, the BSF cross-sections violate unitarity for $\lambda <0$ and $0<\lambda < 1$ at
\begin{align}
4 \arad \, \zetaB \gtrsim  1 + \order{\lambda}.   
\label{eq:UniViolation_Regime_alphaSneq0}
\end{align}

\subsection{Summary \label{sec:UnitarityViol_Summary}}

\begin{figure}[p!]
\centering
\includegraphics[width=0.42\linewidth]{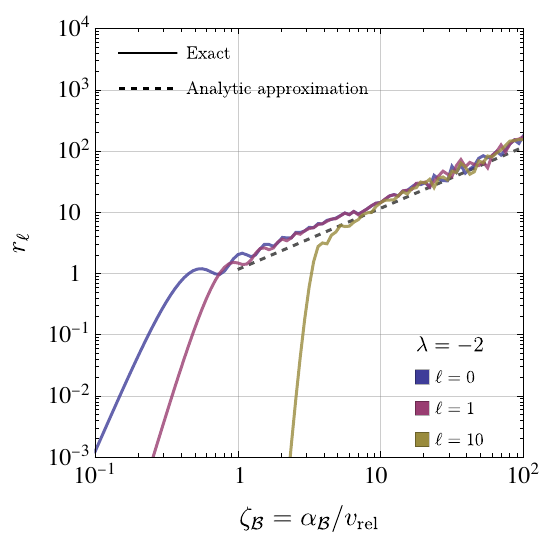}~~~
\includegraphics[width=0.42\linewidth]{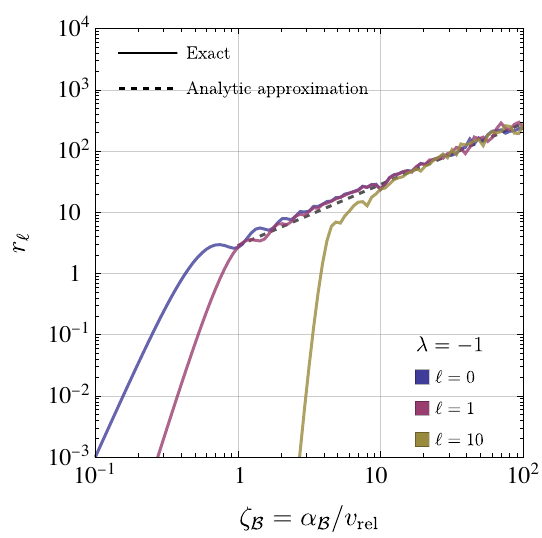}
\\[0.5em]
\includegraphics[width=0.42\linewidth]{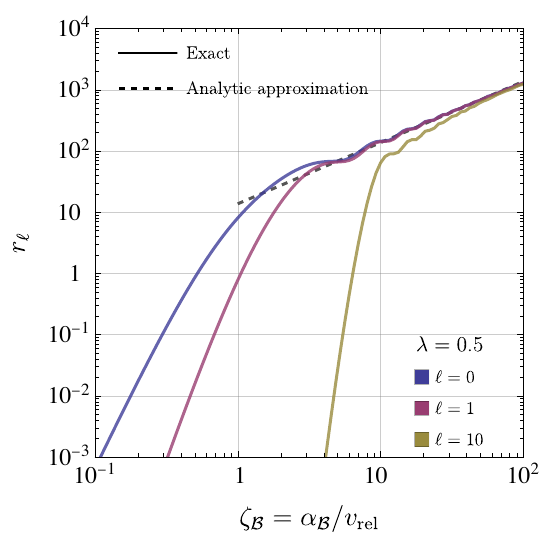}~~~
\includegraphics[width=0.42\linewidth]{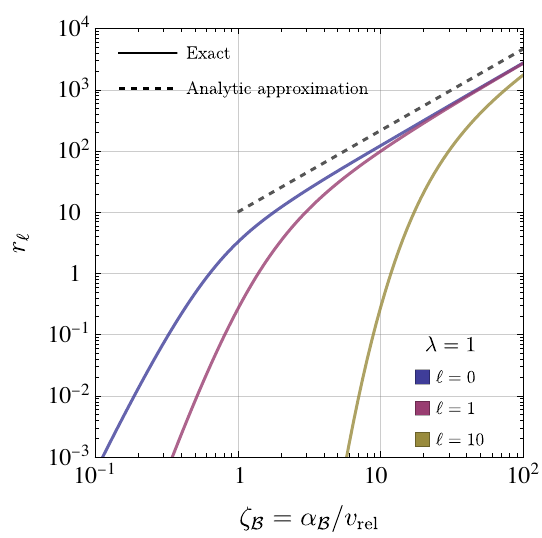}
\\[0.5em]
\includegraphics[width=0.42\linewidth]{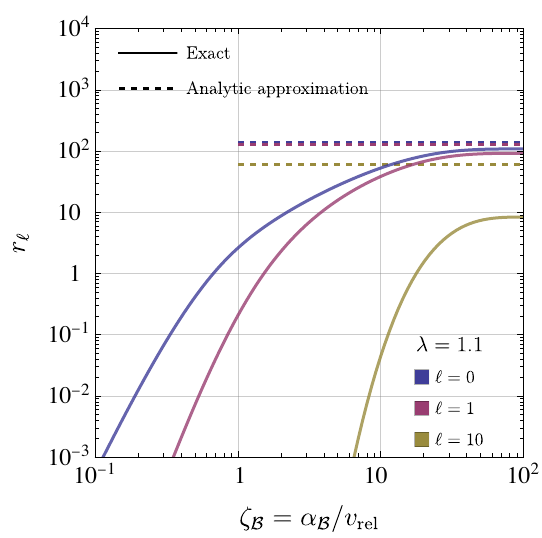}~~~
\includegraphics[width=0.42\linewidth]{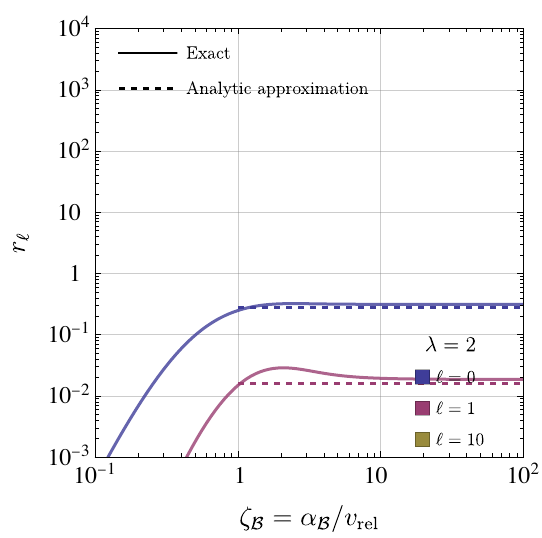}
\caption{\label{fig:rl} 
$r_{\ell}$ as a function of $\zetaB$, shown for different values of $\lambda=\aS/\aB$ and $\ell$, as indicated by the labels (see \cref{fig:rl_lambda=0} for the case $\lambda=0$). The analytic approximation \eqref{eq:rl_zetaSneq0_together} is also shown. It accurately predicts the slope of the growth at low velocities, although ${\cal O}(1)$ discrepancies with the numerical result appear mostly for $|\lambda|\in (0,1)$. Here, the approximation also overestimates $r_\ell$ for $\lambda=1.1$ and $\ell=10$ (lower-left panel), as these parameters lie in the transition region between the first two regimes discussed in \cref{sec:UnitarityViol_alphaSnot0_Kramers}.
}
\end{figure}

The main conclusions drawn from the above analysis are as follows:
\begin{itemize}
\item 
\Cref{eq:rl_alphaS=0_together,eq:rl_zetaSneq0_together} demonstrate the monotonic growth of the partial-wave BSF cross-sections summed over bound levels, at low velocities, for $\lambda \leqslant 1$. Remarkably, this growth persists for $\lambda < 0$, despite the Sommerfeld suppression. Considering the BSF cross-section per partial wave, given by \cref{eq:BSF_sigma_lS_total}, 
the Kramers-like  \cref{eq:rl_alphaS=0_together,eq:rl_zetaSneq0_together} establish the apparent unitarity violation for arbitrarily low couplings $\arad$, at low enough velocities.

Note that summation over partial waves, as in Kramers formula for QED~\cite{Kramers1923XCIIIOT}, results in even steeper growth at low velocities for $\lambda \leqslant 1$. This is, however, not constrained by unitarity. 

\item 
For $\lambda > 1$, the excited levels are subdominant or completely negligible; this behavior resembles the QED dipole transitions, where $\lambda =1$. In such cases, the unitarity limit \eqref{eq:UnitarityLimit_Inelastic} is saturated or violated only at sufficiently large couplings, $\arad$. Higher partial waves are also generally suppressed. 

\item 
\Cref{eq:rl_alphaS=0_together,eq:rl_zetaSneq0_together}, and the numerical results of \cref{fig:rl_lambda=0,fig:rl}, show that $r_\ell$ is insensitive to $\ell$ for $\zetaB \gtrsim 1+\ell$ and $\lambda \leqslant 1$, i.e.~where unitarity violation is severe.
\end{itemize}


\clearpage
\section{Unitarity restoration \label{sec:UnitarityRestore}}

Having firmly established the severe unitarity violation at low velocities in current BSF computations, we now demonstrate how to reconcile them with unitarity. In \cref{sec:UnitarityRestore_Formalism}, we review the formalism of Ref.~\cite{Flores:2024sfy, Flores:2025uoh} adapted to (monopole) BSF processes, before computing the unitarized BSF amplitudes and cross-sections in \cref{sec:UnitarityRestore_BSF}.

\subsection{Formalism \label{sec:UnitarityRestore_Formalism}} 

The unitarization scheme of Refs.~\cite{Flores:2024sfy,Flores:2025uoh} relates the regulated wavefunctions, amplitudes and cross-sections to their unregulated counterparts. The regulated quantities incorporate the absorptive potential from on-shell inelastic channels, whereas the unregulated ones do not. Both include the Hermitian potential associated with purely elastic interactions. 

\subsubsection{Hermitian potential, unregulated wavefunctions and Green's function \label{sec:UnitarityRestore_Formalism_RealPotential}}

The unregulated  partial-wave radial wavefunctions for the incoming pair, rescaled for scattering and bound states as\footnote{
Note that these bound-state solutions are \emph{not} the bound states formed in the BSF processes of interest, which in general obey a different potential, as described in \cref{sec:Models}. They are, rather, the bound eigenstates of the Hermitian potential that governs the incoming scattering state, and are needed in our analysis only for the completeness of the unitarization formalism~\cite{Flores:2025uoh}. We label them with the index ${\cal S}$ to distinguish them from the bound states produced in the BSF processes under consideration.}
%
\begin{align}
u_{k,\ell}^{(0)} (r) \equiv k r \ \psi_{k,\ell}^{(0)} (r),
\qquad
u_{\nS,\ell}^{(0)} (r) \equiv r \ \psi_{\nS\ell}^{(0)} (r),
\label{eq:RadialWFs_def_unreg}
\end{align}
obey the Schr\"odinger equation
\begin{align}
{\cal S}_{\ell}(r) u_{\ell}^{(0)} (r) 
= {\cal E}  u_{\ell}^{(0)} (r) ,
\label{eq:SchroedingerEq_0}
\end{align}
where ${\cal S}_{\ell}$ is the differential operator,
\begin{align}
{\cal S}_{\ell}(r) \equiv 
-\dfrac{1}{2\mu}\dfrac{\dd^2}{\dd r^2} 
+ \dfrac{\ell(\ell+1)}{2\mu \, r^2} 
+\VS(r) ,
\label{eq:SchroedingerOperator}
\end{align}
with the potential $\VS (r) = -\aS/r$ (cf.~\cref{eq:Potentials_SSandBS}). 
${\cal E}$ is the energy in excess of the total mass of the two interacting particles, in the CM frame. Following \cite{Flores:2024sfy, Flores:2025uoh}, we denote the two independent scattering-state solutions to \cref{eq:SchroedingerEq_0} as ${\cal F}_{k,\ell}(r)$ and ${\cal G}_{k,\ell}(r)$, corresponding to the regular and irregular family of solutions, for momentum $k = |\vb{k}|$, with energy eigenvalue ${\cal E}_{\vb{k}} = \vb{k}^2/(2\mu)$. We also introduce the linear combinations, ${\cal H}_{k,\ell}^{(\pm)}(r) \equiv 
{\cal F}_{k,\ell}(r) \pm \im {\cal G}_{k,\ell}(r)$ which represent outgoing and incoming wave solutions. For $\aS >0$, the potential  $\VS(r)$ supports bound-state solutions that we denote by ${\cal B}_{\nS \ell}(r)$, where $\nS$ stands for the principal quantum number associated with the binding energy 
${\cal E}_{\nS} 
= -\kappaS^2/(2\mu \nS^2) 
= -\mu \aS^2/(2\nS^2) < 0$, 
where $\kappaS \equiv \mu \aS$ (cf.~\cref{eq:kappas_def}). 
The exact form of the scattering-state and bound-state solutions is determined by their limiting behavior at the origin and infinity. In the case of identical particles, ${\cal F}_{k,\ell}$, ${\cal G}_{k,\ell}$ and ${\cal B}_{k,\ell}$ carry the appropriate symmetry factors. Further details, including the exact orthonormality conditions, can be found in~\cite{Flores:2025uoh}.

Fundamental to our analysis is the Green's function, defined by
\label{eq:RealPotential_SE}
\begin{align}
[{\cal S}_\ell (r)-{\cal E}_{\vb{k}}] \, G_{k,\ell} (r,r') &= \delta(r-r') ,
\label{eq:RealPotential_SE_Greens}
\end{align}
with boundary conditions selecting the solution that is regular at $r\to 0$ and behaves as an outgoing spherical wave at $r\to \infty$.  The resulting solution can be written in two equivalent ways, as~\cite{Flores:2025uoh} 
\begin{subequations}
\label{eq:GreensFunction}
\label[pluralequation]{eqs:GreensFunction}
\begin{empheq}[box=\myshadebox]{align}
G_{k,\ell} (r,r') 
&= \dfrac{2\mu}{\sym_\ell} \left[
\dfrac{2}{\pi}
\int_{0}^\infty \dd q \ 
\dfrac
{{\cal F}_{q,\ell}(r) \, {\cal F}_{q,\ell}^*(r')}
{q^2-k^2-\im \epsilon}     
- \sum_{\nS} 
\dfrac
{{\cal B}_{\nS\ell}^{} (r) {\cal B}_{\nS\ell}^* (r') }
{\kappaS^2/\nS^2+k^2}
\right]
\label{eq:GreensFunction_SpectralDecomp}
\\
&=
\dfrac{2\mu \im}{\sym_\ell k} 
\, {\cal F}_{k,\ell}^*(r_<)
\, {\cal H}_{k,\ell}^{(+)}( r_>) ,
\label{eq:GreensFunction_FH}
\end{empheq}
\end{subequations}
where $r_< \equiv \min\{r,r'\}$, $r_> \equiv \max\{r,r'\}$ and  $\epsilon \to 0^+$. \Cref{eq:GreensFunction_SpectralDecomp} is the spectral decomposition of the Green's function in the complete set of scattering and bound states compatible with the desired boundary conditions, whereas \cref{eq:GreensFunction_FH} follows from standard Sturm-Liouville methods~\cite{Arfken:2011mat}. Their derivation and proof of equivalence can be found in~\cite{Flores:2025uoh}. Evidently, in a scattering process where $k$ represents the on-shell momentum of a scattering state, the spectral decomposition \eqref{eq:GreensFunction_SpectralDecomp} introduces off-shell scattering states, $q\neq k$, and off-shell bound states, ${\cal E}_{\nS} \neq {\cal E}_{\vb{k}}$. This fact will become important in the following.

For finite-range potentials, the integrand in \cref{eq:GreensFunction_SpectralDecomp} is even in $q$. It is therefore possible to extend the integration domain to $q\in(-\infty,\infty)$ and include an overall factor of $1/2$~\cite{Flores:2025uoh}. For long-range (Coulombic) potentials, which is the case of interest here, logarithmic phase factors can break this evenness, so the continuation to $q<0$ is nontrivial and requires a carefully defined extension. We return to this point in \cref{sec:UnitarityRestore_BSF,sec:UnitarityRestore_alphaS=0,sec:UnitarityRestore_alphaSnot0}.

\subsubsection{Anti-Hermitian potential and regulated wavefunctions \label{sec:UnitarityRestore_Formalism_ImagPotential}}

On-shell inelastic processes generate anti-Hermitian potentials. In a Hermitian theory, these terms can be determined by integrating out inelastic channels~\cite{Feshbach:1958nx,Feshbach:1962nra}, and expressed uniquely in terms of the irreducible inelastic amplitudes, yielding a generally non-local but separable kernel~\cite{Flores:2024sfy,Flores:2025uoh}. 

Following Refs.~\cite{Flores:2024sfy,Flores:2025uoh}, the momentum-space kernel for the scattering states, generated by the BSF processes of interest, is 
\begin{align}
{\cal K}_{A,\ell} (p',p) 
&= \dfrac{\im}{(8\pi)^3} 
\sum_{n = 1}^{\infty}
\sum_{\ellB=0}^{n-1}
\sum_{\mB=-\ellB}^{\ellB}
\int \dd_\subsqrts\tau^{(n)}
\int \dd\Omega_{\vb{p'}} \dd\Omega_{\vb{p}}
\, P_\ell (\hat{\vb{p'}} \cdot \hat{\vb{p}})
\, {\cal A}_{{\bf p} \to n\ellB\mB}^{\BSF \, *} 
\, {\cal A}_{{\bf p'} \to n\ellB\mB}^{\BSF} 
\nn \\
&=
\im (8\pi)
\sum_{n=1+\ell}^\infty
\int \dd_\subsqrts\tau^{(n)}
\, {\cal A}_{n\ell}^{\BSF*} (p) 
\, {\cal A}_{n\ell}^{\BSF} (p') 
= \im
\sum_{n=1+\ell}^\infty
\dfrac{2\prad_n(s)}{\sqrt{s}}
{\cal A}_{n\ell}^{\BSF\,*} (p) 
\, {\cal A}_{n\ell}^{\BSF} (p') ,
\label{eq:AntiHermitianKernel_mom}
\end{align}
where 
${\cal A}_{\vb{p} \to n\ellB\mB}^{\BSF}$ and 
${\cal A}_{n\ell}^{\BSF} (p)$ 
are defined in \cref{eq:BSF_Acal,eq:BSF_Acal_PW}. 
The first Mandelstam variable $s$ parametrizes the total energy of the system. The scattering-state momenta $p$ and $p'$ are in general off-shell, while $\prad_n(s) = \omega_n (s)$ is the momentum or energy of the on-shell boson radiated in the BSF process, which we take to be massless. $d_\subsqrts\tau^{(n)}$ stands for the phase-space element of the products (i.e.,~bound state plus emitted boson), with the energy-momentum conservation included,
\begin{align}
\dd_\subsqrts\tau^{(n)} = 
\dfrac{\prad_n(s)}{16\pi^2 \sqrt{s}} 
\, \dd\Omega^{\rm rad},
\end{align}
where $\Omega^{\rm rad}$ is the solid angle of the radiated boson. In position space, \cref{eq:AntiHermitianKernel_mom} gives the potential\footnote{
Reference~\cite{Flores:2025uoh} considered also Hermitian counterparts of the anti-Hermitian separable potentials. Such potentials are generated by off-shell contact inelastic interactions, and are required to absorb UV divergences of the anti-Hermitian terms. However, BSF processes are ultrasoft, thus highly convergent at large momenta, and no such divergences arise. Although Hermitian contributions may still be generated, they are in general subdominant, and irrelevant for renormalization. We thus do not consider them here, and absorb the couplings of the anti-Hermitian terms in the $\nu_{n\ell}(r)$ factors, hence also in $\WW_\ell$ defined in \cref{eq:Wmatrix_def}. The associated imaginary factors enter \cref{eq:AntiHermitianKernel_pos,eq:Nmatrix_def} explicitly.
}
%
\begin{align} 
{\cal V}_{A, \ell} (r',r) 
&=
-\im \sum_{n=1+\ell}^\infty 
\nu^{}_{n\ell} (r') \, 
\nu_{n\ell}^{*} (r) ,
\label{eq:AntiHermitianKernel_pos}
\end{align}
where~\cite{Flores:2024sfy,Flores:2025uoh}
\begin{subequations}
\label{eq:AntiHermitianKernel_nu-A}
\label[pluralequation]{eqs:AntiHermitianKernel_nu-A}
\begin{align}
\nu_{n\ell} (r)  
&\equiv
(-\im)^\ell
\sqrt{\dfrac{8 \prad_n(s)}{\pi^2 \mu \, s}} 
\int_0^\infty \dd p 
\, p^2 
\, j_\ell^{} (pr) 
\, {\cal A}_{n\ell}^{\BSF} (p)
\label{eq:AntiHermitianKernel_nufactors_def}
\\
&= - \yrad
\sqrt{\dfrac{ 
\, \prad_n(s)}{16 \pi}}
\, \Psi_{n\ell}^*(r;~\kappaB),
\label{eq:AntiHermitianKernel_nufactors}
\end{align}
with the inverse being 
\begin{align}
{\cal A}_{n\ell}^{\BSF}(p)
&= \im^\ell
\sqrt{\frac{\mu\, s}{2\,\prad_n(s)}}
\int_{0}^{\infty} \dd r 
\, r^{2}
\, j_\ell(pr)
\, \nu_{n\ell}(r)\, .
\label{eq:AntiHermitianKernel_A_wrt_nu}
\end{align}
\end{subequations}
In deriving \cref{eqs:AntiHermitianKernel_nu-A}, we made the standard non-relativistic approximation $s\simeq \mT^2$ in all factors that do not exhibit threshold sensitivity, took into account ${\cal A}_{n\ell}^{\BSF}(p)$ as given by \cref{eq:BSF_nl_Acal} and the wavefunction Fourier transforms~\eqref{eqs:PW_Wavefun}.

Considering the contributions \eqref{eq:AntiHermitianKernel_pos} to the potential, Schr\"odinger's equation for the partial-wave radial wavefunction for the scattering states,
$u_{k,\ell} (r) \equiv k r\,\psi_{k,\ell} (r)$,  
reads
\begin{align}
({\cal S}_{\ell}- {\cal E}_{\vb{k}}) \, u_{k,\ell} (r) =
+
\im 
\sum_{n=1+\ell}^\infty
r \, \nu^{*}_{n\ell} (r) 
\int_0^\infty \dd r' r' \nu_{n\ell} (r') 
\, u_{k,\ell} (r') .
\label{eq:SchroedingerEq}
\end{align}
Defining the matrix $\mathbb{N}_\ell (k)$,
\begin{empheq}[box=\myshadebox]{align}
[\mathbb{N}_{\ell} (k)]^{nn'} 
&\equiv 
\delta^{nn'}  
- 
\ii\, 
\int_0^\infty \dd r \ r \int_0^\infty \dd r' \ r'
\qty[
\nu_{n\ell}^{} (r) 
\, G_{k,\ell} (r,r') 
\, \nu_{n'\ell}^*(r')] ,
\label{eq:Nmatrix_def}
\end{empheq}
with $G_{k,\ell} (r,r')$ given by \cref{eqs:GreensFunction}, the solution to \cref{eq:SchroedingerEq} is~\cite{Flores:2024sfy, Flores:2025uoh}
\begin{align} 
u_{k,\ell} (r) 
&=
{\cal F}_{k,\ell} (r) 
\label{eq:Wavefunction_sol}
\\[1ex]
&+ \ii   
\sum_{n =1+\ell}^{\infty} 
\sum_{n'=1+\ell}^{\infty} 
\qty(
\int_0^\infty \dd r'' 
\, r'' \, G_{k,\ell}^{}(r,r'') 
\, \nu_{n\ell}^*(r'')
)
[\mathbb{N}_\ell^{-1}(k)]^{nn'} 
\qty(
\int_0^\infty\dd r' 
\, r'\, {\cal F}_{k,\ell}^{} (r')
\, \nu_{n'\ell}^{}(r')
) .
\nn 
\end{align}

\subsubsection{Regularization \label{sec:UnitarityRestore_Formalism_Regularization}} 

\subsubsection*{Amplitudes and phase shift}

Considering \cref{eq:BSF_nl_McalvsAcal,eq:RadialWFs_def_unreg,eq:AntiHermitianKernel_A_wrt_nu}, we define the unregulated and regulated BSF amplitudes, 
\begin{subequations}
\label{eq:McalBSF_def}
\label[pluralequation]{eqs:McalBSF_def}
\begin{align}
{\cal M}_{n\ell,\unreg}^{\BSF} (k)
&\equiv
\dfrac{1}{2\pi^2}
\int_0^\infty \dd p \, p^2 
\, \tilde\psi_{k,\ell}^{(0)} (p)
\, {\cal A}_{n\ell}^{\BSF} (p)
=
\dfrac{1}{k} 
\sqrt{\dfrac{\mT^2\mu}{2 \prad_n(s)}}
\int_0^\infty \!\! \dd r \ r\ {\cal F}_{k,\ell}(r)\nu_{n\ell}(r),
\label{eq:McalBSF_def_unreg}
\\[1ex]
{\cal M}_{n\ell,\reg}^{\BSF}(k)
&\equiv
\dfrac{1}{2\pi^2}
\int_0^\infty \dd p \, p^2 
\, \tilde\psi_{k,\ell} (p)
\, {\cal A}_{n\ell}^{\BSF} (p)
=
\dfrac{1}{k} \sqrt{\dfrac{\mT^2\mu}{2 \prad_n(s)}}
\int_0^\infty \dd r\ r\ u_{k,\ell}(r)\nu_{n\ell}(r),
\label{eq:McalBSF_def_reg}
\end{align}
\end{subequations}
which neglect and consider the anti-Hermitian potential \eqref{eq:AntiHermitianKernel_pos}, respectively. Both incorporate the effect of the real central potential. ${\cal M}_{n\ell,\unreg} (k)$ correspond to the amplitudes introduced in \cref{sec:Models}. 
Note that in \cref{eqs:McalBSF_def}, the momentum $k$ may be in general off-shell. 
When on-shell, $k$ is related to $s$ via \cref{eq:ScatteringStates_OnShellRelation}, otherwise it is independent.   
We also define the rescaled versions of the above amplitudes,
\begin{align}
M_{n\ell,{\rm (un)reg}}^{\BSF} (k)  = 
\sqrt{\dfrac{4\,k\,\prad_n(s)}{\sym_\ell\,s}}
{\cal M}_{n\ell,{\rm (un)reg}}^{\BSF} (k) . 
\label{eq:Mrescaled_def}
\end{align}
Regulated and unregulated amplitudes, in both their original and rescaled versions, are related via~\cite{Flores:2024sfy, Flores:2025uoh} 
\begin{subequations}
\label{eq:Regularization_general}
\label[pluralequation]{eqs:Regularization_general}
\begin{align}
M_{\ell,\reg}^{\BSF} (k) 
&= 
\qty[\mathbb{N}_\ell (k)]^{-1} 
M_{\ell,\unreg}^{\BSF} (k) ,  
\label{eq:MregToMunreg_general}
\end{align}
where in the above and in the following, we are using matrix notation: 
$M_{\ell,{\rm (un)reg}}^{\BSF}$ are the vectors formed by the (un)regulated amplitudes, with components $M_{n\ell,{\rm (un)reg}}^{\BSF}$, where $n\in[1+\ell,\infty)$. 
The phase shift corresponding to the wavefunction \eqref{eq:Wavefunction_sol} is $\Delta_\ell (k) = \theta_\ell (k) + \delta_\ell (k)$, with~\cite{Flores:2024sfy, Flores:2025uoh}
\begin{align}
e^{2\im \delta_\ell (k) } &= 
1 
- 
2
M_{\ell,\unreg}^{\BSF \, \dagger}(k) 
\, \NN_\ell^{-1}(k) 
\, M_{\ell,\unreg}^{\BSF}(k).
\label{eq:PhaseShift_general}
\end{align}
\end{subequations}

From \cref{eqs:Regularization_general}, Refs.~\cite{Flores:2024sfy,Flores:2025uoh} derived simpler expressions for the regulated elastic and inelastic cross-sections. Essential for this purpose is the $\NN_\ell(k)$ matrix, defined in \cref{eq:Nmatrix_def}. Considering \cref{eq:GreensFunction_FH,eq:McalBSF_def_unreg,eq:Mrescaled_def}, it takes the form~\cite{Flores:2025uoh}
\begin{empheq}[box=\myshadebox]{align}
\mathbb{N}_\ell (k) = 
\ID 
+
M_{\ell,\unreg}^{\BSF} (k) M_{\ell,\unreg}^{\dagger\, \BSF}(k)    
+ 
\im \WW_\ell (k) ,
\label{eq:Nmatrix_GeneralForm}
\end{empheq}
where $\WW_\ell (k)$ is the Hermitian matrix
\begin{empheq}[box=\myshadebox]{align}
\label{eq:Wmatrix_def}
[\WW_\ell(k)]^{nn'}
\equiv
\frac{2\mu}{\sym_\ell k}
\int_0^\infty \dd r\ r\int_0^\infty \dd r'\ r'
{\cal F}_{k,\ell}^*  (r_<)
{\cal G}_{k,\ell}^{} (r_>)
\nu_{n\ell}^{}(r)
\nu_{n'\ell}^{*}(r').
\end{empheq}
As shown in \cite{Flores:2025uoh}, $\WW_\ell$ encodes the non-analytic and non-convergent behavior of the unregulated inelastic amplitudes. This quantity will be crucial to our later discussion.

\subsubsection*{Cross-sections}

For convenience, we define the cross-sections normalized to the unitarity limit,
\begin{subequations}
\label{eq:xyw_def}
\label[pluralequation]{eqs:xyw_def}
\begin{align}
x_{\ell, {\rm (un)reg}}^{} (k) &\equiv 
\dfrac{\sigma_{\ell, {\rm (un)reg}}^\elas  (k)}
{\sigma_\ell^U  (k)} ,
\label{eq:x_def}
\\
y_{n\ell, {\rm (un)reg}}^{} (k) &\equiv 
\dfrac{\sigma_{n\ell, {\rm (un)reg}}^{\BSF} (k)}
{\sigma_\ell^U (k)} ,
\label{eq:yj_def}
\\
y_{\ell, {\rm (un)reg}}^{} (k) 
&\equiv \sum_{n=1+\ell}^\infty 
y_{n\ell, {\rm (un)reg}}^{} (k) ,
\label{eq:y_def}
\end{align}
\end{subequations}
as well as the related quantities
\begin{subequations}
\label{eq:wParameters_def}
\label[pluralequation]{eqs:wParameters_def}
\begin{align}
w_{n\ell}^{} (k) &\equiv
\left|\left[
[\ID + \ii\, \WW_\ell (k)]^{-1} 
\, M_{\ell,\unreg}^\BSF (k)
\right]_n \right|^2 ,
\label{eq:wAj_def}
\\
w_\ell^{} (k) &\equiv
M_{\ell,\unreg}^{\BSF \, \dagger} (k)
\, \qty[\ID + \WW_\ell^2 (k)]^{-1}
M_{\ell,\unreg}^{\BSF} (k) ,
\label{eq:w_def}
\\
\tilde{w}_\ell^{} (k) &\equiv
M_{\ell,\unreg}^{\BSF \, \dagger} (k)
\ \WW_\ell (k)
\qty[\ID + \WW_\ell^2 (k)]^{-1}
\, 
M_{\ell,\unreg}^\BSF  (k),
\label{eq:wtilde_def}
\end{align}
\end{subequations}
where $w_{n\ell} (k)$, $w_\ell (k)$, $\tilde{w}_\ell (k) \in \RR$ and $w_\ell (k) = \sum_{n=1+\ell}^\infty w_{n\ell} (k)$ (c.f.~\cite[Appendix A.1]{Flores:2025uoh}). In \cref{eq:xyw_def,eq:wParameters_def}, $k$ is the on-shell momentum of the incoming scattering state.

The regulated elastic and BSF cross-sections can be expressed as~\cite{Flores:2025uoh},
\begin{subequations}
\label{eq:Regularization_xy}
\label[pluralequation]{eqs:Regularization_xy}
\begin{empheq}[box=\myshadebox]{align}
\label{eq:Regularization_x}
x_{\ell, \reg}^{} 
&= 
\frac{
x_{\ell, \unreg}^{} + (1 - x_{\ell, \unreg}^{})(w_\ell^2 + \tilde{w}_\ell^2) 
\pm
2\tilde{w}_\ell^{}
\sqrt{x_{\ell,\unreg}^{} (1 - x_{\ell,\unreg}^{})}
}{(1 + w_\ell^{})^2 + \tilde{w}_\ell^2},
\\
\label{eq:yj_reg}
y_{n\ell,\reg}^{}
&= 
\dfrac{w_{n\ell}^{}}
{(1 + w_\ell^{})^2 + \tilde{w}_\ell^2},
\\[1ex]
\label{eq:Regularization_y}
y_{\ell,\reg}^{} 
&= 
\dfrac{w_\ell^{}}
{(1 + w_\ell^{})^2 + \tilde{w}_\ell^2} ,
\end{empheq}
\end{subequations}
where the sign of the square root term in \eqref{eq:Regularization_x} is fixed by
$\mathrm{sgn}(\Re M_{\ell,\unreg}^{\elas})$, and cannot be inferred from $\sigma_{\ell,\unreg}^{\elas}(k)$ alone.
In the limit $\WW_\ell \to 0$, the above reduce to~\cite{Flores:2024sfy,Flores:2025uoh}
\begin{align}
\label{eq:Regularization_xy_W=0}
x_{\ell, \reg}^{} = 
\dfrac{
x_{\ell, \unreg}^{} + (1 - x_{\ell, \unreg}^{}) y_{\ell,\unreg}^2 
}{(1 + y_{\ell,\unreg}^{})^2}, 
\quad
y_{n\ell,\reg}^{} = 
\dfrac{y_{n\ell,\unreg}^{}}
{(1 + y_{\ell,\unreg}^{})^2},
\quad
y_{\ell,\reg}^{} = 
\dfrac{y_{\ell,\unreg}^{}}
{(1 + y_{\ell,\unreg}^{})^2} .
\end{align}


\subsection{Bound-state formation \label{sec:UnitarityRestore_BSF}}

In the following, we shall use $k$ as the on-shell momentum of the scattering state, related to the CM energy of the system via \cref{eq:ScatteringStates_OnShellRelation}.

We return to the integral representation \eqref{eq:GreensFunction_SpectralDecomp} of the Green's function and express the $\NN_\ell(k)$ matrix, defined in \cref{eq:Nmatrix_def}, as follows
\begin{align}
[\mathbb{N}_\ell (k)]^{nn'} = 
\delta^{nn'} 
&-\dfrac{\im}{\pi} 
\dfrac
{8 \left[
\prad_n (s) \, \prad_{n'}(s) 
\right]^{1/2}}
{\sym_\ell \mT^2}
\int_{0}^{\infty} \dd q \ q^2 \
\dfrac{
{\cal M}^{\BSF}_{n\ell,\unreg} (q) \ 
{\cal M}^{\BSF \star}_{n'\ell,\unreg} (q)}
{q^2-k^2-\im \epsilon} 
\nn \\[1ex]
&+\ii 
\, \arad
\, 2\mu \left[\prad_n (s) \, \prad_{n'}(s) \right]^{1/2}
\sum_{\nS = 1+\ell}^{\infty}
\dfrac{
{\cal R}_{\ell;~\nS n }^{} (\kappaS,\kappaB) \,
{\cal R}_{\ell;~\nS n'}^* (\kappaS,\kappaB)
}
{\kappaS^2/\nS^2 +k^2} ,
\label{eq:N_IntegralForm}
\end{align}
where ${\cal M}_{n\ell,\unreg}^{\BSF} (q)$ is the unregulated BSF amplitude given by \cref{eq:BSF_nl_Mcal}, and ${\cal M}^{\BSF\star}_{n\ell,\unreg} (q)$ is obtained from ${\cal M}^{\BSF}_{n\ell,\unreg} (q)$ with the replacement $\im\to-\im$ without conjugating $q$, such that the integrand in \cref{eq:N_IntegralForm} is a holomorphic function of $q$ except for singularities, that can be analytically continued on the complex $q$ plane. ${\cal R}_{\ell;~\nS n}(\kappaS,\kappaB)$ is the bound-bound overlap integral defined in \cref{eq:Rcal_BoundBound_def} and computed in \cref{app:OverlapIntegrals}; the corresponding term in \cref{eq:N_IntegralForm} has been derived taking into account the $\nu_{n\ell} (r)$ factors given by \cref{eq:AntiHermitianKernel_nufactors_def}.

Considering that the amplitudes ${\cal M}_{n\ell,\unreg}^{\BSF} (q)$, given by \cref{eq:BSF_nl_Mcal}, have no singularities on the real $q$ axis, and that 
\begin{align}
\dfrac{1}{q^2-k^2-\im \epsilon} = 
+\im \pi \delta(q^2-k^2)
+\PV \dfrac{1}{q^2-k^2},
\label{eq:propagator=delta+PV}
\end{align}
where $\PV$ stands for the Principal Value, we may write \cref{eq:N_IntegralForm} in the form of \cref{eq:Nmatrix_GeneralForm} with 
\begin{empheq}[box=\myshadebox]{align}
\WW_{\ell}^{nn'}  =  
&-\dfrac{8 \left[
\prad_n (s) \, \prad_{n'}(s) 
\right]^{1/2}}
{\pi \sym_\ell \mT^2}
\ \PV
\int_{0}^{\infty} \dd q \ q^2 \
\dfrac{
{\cal M}^{\BSF}_{n\ell,\unreg} (q) \ 
{\cal M}^{\BSF \star}_{n'\ell,\unreg} (q)}
{q^2-k^2} 
\nn \\[1ex]
&+
\arad
\, 2\mu \left[\prad_n (s) \, \prad_{n'}(s) \right]^{1/2}
\sum_{\nS = 1+\ell}^{\infty}
\dfrac{
{\cal R}_{\ell;~\nS n }^{} (\kappaS,\kappaB) \,
{\cal R}_{\ell;~\nS n'}^* (\kappaS,\kappaB)
}
{\kappaS^2/\nS^2 +k^2} .
\label{eq:Wmatrix_dq}
\end{empheq}
Alternatively, defining in analogy to \cref{eq:zetas_def}, the dimensionless parameters that entail the off-shell momentum $q$, 
\begin{align}
\zS\equiv \kappaS/q , \qquad
\zB\equiv \kappaB/q, \qquad
\zn\equiv \zB/n,
\label{eq:zs_def}
\end{align}
and noting that 
$\int_{0}^\infty dq \, q^2 / (q^2-k^2)= 
-\kappaB \, \zetaB^2 \int_{0}^\infty d\zB / [\zB^2(\zB^2-\zetaB^2)]$, \cref{eq:Wmatrix_dq} becomes
\begin{empheq}[box=\myshadebox]{align}
\WW_{\ell}^{nn'}  
&=  
\dfrac{\arad}{2\pi}
\left(1-\lambda\right)^2
(nn')^\frac32
\, [(1+\zetan^2)(1+\zetanp^2)]^\frac12
\ \PV \!
\int_{0}^{\infty} 
\dd \zB \, 
\dfrac{
\rM_{n\ell} (\lambda\zB,\zB) \, 
\rMstar_{n'\ell} (\lambda\zB,\zB)
}
{\zB^2 \, (\zB^2-\zetaB^2)}
\nn \\[1ex]
&+ 
\arad
\, [(1+\zetan^2)(1+\zetanp^2)]^\frac12
\sum_{\nS = 1+\ell}^{\infty}
\dfrac{
{\cal R}_{\ell;~\nS n }^{} (\lambda\kappaB,\kappaB) \,
{\cal R}_{\ell;~\nS n'}^* (\lambda\kappaB,\kappaB)
}
{\lambda^2\zetaB^2/\nS^2 + 1}  ,
\label{eq:Wmatrix_dzB}
\end{empheq}
where $\lambda\equiv\kappaS/\kappaB \in \RR$ as per~\cref{eq:lambda_def}, and $\rM_{n\ell}$ has been defined in \cref{eq:BSF_rMnl}. Recalling from \cref{app:OverlapIntegrals} that the overlap integrals ${\cal R}_{\ell;~\nS n }^{} (\kappaS,\kappaB)$ are functions of $\kappaS/\kappaB$ only, it is evident from \cref{eq:Wmatrix_dzB} that $\WW_\ell^{nn'} / \arad$ depends only on $\zetaB$ and $\lambda$, besides the discrete numbers $\ell$ and $n,n'$.

We may perform the integrations in \cref{eq:Wmatrix_dq,eq:Wmatrix_dzB} in two alternative ways~\cite{Flores:2025uoh}: (i) if the integrand is even under $q\leftrightarrow -q$ (equivalently $\zB\leftrightarrow -\zB$), we may extend the integration to the entire real axis, multiplying by (1/2), and close the contour appropriately in the complex $q$ or $\zB$ plane, considering  the analytic structure and accounting for the singularities of the BSF amplitudes; or
(ii) directly along the positive real ($q$ or $\zB$) axis, removing the contribution from the $q=\pm k$ poles, as follows
\begin{align}
\PV \int_{0}^\infty \dd q \dfrac{f(q)}{q^2-k^2} 
= \PV \left[
\int_0^\infty \dd q \, \dfrac{f(q)-f(k)}{q^2-k^2} 
+ \int_0^\infty \dd q \, \dfrac{f(k)}{q^2-k^2} 
\right]
= \int_0^\infty \dd q \, \dfrac{f(q)-f(k)}{q^2-k^2} ,
\label{eq:PVevaluation}
\end{align}
where we took into account that $\PV \int_0^\infty \dd q / (q^2-k^2) = 0$. Since the integrand in the remaining term is finite as $q\to k$, no $\PV$ is needed.  We will now see cases where each of these methods is preferable.

\subsection[$\alpha_{\scriptscriptstyle{\cal S}}=0$]
{$\bm{\aS=0}$  \label{sec:UnitarityRestore_alphaS=0}}

\subsubsection{Analytical calculation \label{sec:UnitarityRestore_alphaS=0_AnalCalc}}

For $\aS=0$, the absence of a Hermitian scattering-state potential implies that no bound states are present, and only the continuum integral contributions in \cref{eq:Wmatrix_dq,eq:Wmatrix_dzB} remain. Moreover, the integrands are symmetric under $q\leftrightarrow -q$, as follows from the general properties of wavefunctions for finite-range potentials~\cite{Flores:2025uoh} and can be verified explicitly below. Both methods described above may therefore be used to perform the integrations.

Using the quadratic hypergeometric transformation \eqref{eq:2F1_QuadraticTransf}, \cref{eq:BSF_rMnl} can be written in the following equivalent forms, 
\begin{align}
\label{eq:BSF_rMnl_alphaS=0}
\rM_{n\ell} (0,\zB) 
&\equiv
\im^\ell (-1)^{n-\ell} 
\frac{2^{2\ell+3} \ell!}{(2\ell+1)!} 
\left[\frac{n (n+\ell)!}{(n-\ell-1)!}\right]^{1/2} 
\nn 
\\[1ex]
&\times 
\begin{cases}
&\dfrac
{\zn^{\ell+4}}
{(\zn\mp\im)^{2\ell+4}}
\left(
\dfrac
{\zn\mp\im}
{\zn\pm\im}
\right)^{n+1}
{}_2F_1 \left( 
1+\ell-n, ~ 
1+\ell; ~
2\ell+2; ~
\mp\dfrac{4 \im \zn}{(\zn \mp \im)^2}
\right) ,
\\[1em]
&\text{or}
\\
& 
\chi_{n\ell}(\zn)
\ \dfrac
{\zn^{\ell+4}}
{(1+\zn^2)^{\ell+2}}
\ {}_2F_1 \left(
-\dfrac{n-\ell-1}{2},
\dfrac{n+\ell+1}{2};~
\ell+\dfrac{3}{2};~
\left(\dfrac{2\zn}{1+\zn^2}\right)^2
\right), 
\end{cases}
\end{align}
where the $\chi_{n\ell} (\zn)$ factor ensures that the logarithms are evaluated in their principal branches,\footnote{
In performing the quadratic transformation that relates the two forms of \cref{eq:BSF_rMnl_alphaS=0}, we encounter the factor $([(\zn+\im)/(\zn-\im)]^2)^{(n-\ell-1)/2}$. We set
\begin{align*}
a \equiv \Arg 
\qty(\dfrac{\zn+\im}{\zn-\im}) 
=2 \arctan \qty(\dfrac{1}{\zn}) \in
\begin{cases}
(0,\pi/2], & 1\leqslant\zn, \\
(\pi/2,\pi], & 0\leqslant\zn<1,     
\end{cases}
\end{align*}
where $\Arg (\cdot) \in (-\pi,\pi]$ is the argument in the principal branch. Then
\begin{align*}
\Arg\qty[\qty(\dfrac{\zn+\im}{\zn-\im})^2] 
= \begin{cases}
2a, & 1\leqslant\zn, \\
2a-2\pi, &0\leqslant\zn<1,
\end{cases}
\end{align*}
which implies
\begin{align*}
\qty[\qty(\dfrac{\zn+\im}{\zn-\im})^2]^{1/2} 
= \begin{cases}
e^{\im a}, & 1\leqslant\zn, \\
e^{\im (a-\pi)} = 
-e^{\im a}, &0\leqslant\zn<1,
\end{cases}
\qquad
\Rightarrow
\qquad
\qty[\qty(\dfrac{\zn+\im}{\zn-\im})^2]^{(n-\ell-1)/2} = 
\qty(\dfrac{\zn+\im}{\zn-\im})^{n-\ell-1}
\chi_{n\ell}^{} (\zn),
\end{align*}
where $\chi_{n\ell}^{} (\zn)$ is defined in \cref{eq:chi}.
}
\begin{align}
\chi_{n\ell}(\zn) \equiv 
\begin{cases}
1, & \zn >1, \\
(-1)^{n-\ell-1}, &0\leqslant \zn <1.
\end{cases}
\label{eq:chi}
\end{align}

\Cref{eq:BSF_rMnl_alphaS=0} implies that the integrand in \cref{eq:Wmatrix_dq} falls off at $|q|\to\infty$ as $q^{-2(\ell+4)}$. Considering that it is even in $q$, we may extend the integration to $q\in(-\infty,\infty)$ multiplying by (1/2), close the contour in the upper half complex-$q$ plane and evaluate the residues at the enclosed poles. The pole at $q=k+i\epsilon$ due to the propagator yields the $M_{\ell,\unreg}^{\BSF}(k)M_{\ell,\unreg}^{\BSF\,\dagger}(k)$ term in \cref{eq:Nmatrix_GeneralForm}, already obtained from the non-integral representation \eqref{eq:GreensFunction_FH} of the Green's function, or from the expansion \eqref{eq:propagator=delta+PV} of the propagator; it is not included in the $\PV$ of the integral. Considering the first version of \cref{eq:BSF_rMnl_alphaS=0}, we see that
the off-shell amplitudes introduce additional poles in the upper half-plane, at the bound-state momenta $q=\im\kappa_B/n$ and $q=\im\kappa_B/n'$. For $n\neq n'$ these are poles of order $n+1$ and $n'+1$, respectively; for $n=n'$ they coincide into a single pole of order $2n+2$. 
We may instead close the contour in the lower half-plane. The propagator pole at $q=-k-i\epsilon$ gives $M_{\ell,\unreg}^{\BSF}(-k)M_{\ell,\unreg}^{\BSF\,\dagger}(-k)=M_{\ell,\unreg}^{\BSF}(k)M_{\ell,\unreg}^{\BSF\,\dagger}(k)$ by analytic continuation, not to be included in the $\PV$, and the additional poles lying at $q=-\im\kappa_B/n$ and $q=-\im\kappa_B/n'$, are of order $n+1$ and $n'+1$ (or $2n+2$ if $n=n'$).

Equivalently, we may perform the contour integration with respect to $\zB$, as in \cref{eq:Wmatrix_dzB}. The integrand scales as $\zB^{-2(\ell+2)}$ at $|\zB|\to \infty$, we may therefore close the contour either on the lower or upper complex-$\zB$ plane, and pick up the poles at $\zB=\{-\im n,-\im n'\}$, or $\zB=\{\im n,\im n'\}$, of order $\{n+1,n'+1\}$ (or $2n+2$ for $n=n'$). Concretely, selecting the lower $\zB$-plane, we find
\begin{align}
&\WW_\ell^{nn'} 
= \im 
\, 2\arad
\left[
\frac{2^{2\ell+2} \ell!}{(2\ell+1)!} 
\right]^2 
(-1)^{n+n'+1}
(nn')^2
\left[
\frac{(n+\ell)!}{(n-\ell-1)!}
\frac{(n'+\ell)!}{(n'-\ell-1)!}
\right]^{\frac12}
\sqrt{(1+\zetan^2)(1+\zetanp^2)}   
\nn \\
&\times 
\!\! 
\sum_{\zB=\left\{
-\im n,\,-\im n'
\right\}} 
\!\!
{\rm Res}_{\zB} \Bigg[ \dfrac{1}{\zB^2(\zB^2-\zetaB^2)}
\dfrac{\zn^{\ell+4}}{(\zn-\im)^{2\ell+4}}
\left(\dfrac{\zn-\im}{\zn+\im}\right)^{n+1}
{}_2F_1 \left( 
1+\ell-n, ~ 
1+\ell; ~
2\ell+2; ~
-\frac{4 \im \zn}{(\zn - \im)^2}
\right)
\nn \\
&
\times
\dfrac{\znp^{\ell+4}}{(\znp-\im)^{2\ell+4}}
\left(\dfrac{\znp-\im}{\znp+\im}\right)^{n'+1}
{}_2F_1 \left( 
1+\ell-n', ~ 
1+\ell; ~
2\ell+2; ~
-\frac{4 \im \znp}{(\znp - \im)^2}
\right)
\Bigg],
\label{eq:Wmatrix_alphaS=0_PoleSum}
\end{align}
where we emphasize that in \cref{eq:Wmatrix_alphaS=0_PoleSum}, the residues must be calculated with respect to $\zB$. For $\rM_{n\ell}(0,\zetaB)$ and $\rM_{n'\ell}(0,\zetaB)$, given by \cref{eq:BSF_rMnl}, we have used the Pfaff transforms that allow us to pull the singular factors outside the hypergeometric functions. Note that the Pfaff transforms give equal expressions and do \emph{not} change the location of the poles. We evaluate \cref{eq:Wmatrix_alphaS=0_PoleSum} recalling that the residue of a function $f(x)$ at an isolated pole of order $n+1$ at $x=x_0$ is
\begin{align}
\Res_{x=x_0} f(x)
&= \frac{1}{n!}\,\lim_{x\to x_0}\frac{d^{\,n}}{dx^{\,n}}
\Big[(x-x_0)^{n+1} f(x)\Big].
\label{eq:Residue}
\end{align}
Considering this, we deduce from \cref{eq:Wmatrix_alphaS=0_PoleSum} that $\WW_\ell^{nn'}$ are ratios of polynomials in $\zetaB$. \Cref{fig:Welements_lambda=0} shows indicative examples of $\WW_\ell^{nn'}$ vs. $\zetaB$.

\subsubsection{Numerical evaluation \label{sec:UnitarityRestore_alphaS=0_numerics}}

\begin{figure}[t!]
\centering
\includegraphics[width=0.45\linewidth]{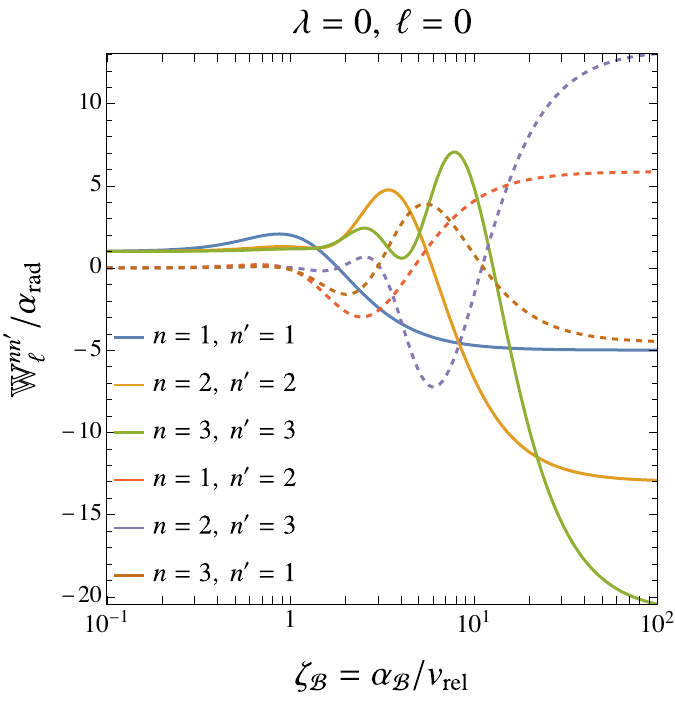}~~~~
\includegraphics[width=0.45\linewidth]{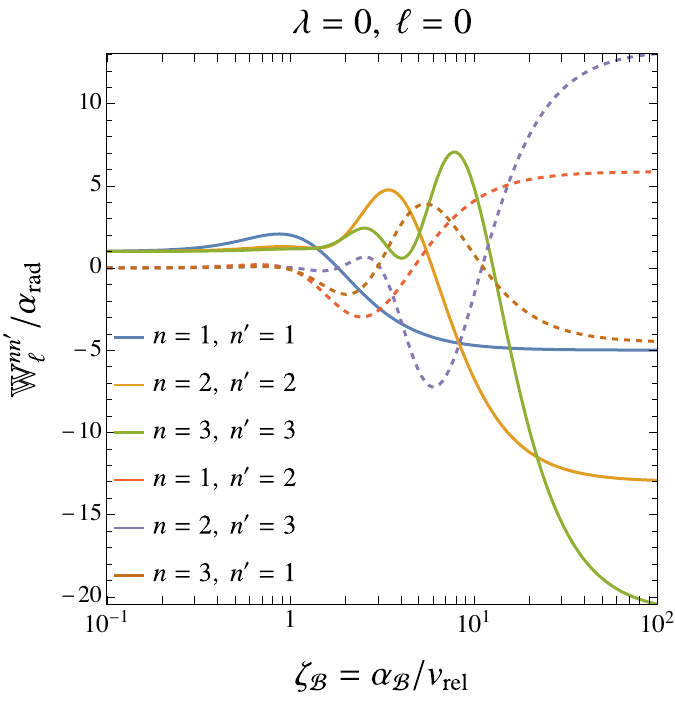}
\caption{Representative examples of $\WW_\ell^{nn'}$ for $\aS=0$, defined in \cref{eq:Wmatrix_alphaS=0_PoleSum}. The $\WW_\ell^{nn'}$ elements tend to constant values at large and low momenta, given by \cref{eqs:Wnn_Limits}. 
\label{fig:Welements_lambda=0}
}
\end{figure}

For small $n$ and $n'$, analytic expressions for $\WW_\ell^{nn'}$ are readily obtained. However, since we ultimately require integer values up to $n,n'\sim 10\zetaB$, as suggested by the estimations of \cref{sec:UnitarityViol} (cf.~\cref{eq:Nmax}), evaluating \cref{eq:Wmatrix_alphaS=0_PoleSum} rapidly becomes impractical. We therefore compute the integral in \cref{eq:Wmatrix_dzB} numerically, along the real axis. Using the second form of \cref{eq:BSF_rMnl_alphaS=0}, which renders the integrand manifestly real and numerically more stable, 
\begin{align}
&\WW_{\ell}^{nn'}  =  
(-1)^{n+n'} 
\ \dfrac{\arad}{\pi}
\ \dfrac{2^{4\ell+5} (\ell!)^2}{[(2\ell+1)!]^2}
\Bigg[
\dfrac{(n+\ell)!}{(n-\ell-1)!}
\dfrac{(n'+\ell)!}{(n'-\ell-1)!}
\Bigg]^{1/2}
\ \dfrac
{[(1+\zetan^2)(1+\zetanp^2)]^{1/2}}
{(nn')^{\ell+2}}
\nn \\
&\times 
\PV
\int_0^{\infty} 
\dd \zB
\ \dfrac{\zB^{2\ell+6}}{\zB^2-\zetaB^2}
\ \dfrac{
\chi_{n\ell}^{} (\zn) \ 
\chi_{n'\ell}^{} (\znp)}
{[(1+\zn^2)(1+\znp^2)]^{\ell+2}}
\label{eq:Wmatrix_alphaS=0_z-integr}
\\
&\times 
{}_2F_1 \left(
-\dfrac{n-\ell-1}{2},
\dfrac{n+\ell+1}{2};
\ell+\dfrac{3}{2};
\left(\dfrac{2\zn}{1+\zn^2}\right)^2
\right) 
{}_2F_1 \left(
-\dfrac{n'-\ell-1}{2},
\dfrac{n'+\ell+1}{2};
\ell+\dfrac{3}{2};
\left(\dfrac{2\znp}{1+\znp^2}\right)^2
\right) .
\nn
\end{align}
\Cref{eq:Wmatrix_alphaS=0_z-integr} makes manifest that $\WW_\ell \in \RR$ for $\aS=0$. While direct integration is faster, the computation still becomes costly at large $\zetaB$. We show results of the numerical evaluation in \cref{fig:Welements_lambda=0,fig:Regularization_lambda=0_inclusive,fig:Regularization_lambda=0_exclusive} and discuss it in \cref{sec:UnitarityRestore_alphaS=0_discussion}.

\subsubsection{Analytical estimates \label{sec:UnitarityRestore_alphaS=0_estimates}}

To gain insight on the impact of the $\WW_\ell\neq 0$ corrections on unitarization formulas \eqref{eq:Regularization_xy}, we first develop analytical estimates. We consider the limits of the $\WW_\ell$ elements at low and high momenta, focusing on the diagonal entries for simplicity. 

For $n = n'$, we use Clausen's formula, \eqref{eq:ClausensFormula}, to re-express \cref{eq:Wmatrix_alphaS=0_z-integr} as
\begin{align}
	\WW_{\ell}^{nn}(k)  =  
	+
    \ \dfrac{\arad}{\pi} 
	\ \dfrac{2^{4\ell+5} (\ell!)^2}{[(2\ell+1)!]^2}
	\Bigg[
	\dfrac{n(n+\ell)!}{(n-\ell-1)!}
	\Bigg]
	\ (1+\zetan^2)\ 
    {\cal I}_{n\ell}(\zetan^2) ,
	\label{eq:Wnn_Clausen} 
\end{align}
where
\begin{align}
{\cal I}_{n\ell}(u)
\equiv
\PV
\int_0^{\infty} 
\dd t
&\ \dfrac{t^{2\ell+6}}{t^2-u}
\ \dfrac{1}{(1+t^2)^{2\ell+4}}
 \nn\\ 
&\times 
{}_3F_2
\qty(
1+\ell - n, 1+\ell + n, \ell + 1;
\ell + \frac32, 2\ell + 2;
\left(\dfrac{2t}{1+t^2}\right)^2
)
.
\label{eq:Wnn_Inell}
\end{align}
As discussed in \cref{app:NonAnalyticStructure_alphaS=0}, \cref{eq:Wnn_Inell} is a ratio of polynomials in $u$,
\begin{align}
{\cal I}_{n\ell}(u) 
\equiv 
\frac{\pi}{2^{4\ell+5}}\frac{(n-\ell - 1)!}{n(n+\ell)!}
\left[\frac{(2\ell+1)!}{\ell!}\right]^2\ 
\frac{{\cal P}_{n\ell}(u)}{(1 + u)^{2n + 2}},
\end{align}
where ${\cal P}_{n\ell}(u)$ is an order $2n + 1$ polynomial. In particular, for
\begin{align}
{\cal P}_{n\ell}(u)
=
\sum_{j = 0}^{2n + 1}
C_{n\ell;j}\ u^j,
\end{align}
the coefficients are determined by
\begin{align}\label{eq:Cnell_Def}
C_{n\ell;j} 
\equiv 
\frac{2^{4\ell+5}n(n+\ell)!}{\pi(n-\ell - 1)!}
\left[\frac{\ell!}{(2\ell+1)!}\right]^2\ 
\cdot
\eval{
\frac{1}{j!}\dv[j]{u}\ [(1 + u)^{2n + 2}{\cal I}_{n\ell}(u)]
}_{u = 0}
.
\end{align}
This expression has been evaluated in \cref{app:NonAnalyticStructure_alphaS=0} and found to be,
\begin{align}\label{eq:Wnn_Coeffs}
C_{n\ell;j}	
=
\frac{2^{4\ell+5}n(n+\ell)!}{\pi(n-\ell - 1)!}
&\left[\frac{\ell!}{(2\ell+1)!}\right]^2\ 
\sum_{i = 0}^{j}
\binom{2n+2}{j-i}
\frac{\Gamma(\ell  + i + \frac32)\Gamma(\ell - i + \frac52)}{2\Gamma(2\ell + 4)}\times\nn \\[1ex]
&\times
{}_5F_4
\left(
\begin{matrix}
	1 + \ell-n, & 1 + \ell + n, & \ell + 1, &  \ell+ i + \frac32, & \ell  - i + \frac52\\
	{} & 2\ell+ 2, & \ell + 2, & \ell + \frac32, & \ell + \frac52
\end{matrix}; 1
\right)
.
\end{align}
In general, for fixed $j$, $C_{n\ell,j}$ does not admit a simple polynomial expression in $n$ and $\ell$. However, for certain values of $j$, \cref{eq:Wnn_Coeffs} simplifies significantly. Specifically,
\begin{align}\label{eqs:Specific_Cnells}
C_{n\ell; 0} = 1,
\qquad
C_{n\ell; 1} = 2n + 3,
\qquad
C_{n\ell; 2n + 1} = -\frac{8n - 6\ell - 3}{2\ell + 1}.
\end{align}
Thus, we arrive at the simple expression,
\begin{align}
\WW_\ell^{nn}(k)
&=  
\arad
\ \dfrac{{\cal P}_{n\ell}(\zetan^2)}{(1 + \zetan^2)^{2n + 1}}.
\label{eq:Wnn_Clausen_expansion} 
\end{align}
In combination with \cref{eqs:Specific_Cnells}, the limiting behavior of \cref{eq:Wnn_Clausen_expansion} yields
\begin{subequations}
\label{eq:Wnn_Limits}
\label[pluralequation]{eqs:Wnn_Limits}
\begin{empheq}[box=\myshadebox]{align}
\lim_{\zetan\to 0}
\WW_{\ell}^{nn}(k) 
&=  
\arad
\ \qty(1 
+ 2\zetan^2
+\order*{\zetan^4}
) 
,
\label{eq:Wnn_Limit_LargeMomenta}
\\[1ex]
\lim_{\zetan\to \infty}
\WW_{\ell}^{nn}(k) &=  
\arad
\ \dfrac{6\ell + 3 - 8n}{2\ell + 1} .
\label{eq:Wnn_Limit_LowMomenta}
\end{empheq}
\end{subequations}
The limiting scalings of \cref{eqs:Wnn_Limits} can be observed in \cref{fig:Welements_lambda=0}. 

The estimation of the non-diagonal elements is less reliable, we thus do not attempt it here. Generally, they tend to be suppressed with respect to the diagonal elements if we truncate $\WW_\ell$ to a finite-size square matrix (i.e., typically $|\WW_\ell^{nn'}| \lesssim |\WW_\ell^{nn}|$ for $n'<n$), in part due to destructive interference between the oscillatory factors of the hypergeometric functions, which carry different phases. However, we have checked numerically that the non-diagonal elements play an important role in the unitarization of the cross-sections (see below).

\Cref{eqs:Wnn_Limits}  allow us to assess the regime where the  $\WW_\ell$ corrections become significant. \Cref{eq:Wnn_Limit_LargeMomenta} shows that at large momenta, the $\WW_\ell$ corrections are important only for very large couplings, and can be safely neglected if $\arad \ll 1$. However, $\WW_\ell^{nn}$ grow in magnitude with decreasing velocity (increasing $\zetaB$), with the excited states playing an important role, as suggested by \eqref{eq:Wnn_Limit_LowMomenta}. While it is not straightforward to deduce it from the asymptotic behaviors of \cref{eqs:Wnn_Limits}, we find numerically that the singularities affect significantly the unitarization scheme  when the unregulated inclusive BSF cross-sections approach the unitarity limit; considering \cref{eq:UniViolation_Regime_alphaS=0}, this occurs roughly at 
\begin{empheq}[box=\myshadebox]{align}
4\arad \, \zetaB 
\ln (\zetaB/b_\ell)
\gtrsim 1.     
\label{eq:Wimportance_condition}
\end{empheq}
For $\lambda \neq 0$, a milder condition without the logarithmic enhancement holds, but with $\lambda$-dependent factors (cf.~\cref{eq:UniViolation_Regime_alphaSneq0}).

\subsubsection{Discussion of effect of non-analytic behavior \label{sec:UnitarityRestore_alphaS=0_discussion}}

\begin{figure}[t!]
\centering
\includegraphics[height=0.7\linewidth]{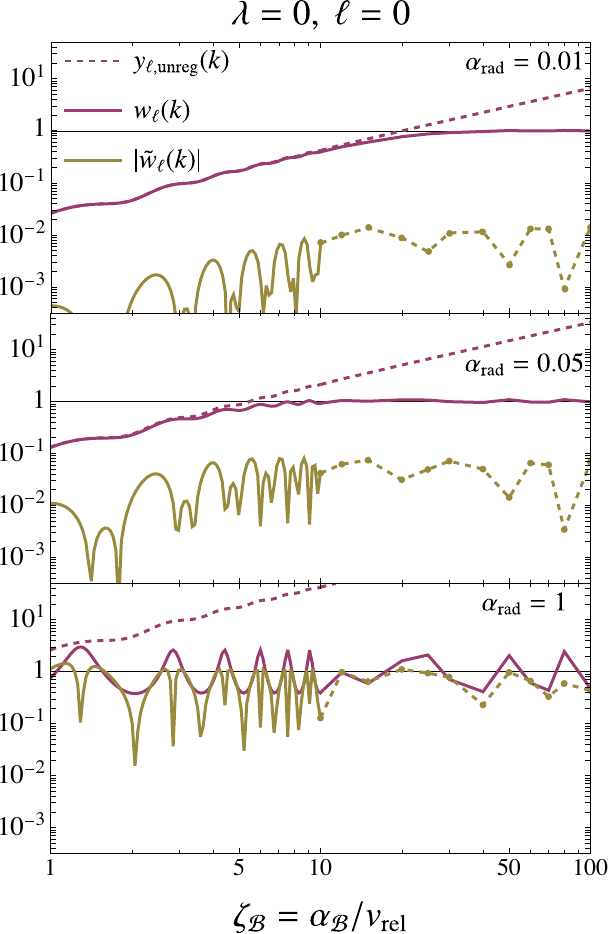}
~~~
\includegraphics[height=0.7\linewidth]{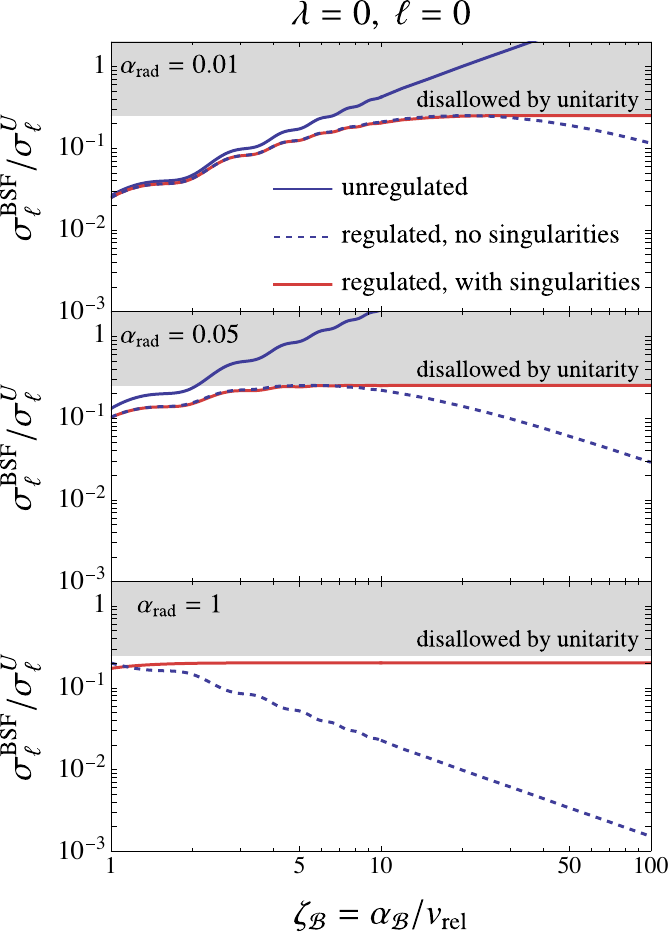}
\caption{
\label{fig:Regularization_lambda=0_inclusive}
$\ell=0$ and $\aS=0$.  
\emph{Left panels:} Unregulated inclusive BSF cross-section normalized to the unitarity limit, $y_{\ell,\unreg}$, and  regularization parameters, $w_\ell$ and $\tilde{w}_\ell$, vs $\zetaB=\aB/\vrel$. 
\emph{Right panels:} Inclusive BSF cross-sections normalized to the unitarity limit. We compare the unregulated version \eqref{eq:BSF_sigma_lS_total}, with the cross-sections regulated according to the prescription \eqref{eq:Regularization_xy_W=0} that neglects the singularities of the BSF amplitudes in the complex momentum plane, and the full formulas \eqref{eq:Regularization_xy} that take into account the singularities.
\emph{Top, middle, bottom rows:} $\arad = 0.01,~0.05,~1$, respectively. 
See \cref{sec:UnitarityRestore_alphaS=0_discussion} for discussion of results. 
The evaluation of the regularization matrix becomes computationally expensive at large $\zetaB$, the resolution is thus limited at low velocities.
}
\end{figure}

\begin{figure}[h!]
\centering
\includegraphics[height=0.9\linewidth]{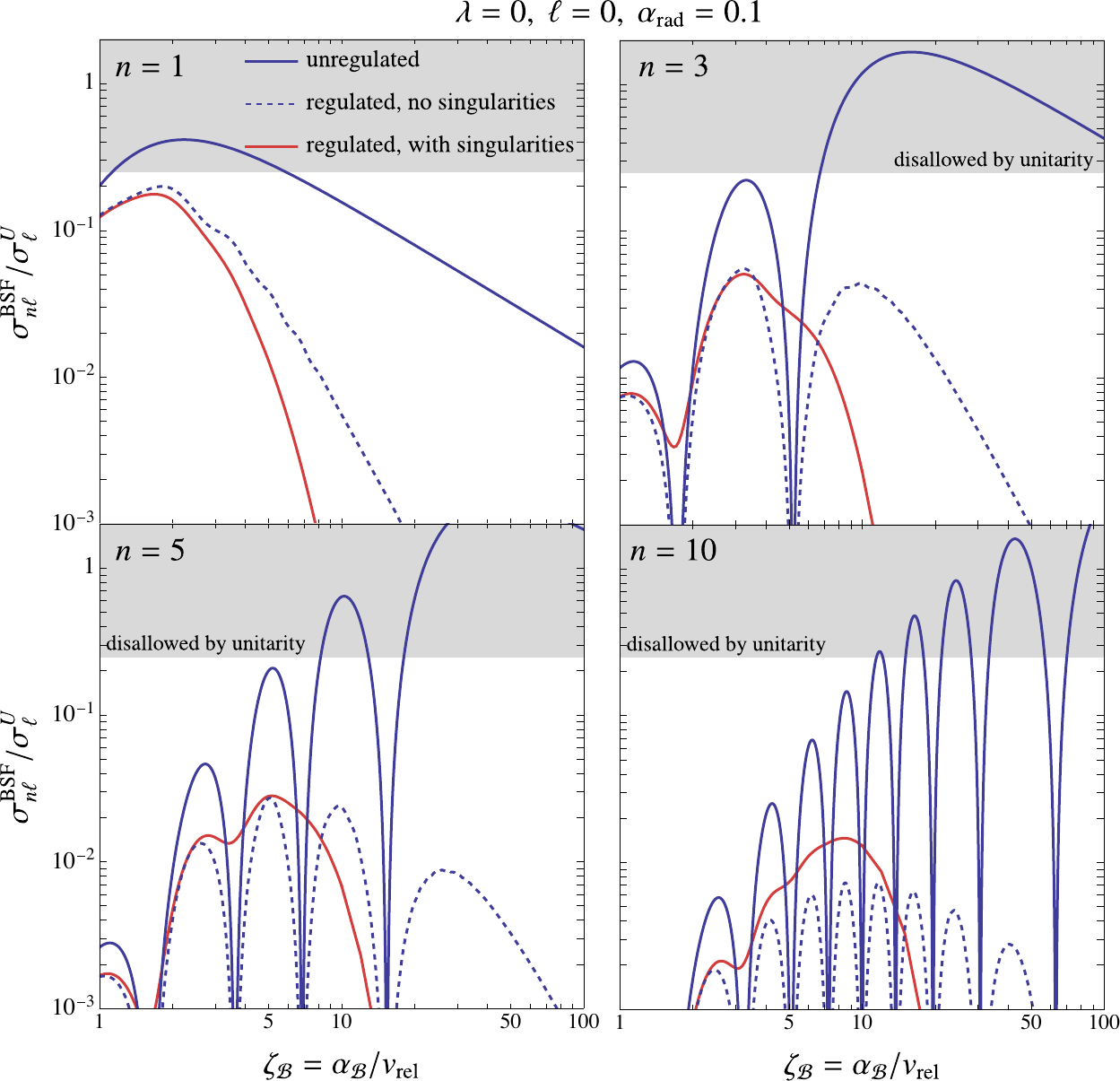}
\caption{
\label{fig:Regularization_lambda=0_exclusive}
$\ell=0$, $\aS=0$ and $\arad = 0.1$.  
BSF cross-sections normalized to the unitarity cross-section, for individual bound levels, $n=1$ (\emph{upper left}), $n=3$ (\emph{upper right}), $n=5$ (\emph{lower left}), $n=10$ (\emph{lower right}). We compare the unregulated cross-sections \eqref{eq:BSF_sigma_lS_total} (\emph{solid blue lines}) with the regulated ones under the prescription  \eqref{eq:Regularization_xy_W=0} that neglects singularities (\emph{dashed blue lines}) and under the full prescription \eqref{eq:Regularization_xy} that accounts for singularities (solid red lines). See \cref{sec:UnitarityRestore_alphaS=0_discussion} for discussion of results. 
}
\end{figure}

The full numerical results for the regularization parameters and the regulated BSF cross-sections, for $\aS=0$, are shown in \cref{fig:Regularization_lambda=0_inclusive,fig:Regularization_lambda=0_exclusive,fig:Regularization_vs_arad}. 

To isolate the impact of the singularities of the BSF amplitudes, in \cref{fig:Regularization_lambda=0_inclusive} (left panels), we compare $y_{\ell,\unreg}$ with $w_\ell$: in the absence of these singularities, $w_\ell\to y_{\ell,\unreg}$ (cf.~\cref{eq:wParameters_def}), whereas the full calculation yields $w_\ell<y_{\ell,\unreg}$ at sufficiently large $\zetaB$. The effect of $\WW_\ell$ can be interpreted as dispersive mixing of the on-shell incoming scattering state with a tower of off-shell intermediate bound states. These dispersive terms interfere destructively with the direct capture contribution to the amplitude, thereby taming the low-velocity growth  exhibited by $y_{\ell,\unreg}$. The effect becomes significant once  $y_{\ell,\unreg} \gtrsim 1$, whereupon $w_\ell$ tends to saturate at $\sim 1$, for perturbative couplings, $\arad < 1$. In this regime, the second regularization parameter, $\tilde{w}_\ell$, is suppressed by $\arad$ with respect to $w_\ell$, thus typically negligible.

Crucially, the off-diagonal elements of $\WW_\ell$, which encode transitions via different off-shell bound levels, are essential for curtailing the low-velocity growth. We have checked numerically that keeping only the diagonal part of $\WW_\ell$ reduces the enhancement seen in $y_{\ell,\unreg}$ but does not eliminate it. Only the full matrix results in the saturation of $w_\ell$ at low velocities, seen in \cref{fig:Regularization_lambda=0_inclusive} (left panels).

The suppression of $w_\ell$ relative to $y_{\ell,\unreg}$ implies lower depletion of the incoming flux into inelastic channels than the fully unregulated result suggests. Through \cref{eq:Regularization_y}, this translates into a \emph{larger} fully regulated inclusive BSF cross-section at low velocities. Indeed, $\sigma_{\ell,\reg}^{\BSF}/\sigma_\ell^U$ saturates to a constant as $\zetaB\to\infty$. For perturbative couplings, $\arad\lesssim 1$, the saturation value approaches the unitarity ceiling for inelastic capture, $\sigma_{\ell,\reg}^{\BSF}/\sigma_\ell^U \sim 1/4$. In contrast, neglecting the singularities of the BSF amplitudes in the unitarization scheme would imply that $\sigma_{\ell,\reg}^{\BSF}/\sigma_\ell^U$ decreases at low velocities, once $y_{\ell,\unreg}$ exceeds unity. This is depicted in \cref{fig:Regularization_lambda=0_inclusive} (right panels). 

\Cref{fig:Regularization_lambda=0_exclusive} displays the BSF cross-sections for capture into individual bound states. As expected, the regulated cross-sections are suppressed relative to the unregulated ones. For small $n$, properly accounting for the singularities of the BSF amplitudes in the regularization procedure further suppresses the corresponding exclusive cross-sections. In contrast, for larger $n$ this behavior is reversed: within a certain velocity range around $\zetaB \sim n$, the contributions from higher-lying levels are enhanced, which in turn increases the inclusive BSF cross-section at all velocities, as discussed above. The exclusive BSF cross-sections exhibit smoothed oscillatory features, and, eventually, a steeper decline at low velocities, $\zetaB \gg n$.

In \cref{fig:Regularization_vs_arad}, we investigate further the effect of the coupling at low velocities. Selecting $\zetaB =10^2$, we see that $w_\ell$ and  $\sigma_{\ell,\reg}^{\BSF}/\sigma_\ell^U$ have reached their asymptotic values for $\arad \gtrsim 1/(4\zetaB) = 2.5 \times 10^{-3}$ (cf.~\cref{eq:Wimportance_condition}). Remarkably, the saturation values,  $w_\ell \sim 1$ and $\sigma_{\ell,\reg}^{\BSF}/\sigma_\ell^U \sim 1/4$ are insensitive to the couplings, as long as the coupling is not too large. For very large couplings, however, $w_\ell$ saturates at values scaling as $w_\ell\propto 1/\arad$, while $\tilde{w}_\ell$ dominates over $w_\ell$ and approaches an ${\cal O}(1)$ value largely independent of $\arad$; these scalings are justified by \cref{eq:w_def,eq:wtilde_def}, considering that $M_{\ell,\unreg}^\BSF \propto \arad^{1/2}$ and $\WW_\ell \propto\arad$.  
Consequently, $\sigma_{\ell,\reg}^{\BSF}/\sigma_\ell^U \propto 1/\arad$, as implied by \cref{eq:Regularization_y}.

\subsection[$\alpha_{\scriptscriptstyle{\cal S}}\neq0$]
{$\bm{\aS\neq0}$  \label{sec:UnitarityRestore_alphaSnot0}}

For $\aS\neq 0$, the integrands in \cref{eq:Wmatrix_dq,eq:Wmatrix_dzB} are not even in $q$ (or $\zB$), due to the factor $e^{\pi \lambda \zB}$ inherited from the long-range scattering-state wavefunction (cf.~\cref{eq:BSF_rMnl}). In addition, the singularity structure in the complex momentum plane is considerably richer than in the $\aS = 0$ case. For these reasons, it is easier to evaluate the integrals numerically along the positive real axis.

\begin{figure}[t!]
\centering
\includegraphics[height=0.7\linewidth]{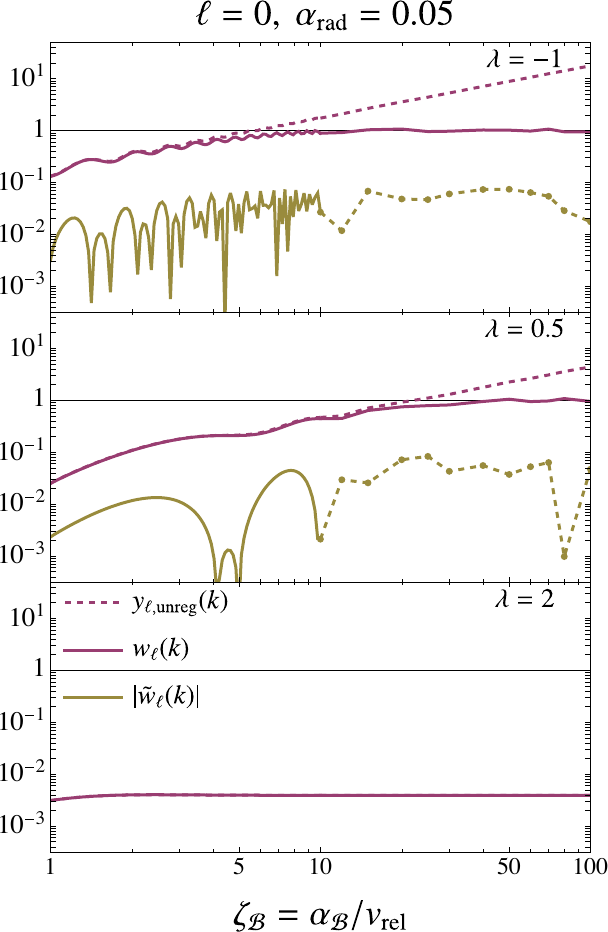}
~~~
\includegraphics[height=0.7\linewidth]{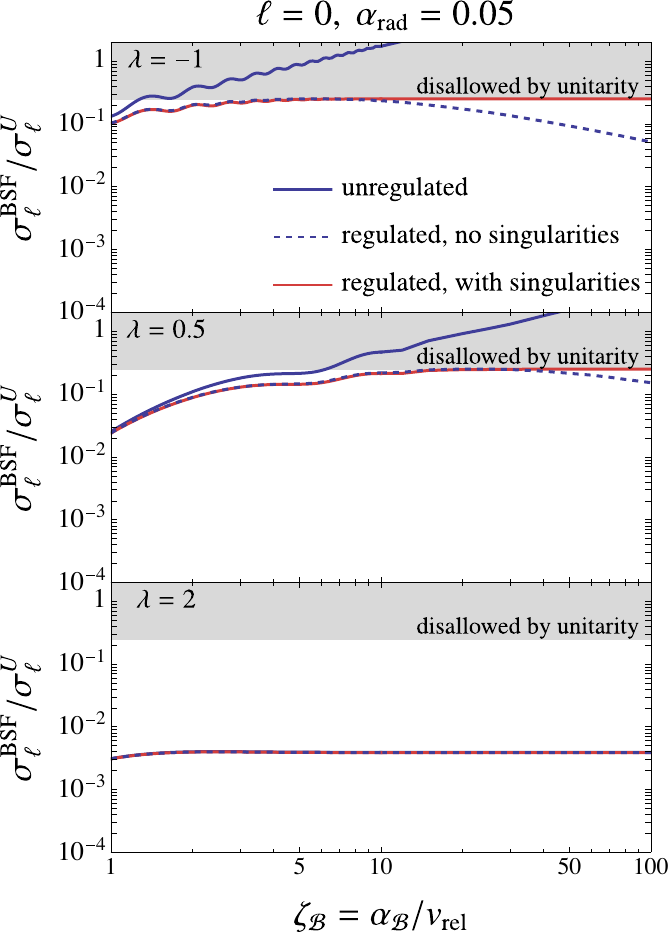}
\caption{Same as in \cref{fig:Regularization_lambda=0_inclusive}, for different values of $\lambda=\kappaS/\kappaB$:
$\lambda=-1$ (\emph{top}),  
$\lambda=0.5$ (\emph{center}),
$\lambda=2$ (\emph{bottom}). 
See \cref{sec:UnitarityRestore_alphaS=0_discussion,sec:UnitarityRestore_alphaSnot0} for discussion of results. 
\label{fig:Regularization_lambdaNot0}
}
\end{figure}

The numerical results are shown in \cref{fig:Regularization_lambdaNot0,fig:Regularization_vs_arad}. 
For $\lambda\equiv \kappaS/\kappaB<1$, where the unregulated BSF cross-sections exhibit the most severe unitarity violation, the qualitative behavior closely parallels the $\lambda=0$ case discussed above. By contrast, for $\lambda > 1$, the regularization, including the singularities, has little to no effect for perturbative couplings.

\subsection{Massive force mediators and radiated particles \label{sec:UnitarityRestore_BeyondCoulomb}}

So far we have assumed Coulombic potentials in both the scattering and bound states, as appropriate when the corresponding force mediators are massless. We have also taken the particle emitted in BSF to be massless. We now comment on how our results are modified when the force mediators and/or the radiated particle are massive. We discern the following:

\begin{enumerate}[label=(\roman*)]
\item
If the potentials in the scattering and/or bound states are generated by mediators of non-zero masses $\mmedS$ and $\mmedB$ respectively, they are Yukawa rather than Coulomb. The Coulomb approximation for the wavefunctions remains valid as long as the mediator Compton wavelength is larger than the characteristic size of the state under consideration. Parametrically, this requires
\begin{subequations}
\label{eq:YukawaPot_CoulombRegime}
\begin{align}
\text{Scattering states:}\qquad
&\mmedS \lesssim \mu v_{\rm rel} \,,
\label{eq:YukawaPot_CoulombRegime_ScattStates}
\\
\text{Bound states:}\qquad
&\mmedB \ll \kappan = \mu \aB / n \,.
\label{eq:YukawaPot_CoulombRegime_BoundStates}
\end{align}
\end{subequations}
Outside these regimes, the wavefunctions are not Coulombic, while bound states do not exist roughly if $\mmedB$ exceeds the right-hand side of \cref{eq:YukawaPot_CoulombRegime_BoundStates}. 
If several mediators with different masses contribute to the potentials, the above conditions may be significantly modified~\cite{Harz:2017dlj,Harz:2019rro}.

\item
If the radiated particle has non-zero mass, $\mrad$, the phase-space factor of \cref{eq:Phi_n} suppresses capture into weakly bound levels and vanishes once the emitted particle becomes kinematically inaccessible. Neglecting $\mrad$ is a good approximation for
\begin{align}
\label{eq:PhaseSpaceBSF_CoulombRegime}
\mrad \ll \frac{\mu}{2}\qty(v_{\rm rel}^2 + \frac{\aB^2}{n^2}) \,.
\end{align}
Equivalently, capture into the $n$th level is kinematically forbidden if $\mrad$ exceeds the right-hand side of the above.
\end{enumerate}

A non-zero $\mmedB$ makes the bound-state spectrum finite, while a non-zero $\mrad$ truncates the set of bound levels that can be formed radiatively. The sums entering the Kramers-like formulae of \cref{sec:UnitarityViol} therefore terminate at some finite $n$, thereby taming the growth at low $\vrel$. In the unitarization procedure, one must accordingly include only the amplitudes, and hence only the associated singularities, of bound levels that both exist and are kinematically accessible.  The Coulombic results derived here remain a good approximation so long as this cutoff is higher than the range of $n$ needed for convergence (estimated in \cref{eq:Nmax} for $\aS=0$).

By contrast, a finite $\mmedS$ does not reduce the number of bound states supported by the bound-state potential, nor the set of levels that are kinematically accessible. In the limit in which the scattering-state interaction becomes contact-type, $\mmedS \gg \mu \aS$ and $\mu \vrel$, the problem reduces effectively to the case $\aS=0$ discussed in \cref{sec:UnitarityViol_alphaS=0,sec:UnitarityRestore_alphaS=0}. For intermediate $\mmedS$, we expect departure from the semi-analytical and numerical results presented here. However, the qualitative features remain largely the same. The non-analytic structure of the BSF amplitudes in the complex momentum plane can be affected, typically simplified, as there is less interference of non-analyticities emanating from the scattering-state and bound-state potentials. 

It is worth noting that, for discrete values of $\mmedS \neq 0$ corresponding to the thresholds at which the scattering-state potential begins to support bound states, the BSF cross-sections exhibit parametric resonances~\cite{Petraki:2016cnz}. This is analogous to the resonant enhancement familiar from hard-scattering inelastic processes (annihilation into relativistic species) when the incoming state is governed by a finite-range potential. Near such resonances, even the exclusive inelastic cross-sections grow rapidly at low velocities and may exceed the partial-wave unitarity bound, especially for $\ell=0$. The unitarization of hard-scattering inelastic processes requires renormalization, because their irreducible inelastic amplitudes do not converge at large momenta; this has been discussed in detail in Ref.~\cite{Flores:2025uoh}.\footnote{For hard-scattering inelastic processes, renormalization is required for unitarization with or without Sommerfeld effect and/or a Sommerfeld resonance. In the absence of a finite-range potential generating resonances, unitarity is violated at sufficiently large, though still conventionally perturbative, values of the coupling.}
For BSF, by contrast, renormalization is not necessary, since the irreducible inelastic amplitude converges rapidly at large momenta, as discussed at the beginning of this section. The unitarization procedure outlined here ensures the unitarization of the BSF processes, including on Sommerfeld resonances.

\clearpage

\begin{figure}[h!]
\centering
\includegraphics[width=0.9\linewidth]{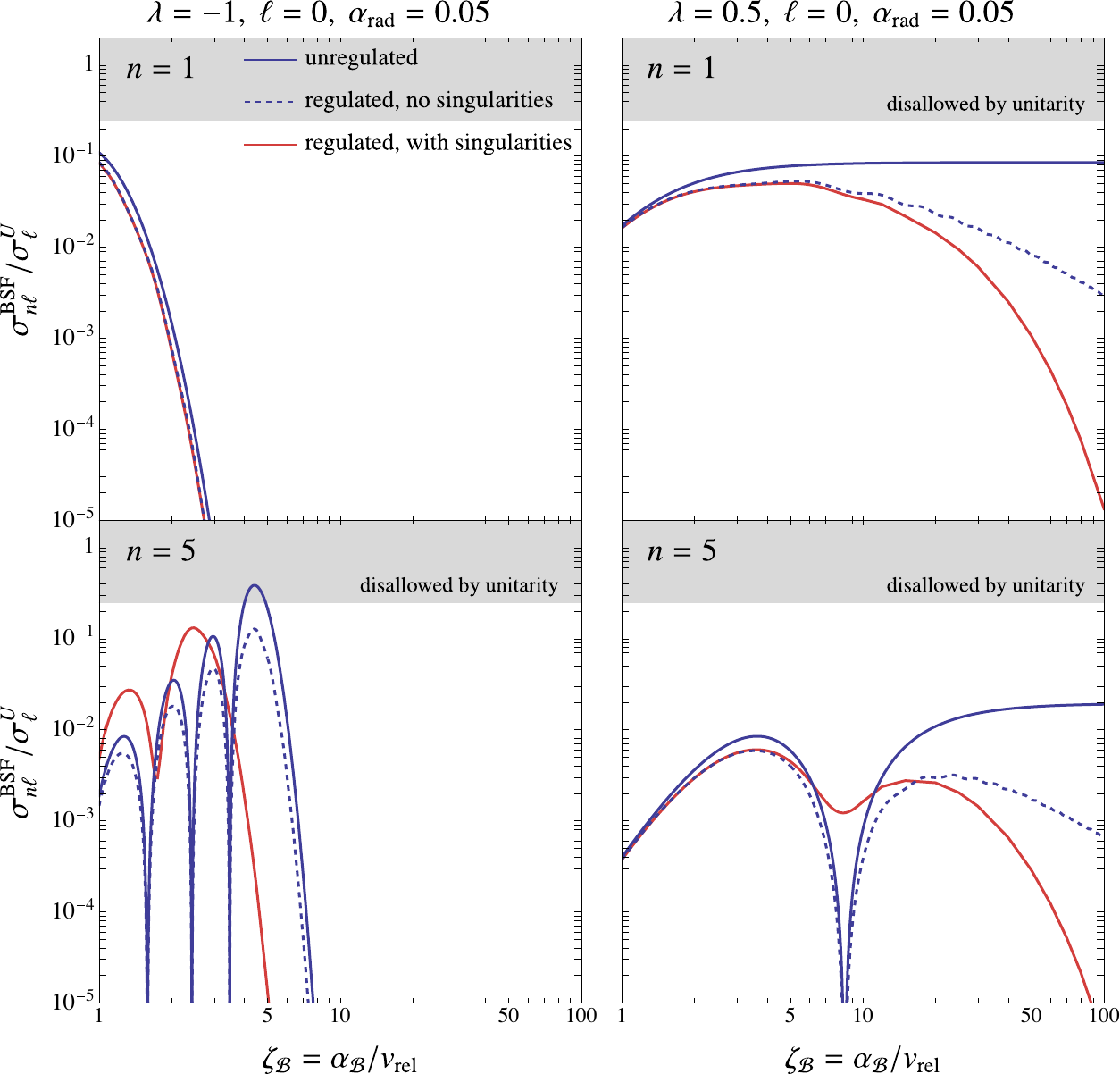}
\caption{
\label{fig:Regularization_lambdaNot0_exclusive}
Same as in \cref{fig:Regularization_lambda=0_exclusive}, for different values of $\lambda=\kappaS/\kappaB$:
$\lambda=-1$ (\emph{left}),  
$\lambda=0.5$ (\emph{right}). 
See \cref{sec:UnitarityRestore_alphaS=0_discussion,sec:UnitarityRestore_alphaSnot0} for discussion of results. 
}
\end{figure}
\begin{figure}[h!]
\centering
\includegraphics[width=0.47\linewidth]{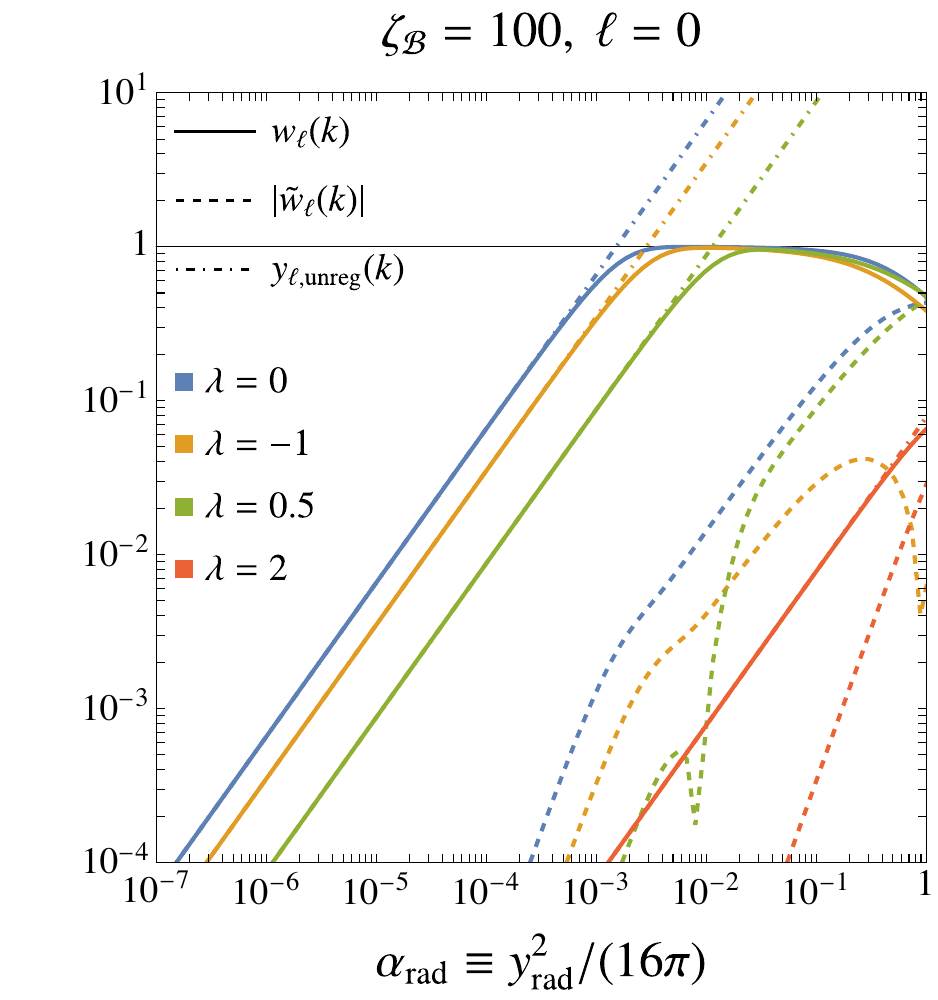}
\quad
\includegraphics[width=0.47\linewidth]{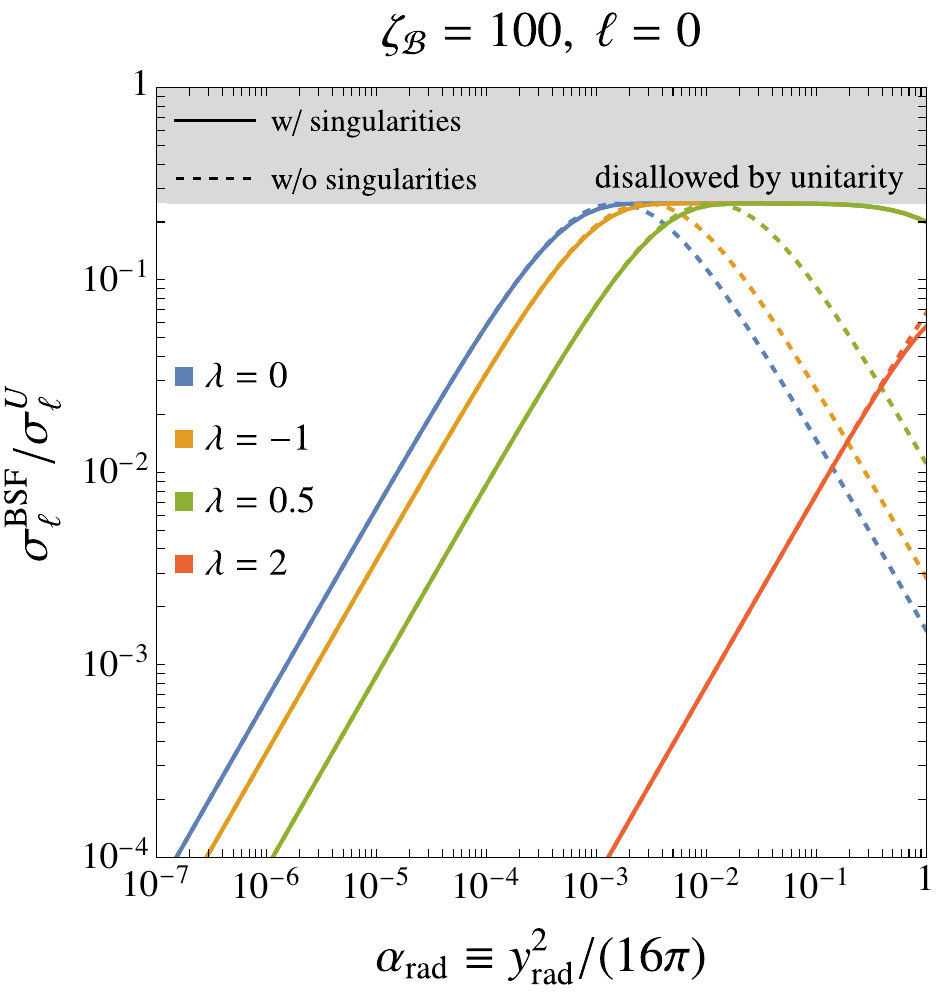}
\caption{
For $\ell=0$ and different values of $\lambda=\kappaS/\kappaB$ shown in the legend, we examine the low-velocity behavior ($\zetaB = 100$) with respect to the coupling $\arad$. 
\emph{Left panel:} 
The unregulated inclusive BSF cross-sections normalized to the unitarity limit, $y_{\ell,\unreg}$, and the regularization parameters $w_\ell$ and $\tilde{w}_\ell$. 
\emph{Right panel:} 
The regulated inclusive BSF cross-sections normalized to the unitarity limit, with and without accounting for the singularities of the BSF amplitudes. 
See \cref{sec:UnitarityRestore_alphaS=0_discussion,sec:UnitarityRestore_alphaSnot0} for discussion of results. 
\label{fig:Regularization_vs_arad}
}
\end{figure}

\clearpage
\section{Conclusions  \label{sec:Concl}}

Radiative BSF is a major inelastic channel in weakly coupled theories with long-range forces, with broad relevance for DM phenomenology. If the emitted boson carries a conserved charge, so that the scattering and bound-state potentials differ, standard non-relativistic computations can yield BSF cross-sections that grow at low velocities, violating partial-wave unitarity even for arbitrarily small couplings~\cite{Oncala:2019yvj,Oncala:2021tkz,Oncala:2021swy,Binder:2023ckj,Beneke:2024nxh}. In this work, we quantified this behavior and, using the methodology of Refs.~\cite{Flores:2024sfy,Flores:2025uoh}, demonstrated explicitly how unitarity is restored once the BSF contributions to the incoming-state self-energy are resummed. Our methodology and results readily generalize to other theories where BSF is important, including a variety of WIMP models, colored coannihilation scenarios, and hidden-sector models.

We focused on monopole capture processes with the selection rule $\Delta\ell=0$, which correspond to the simplest partial-wave structure. Such processes arise in models with charged scalar mediators~\cite{Oncala:2019yvj}, notably the Higgs doublet~\cite{Oncala:2021tkz,Oncala:2021swy}. Depending on parameters, they can also emulate related aspects of both Abelian and non-Abelian gauge dynamics. By summing Coulombic BSF cross-sections over all bound levels accessible from a fixed incoming partial wave, we derived Kramers-like relations for each $\ell$ mode and obtained compact approximations that exposed the role of highly excited levels and the resulting enhancement of the inclusive BSF cross-section at low velocities. We showed that for scattering-to-bound coupling ratios $\lambda=\kappaS/\kappaB=\aS/\aB<1$, the BSF cross-sections per partial wave grow at low velocities significantly faster than in QED, implying violation of partial-wave unitarity even at arbitrarily small couplings (see also Ref.~\cite{Beneke:2024nxh}). By contrast, for $\lambda>1$ the contributions of highly excited levels are exponentially suppressed and the low-velocity behavior remains consistent with unitarity.

We then applied the unitarization framework of Refs.~\cite{Flores:2024sfy,Flores:2025uoh}. The BSF channels generate non-local but separable anti-Hermitian potentials that are uniquely determined by the irreducible BSF amplitudes. Including these potentials amounts to resumming the BSF contributions to the incoming-state self-energy. The resulting wavefunctions and amplitudes yield unitarized elastic and inelastic cross-sections. An important feature of BSF in this context is that it is an ultrasoft process: the corresponding kernels are UV convergent and therefore do not require renormalization in the way contact-type absorptive interactions typically do. Instead, BSF processes possess a nontrivial singularity structure that affects unitarization, as was explored in this work.

Accounting for these singularities in the unitarization procedure exposes the interplay between dispersive and absorptive effects in the incoming-state self-energy, encoded in the regularization matrix $\NN_\ell$ (cf.~\cref{eq:Nmatrix_GeneralForm}). The $\WW_\ell$ matrix describes dispersive mixing of the on-shell scattering state with a tower of off-shell intermediate bound levels, while the absorptive contribution (the $M_{\ell,\unreg}^{\BSF} M_{\ell,\unreg}^{\BSF\dagger}$ term in $\NN_\ell$) accounts for depletion into on-shell BSF channels. This interplay becomes important once the unregulated BSF cross-sections approach or exceed the unitarity limit. If the singularities are neglected, $\sigma_{\ell,\reg}^{\BSF}/\sigma_\ell^U$ decreases at low velocities, since the increasingly large $\sigma_{\ell,\unreg}^{\BSF}/\sigma_\ell^U$ implies stronger depletion of the incoming flux. By contrast, retaining the singularities weakens this depletion through dispersive mixing and drives $\sigma_{\ell,\reg}^{\BSF}/\sigma_\ell^U$ toward an ${\cal O}(1)$ constant, approaching the partial-wave inelastic ceiling $\sim 1/4$ at low velocities (for perturbative couplings).

While resumming the BSF contributions to the incoming-state self-energy unitarizes BSF, a closely related ultrasoft process, bremsstrahlung, generically arises in the same models and is expected to exhibit analogous low-velocity enhancements. A complete description of ultrasoft dynamics should therefore include BSF and bremsstrahlung on the same footing: resumming either process affects the other, and more generally affects any inelastic channel that shares the same initial state, such as annihilation into relativistic species, even if that channel does not itself exhibit unphysical growth~\cite{Flores:2024sfy,Oncala:2021tkz,Oncala:2021swy}. We leave a combined treatment to future work.

In addition to formal consistency, these results may impact phenomenology wherever BSF is important. This includes DM production via thermal freeze-out~\cite{Petraki:2025zvv}, indirect detection where BSF produces new radiative signals, and the formation of stable DM bound states in the early universe.

For standard QED, radiative capture processes such as hydrogen or positronium formation proceed via dipole transitions, with the same potential in the scattering and bound states, and the severe low-velocity unitarity violation discussed in this work is absent. The unitarization mechanism considered here would still induce corrections, but these are expected to be small for $\alpha_{\rm QED}\simeq 1/137$. Given the high precision of QED observables, however, it would be interesting to compare such effects with existing precision calculations and with experiment. We leave this to future work.

\section*{Acknowledgements}
This work was supported by the European Union’s Horizon 2020 research and innovation programme under grant agreement No 101002846, ERC CoG CosmoChart. MMF was also supported through a FRIPRO grant of the Norwegian Research Council (project ID 353561 ‘DarkTurns’).

\clearpage
\appendix
\section*{Appendices}
\section{Monopole overlap integrals \label{app:OverlapIntegrals}}

This appendix follows largely Refs.~\cite[appendix~B]{Oncala:2019yvj} and \cite[appendix~C]{Oncala:2021tkz}. For convenience we collect the scattering-bound and bound-bound monopole overlap integrals used in the main text and fix our conventions.

We consider the Coulomb potentials for scattering and bound states (cf.~\cref{eq:Potentials_SSandBS,eq:kappas_def,eq:zetas_def}),
\begin{align}
\VS = -\aS/r 
\quad\text{and}\quad
\VB = -\aB/r,
\label{eq:Potentials_SSandBS_app}
\end{align}
and define the Bohr momenta and their ratios to the scattering-state momentum,
\begin{align}
&\kappaS \equiv \mu \aS ,&  
\quad
&\kappaB \equiv \mu \aB ,&  
\quad
&\kappan \equiv \mu \aB/n ,& 
\label{eq:kappas_def_app}
\\
&\zetaS \equiv \kappaS/k = \aS/\vrel ,&  
\quad
&\zetaB \equiv \kappaB/k = \aB/\vrel ,&
\quad
&\zetan \equiv \kappan/k = \zetaB/n .&
\label{eq:zetas_def_app}
\end{align}
as well as the space variables
\begin{align}
\xS \equiv k r 
\quad \text{and} \quad
\xB \equiv \kappaB r.
\label{eq:xSandxB_def_app}
\end{align}

\subsection{Wavefunctions}

We consider the wavefunctions in position and momentum space, related via the Fourier transform
\begin{align}
\label{eq:FT_wavefunctions}
\psi(\vb{r})
&=
\int
\dfrac{\dd^3 p}{(2\pi)^3}
e^{\im \vb{p}\cdot \vb{r}}
\tilde{\psi}(\vb{p}) .
\end{align}
We analyze the scattering-state and bound-state wavefunctions in partial waves as follows,
\begin{subequations}
\label{eq:PW_Wavefun}
\label[pluralequation]{eqs:PW_Wavefun}
\begin{align}
\label{eq:PW_Wavefun_mom}
\tilde{\psi}_{\vb{k}}(\vb{p}) 
&= \sum_{\ell}(2\ell + 1)
P_\ell(\hat{\vb{k}}\cdot\hat{\vb{p}})
\tilde{\psi}_{k, \ell}(p),
&\quad
\tilde{\psi}_{n\ell m}(\vb{p}) 
&= \tilde{\psi}_{n \ell}(p)
Y_{\ell m} (\Omega_{\vb{p}}),
\\
\label{eq:PW_Wavefun_pos}
\psi_{\vb{k}}(\vb{r}) 
&= \sum_{\ell}(2\ell + 1)
P_\ell(\hat{\vb{k}}\cdot\hat{\vb{r}})
\psi_{k, \ell}(r),
&\quad
\psi_{n\ell m}(\vb{r}) 
&= \psi_{n \ell}(r)
Y_{\ell m} (\Omega_{\vb{r}}),
\\
\label{eq:PW_Wavefun_pos-mom}
\psi_{k, \ell}(r) 
&= \dfrac{\im^\ell}{2\pi^2} 
\int_0^\infty \dd p \, p^2 \, 
\tilde{\psi}_{k,\ell}(p) 
j_\ell (pr) ,
&\quad
\psi_{n\ell}(r) 
&= \dfrac{\im^\ell}{2\pi^2} 
\int_0^\infty \dd p \, p^2 \, 
\tilde{\psi}_{n\ell}(p) 
j_\ell (pr) ,
\\
\label{eq:PW_Wavefun_pos-mom_inv}
\tilde{\psi}_{k, \ell}(p) 
&= 4\pi(-\im)^\ell 
\int_0^\infty \dd r 
\, r^2 
\, \psi_{k,\ell}(r) 
\, j_\ell (pr) ,
&\quad
\tilde{\psi}_{n\ell}(p) 
&= 4\pi (-\im)^\ell  
\int_0^\infty \dd r 
\, r^2 
\, \psi_{n\ell}(r) 
\, j_\ell (pr) .
\end{align}
\end{subequations}

Next, we focus on \emph{distinguishable particles}, and denote their wavefunctions with capital $\Psi$ for clarity. In position space, the wavefunctions corresponding to the Coulomb potentials \eqref{eq:Potentials_SSandBS_app} are
\begin{subequations}
\label{eq:CoulombWaveFun}
\label[pluralequation]{eqs:CoulombWaveFunctions}
\begin{align}
\Psi_{k,\ell}^{} (r;~\kappaS) 
&=
e^{\pi \zetaS/2} 
\dfrac{\Gamma(1+\ell-\im \zetaS)}{(2\ell+1)! }
\times 
\ (2\im\xS)^{\ell} 
\ e^{\pm\im \xS}
\ {}_1F_1 (1+\ell \mp \im \zetaS; ~ 2\ell + 2; ~ \mp2 \im \xS) ,
\label{eq:psi_SS_PW}
\\[1em]
\Psi_{n\ell} (r;~\kappaB) 
&= 
\dfrac{2 \kappaB^{3/2}}{n^2 (2\ell+1)!} 
\left[\frac{(n+\ell)!}{(n-\ell-1)!}\right]^{1/2} 
\left(\dfrac{2\xB}{n}\right)^{\ell} 
e^{-\xB/n} 
\ {}_1F_1 \left(1 + \ell - n;~ 2\ell+2;~ \dfrac{2\xB}{n} \right) ,
\label{eq:psi_BS_PW}
\end{align}
\end{subequations} 
where the two signs in \cref{eq:psi_SS_PW} arise via the Kummer transformation \eqref{eq:1F1_Kummer}.

\paragraph{Branch convention.}
We define complex powers via the principal logarithm:
$z^\nu \equiv \exp \qty(\nu\,\Log z)$ with 
$\Arg z\in(-\pi,\pi]$.
In particular,
\[
\left(\frac{\zetan-\im}{\zetan+\im}\right)^{n+1-\im\zetaS}
=\exp\!\Big[(n+1-\im\zetaS)\big(\Log(\zetan-\im)-\Log(\zetan+\im)\big)\Big].
\]

\newpage
\subsection{Scattering-bound overlap integral}

We aim to compute the integral \eqref{eq:Rcal_def}
\begin{align}
{\cal R}_{\vb{k}, n\ell m} (\kappaS,\kappaB)
\equiv 
\kappaB^{3/2} \, \int \dd^3 r 
\ \Psi_{n\ell m}^* (\vb{r};~\kappaB) 
\ \Psi_{\vb{k}} (\vb{r};~\kappaS).
\label{eq:Rcal_def_app}
\end{align}
We replace \cref{eqs:CoulombWaveFunctions} into \cref{eq:Rcal_def_app} and perform the angular integral, which picks out $\ellS=\ell$ and $\mS=m$. Setting $t =2\xB/n=2\zetan \xS$, we obtain
\begin{align}
&{\cal R}_{\vb{k}, n\ell m} (\kappaS,\kappaB) =
Y_{\ell m}^* (\Omega_{\vb{k}}) 
\times
\frac{\im^\ell \pi}{[(2\ell+1)!]^2} 
\left[\frac{n^2(n+\ell)!}{(n-\ell-1)!}\right]^{1/2} 
\ e^{\pi\zetaS/2} 
\ \Gamma(1+\ell-\im\zetaS)
\ \zetan^{-\ell}  
\label{eq:Rcal_intermediate}
\\
&\times \int_0^\infty \dd t
\ t^{2\ell+2} 
\ \exp \left(-\dfrac{1+\im/\zetan}{2} t\right)
\ {}_1F_1 ( 1+\ell- n; ~ 2\ell+2; ~ t )
\ {}_1F_1 ( 1+\ell+ \im \zetaS; ~ 2\ell+2; ~ \im t/\zetan ) ,
\nn
\end{align}
where we used the lower sign convention for the scattering-state wavefunction \eqref{eq:psi_SS_PW}. 
The confluent hypergeometric functions ${}_1F_1$ obey the identity~\cite[section~7.622]{Integrals_GradshteynRyzhik}
\begin{align}
&\int_0^\infty \dd t \ t^{c-1} \ e^{-\rho t}
{}_1F_1 \left(a; ~ c; ~ t \right)
{}_1F_1 \left(b; ~ c; ~ \chi t \right) = 
\nn \\ 
&=
\Gamma (c) 
\ \dfrac{\rho^{a+b-c}}{(\rho-1)^a \ (\rho - \chi)^b }
\ {}_2F_1 \left[ a,b;c;~\dfrac{\chi}{(\rho-1) (\rho-\chi)} \right] 
\equiv g(\rho; a,b,c,\chi) , 
\label{eq:Identity} 
\end{align}
for ${\rm Re}(c) > 0$ and ${\rm Re}(\rho) > {\rm Re}(\chi) + 1$,  where ${}_2F_1$ is the ordinary hypergeometric function. Reference~\cite{Oncala:2019yvj} checked numerically that \cref{eq:Identity} remains true for $a$ a non-positive integer, ${\rm Re}(c) > 0$ and ${\rm Re}(\rho) > {\rm Re}(\chi)$, which encompasses the parameter range of interest (see also \cite{Ananthanarayan:2021bqz}).
Differentiating \eqref{eq:Identity} over $\rho$, and setting
\begin{subequations}
\label{eq:HypergeometricParameters}
\label[pluralequation]{eqs:HypergeometricParameters}
\begin{align}
a&=1+\ell-n ,\\
b&=1+\ell+\im \zetaS ,\\  
c&=2\ell+2 , \\
\rho &= 1/2+\im/(2\zetan) , \\   
\chi &= \im/\zetan ,
\end{align}
\end{subequations}
we obtain the integral needed to compute the second line of \cref{eq:Rcal_intermediate},
\begin{align}
-\frac{dg}{d\rho} 
&= 
\, \left(1-\frac{\zetaS}{\zetaB} \right) 
\, (-1)^{n-\ell-1}
\, 2^{2\ell+4} 
\, (2\ell+1)! \,
\, n
\label{eq:dgdrho_1} 
\\  
&\times 
\left(\dfrac{\zetan}{\zetan-\im}\right)^{2\ell+4}
\left(\dfrac{\zetan-\im}{\zetan+\im}\right)^{n+1-\im \zetaS}
{}_2F_1 \left( 
1+\ell-n, ~ 
1+\ell +\im \zetaS; ~
2\ell+2; ~
-\frac{4 \im \zetan}{(\zetan-\im)^2}
\right) .
\nn
\end{align}
Considering \cref{eq:dgdrho_1}, the overlap integral \eqref{eq:Rcal_intermediate} becomes  
\begin{align}
&{\cal R}_{\vb{k}, n\ell m} (\kappaS,\kappaB) =
\label{eq:Rcal_final_app}
\\
&=\im^\ell (-1)^{n-\ell-1} 
\left(1-\dfrac{\kappaS}{\kappaB}\right)
\, Y_{\ell m}^* (\Omega_{\vb{k}}) 
\times
\frac{2^{2\ell+4} \pi}{(2\ell+1)!} 
\ n^2
\left[\frac{(n+\ell)!}{(n-\ell-1)!}\right]^{1/2} 
\ e^{\pi\zetaS/2} 
\ \Gamma(1+\ell-\im\zetaS)
\nn \\
&\times 
\dfrac{\zetan^{\ell+4}}{(\zetan\mp\im)^{2\ell+4}}
\left(\dfrac{\zetan\mp\im}{\zetan\pm\im}\right)^{n+1\mp\im \zetaS}
{}_2F_1 \left( 
1+\ell-n, ~ 
1+\ell \pm \im \zetaS; ~
2\ell+2; ~
\mp\frac{4 \im \zetan}{(\zetan \mp \im)^2}
\right) ,
\nn
\end{align}
where the two different signs in the above are related via a Pfaff transformation, \cref{eq:2F1_Pfaff}.

\subsection{Bound-bound overlap integral}

One can similarly define the bound-to-bound overlap integral~\cite{Oncala:2021tkz},
\begin{align}\label{eq:RcalB2B_def}
{\cal R}_{n\ell m;~n'\ell'm'}^{}
(\kappaB, \kappaB')
\equiv
\int \dd^3r 
\ \Psi_{n'\ell' m'}^*(\vb{r};~\kappaB')
\ \Psi_{n\ell m}(\vb{r};~\kappaB)
=\delta_{\ell\ell'}\delta_{mm'}
\, {\cal R}_{\ell;nn'}^{}(\kappaB, \kappaB') .
\end{align}
We can use the previous result to calculate \cref{eq:RcalB2B_def} by noting that the bound-state and scattering-state wavefunctions are related by
\begin{align}\label{eq:Scattering_to_Bound}
\Psi_{n\ell m}(\vb{r};~\kappaB)
=
\dfrac{\kappaB^{3/2}}{2\pi n^2}
\left[\dfrac{(n+\ell)!}{(n-\ell-1)!}\right]^{1/2}
\lim_{\substack{k\to \im\kappaB/n \\ \kappaS\to\kappaB~~~}} 
\ \dfrac{(-1)^\ell e^{-\pi\zetaS/2}}{\Gamma(1+\ell-\im \zetaS)}
\int \dd \Omega_{\vb{k}} 
\, Y_{\ell m}(\Omega_{\vb{k}})
\, \Psi_{\vb{k}}(\vb{r};~\kappaS),
\end{align}
which implies
\begin{align}
&{\cal R}_{n\ell m;~n'\ell'm'}(\kappaB, \kappaB') =
\nn \\
&=\dfrac{1}{2\pi n^2}
\left(
\frac{\kappaB}{\kappaB'}
\right)^{3/2}
\left[\dfrac{(n+\ell)!}{(n-\ell-1)!}\right]^{\frac{1}{2}}
\lim_{\substack{k\to \im\kappaB/n \\ \kappaS\to\kappaB~~}} 
\ \dfrac{(-1)^\ell  e^{-\pi\zetaS/2}}{\Gamma(1+\ell-\im \zetaS)}
\int \dd \Omega_{\vb{k}} 
Y_{\ell m}(\Omega_{\vb{k}})
{\cal R}_{\vb{k}, n'\ell'm'} (\kappaS,\kappaB'). 
\label{eq:Rcal_BS-SS}   
\end{align}
From this, we obtain
\begin{align}
{\cal R}_{\ell;nn'} (\kappaB,\kappaB')
&=\dfrac{(-1)^{n - \ell - 1}}{(2\ell + 1)!}
\sqrt{
\dfrac{(n + \ell)!(n' + \ell)!}
{nn'(n - \ell-1)!(n' - \ell-1)!}
}
\dfrac{(\kappaB - \kappaB')}
{(\kappaB/n - \kappaB'/n')^{2\ell + 4}}
\left(\dfrac
{4\kappaB\kappaB'}{nn'}
\right)^{\ell + 3/2}
\label{eq:Rcal_nn'}
\\[1ex]
&\times
\left(\dfrac
{\kappaB/n - \kappaB'/n'}
{\kappaB/n + \kappaB'/n'}
\right)^{n + n' + 1} 
{}_2F_1
\left(
1 + \ell - n,
1 + \ell - n';
~2\ell + 2;
~-\dfrac
{4\kappaB\kappaB'/nn'}
{(\kappaB/n - \kappaB'/n')^2}
\right) .
\nn
\end{align}
This agrees with the result of Ref.~\cite[Appendix~C]{Oncala:2021tkz}. Note that 
${\cal R}_{\ell;nn'} (\kappaB,\kappaB')
={\cal R}_{\ell;n'n} (\kappaB',\kappaB)$, 
${\cal R}_{\ell;nn'} (\kappaB,\kappaB')$ is a function of $\kappaB/\kappaB'$ only,  
and $\lim_{\kappaB'\to\kappaB} {\cal R}_{\ell;nn'} (\kappaB,\kappaB')= \delta_{nn'}$.

\clearpage
\section{Identities and approximations \label{app:IdentitiesApprox}}

\subsection{The generalized hypergeometric functions}
The Hypergeometric functions can be expanded as power series as follows~\cite[15.2.1]{NIST:DLMF},
\begin{align}
{}_pF_q (a_1, \cdots, a_p;~~b_1, \cdots, b_q;~~x) =
\sum_{j=0}^\infty 
\dfrac{(a_1)_j \cdots (a_p)_j}{(b_1)_j \cdots (b_q)_j}
\dfrac{x^j}{j!} ,
\label{eq:HypergeometricSeries}
\end{align}
where $(c)_j$, with $j\in \NN_0$, is the rising Pochhammer symbol, with 
\begin{align}
(c)_0 &= 1, \nn \\[1ex]
(c)_j &= \begin{cases}
\dfrac{\Gamma(c+j)}{\Gamma(c)},
& c\neq-1,-2,\cdots,
\\[1.5em]
(-1)^j \, \dfrac{(|c|)!}{(|c|-j)!},
& c=-1,-2,\cdots
\quad\text{and} \quad
0\leqslant j \leqslant |c| ,
\\[0.8em]
0, 
& c=-1,-2,\cdots
\quad\text{and} \quad
|c| < j .
\end{cases}    
\end{align}
Thus, if any of the $a_i$ coefficients is a non-positive integer, the series \eqref{eq:HypergeometricSeries} terminates at $j=|a_i|$. 
The series \eqref{eq:HypergeometricSeries} converges for all $x$ if $p\leqslant q$, and for $|x|<1$ if $p=q+1$ (with convergence on $|x|=1$ depending on the parameters).
\\

\noindent
The following Kummer, Pfaff and quadratic transformations are useful~\cite[13.2.39, 15.8.1, 15.8.14]{NIST:DLMF}:
\begin{align}
{}_1F_1(a;~b;~z) &=
e^{z} \
{}_1F_1(b-a;~b;~-z),
\quad 
b\neq 0,-1,-2,\cdots,
\label{eq:1F1_Kummer}
\\[1ex]
{}_2F_1(a,b;~c;~z) &= 
(1-z)^{-a} 
\ {}_2F_1 \left(a,c-b;~c;~\dfrac{z}{z-1} \right) ,
\quad 
c\neq 0,-1,-2,\cdots, 
\label{eq:2F1_Pfaff}
\\[1ex]
{}_2F_1 (a,b;~2b;~z) &=
(1-z)^{-a/2} 
\ {}_2F_1 \left(
\dfrac{a}{2},~b-\dfrac{a}{2};~~b+\dfrac{1}{2};~~-\dfrac{z^2}{4(1-z)}
\right),
\quad 
b\neq 0,-\dfrac{1}{2},-1,\cdots,
\label{eq:2F1_QuadraticTransf}
\end{align}
where in \cref{eq:2F1_Pfaff,eq:2F1_QuadraticTransf}, a branch choice is needed if $a \notin \ZZ$. 
\\

\noindent
In addition, we will use Clausen's formula~\cite[Theorem 6.7]{Milla:2021dpc},
\begin{align}\label{eq:ClausensFormula}
\left[
{}_2F_1\left(
a, b; a + b + \frac12;
z
\right)
\right]^2
=
{}_3F_2
\left(
\begin{matrix}
2a,& 2b,& a + b\\
{} & 2(a + b),& a+b+\frac12
\end{matrix};
~~
z
\right),
\end{align}
Saalschütz's Theorem~\cite[16.4.3]{NIST:DLMF},
\begin{align}\label{eq:Saalschutz_Theorem}
{}_3F_2
\left(
\begin{matrix}
-n,& a,& b\\
{} & c,& 1+a+b-c-n
\end{matrix}
;
~~1
\right)
=
\dfrac{
(c-a)_n(c-b)_n
}{
(c)_n (c-a-b)_n
},
\qquad n = 0, 1, 2,\cdots
\end{align}
and Gauss's hypergeometric theorem~\cite[15.4.24]{NIST:DLMF},
\begin{align}\label{eq:Gauss_Theorem}
{}_2F_1(-n, b; c; 1) = \frac{(c - b)_n}{(c)_n},
\qquad n = 0, 1, 2, \cdots
.
\end{align}

\newpage

\subsection{The Stirling approximation for the Gamma function}

The Stirling approximation for the Gamma function reads
\begin{align}
\Gamma(z) \approx 
\sqrt{\dfrac{2\pi}{z}} 
\, \left(\dfrac{z}{e}\right)^z ,
\qquad |\arg(z)|<\pi .
\label{eq:Stirling_Gamma}
\end{align}

\bigskip

\noindent \emph{Real argument:} 
For $z,a,b \in \mathbb{R}$ with $|z| \gg |a|,|b|$, \cref{eq:Stirling_Gamma} implies~\cite[5.11.12]{NIST:DLMF}
\begin{align}
\dfrac{\Gamma(z+a)}{\Gamma(z+b)} 
\approx 
\left(\dfrac{z+b}{z+a}\right)^{1/2}
\dfrac
{[(z+a)/e]^{z+a}}
{[(z+b)/e]^{z+b}}     
\approx \left(\dfrac{z}{e}\right)^{a-b} \dfrac{(1+a/z)^z}{(1+b/z)^z}
\approx z^{a-b}.
\label{eq:Stirling_RatioOfGammas}
\end{align}
where we used the fact that 
$\lim_{z\to \infty} (1+a/z)^z = e^a$. 

\bigskip

\noindent\emph{Complex argument:} 
For $a,b\in \mathbb{R}$, \cref{eq:Stirling_Gamma} implies
\begin{align}
\Gamma (a + \im b)
&\approx \sqrt{2\pi} \, (a+\im b)^{a+\im b -1/2} \, e^{-a-\im b} 
\nn \\
&= \sqrt{2\pi} 
\, \left[\sqrt{a^2+b^2} \, e^{\im \atantwo(b,a)} \right]^{a-1/2+\im b}
\, e^{-a-\im b}
\nn \\
&= \sqrt{2\pi}
\, e^{-a-b\atantwo(b,a)}
\  (a^2+b^2)^{(2a-1)/4}
\times  
(a^2+b^2)^{\im b/2}
\, e^{-\im b}
\, e^{\im (a-1/2)\atantwo(b,a)},
\label{eq:Stirling_GammaComplex}
\end{align}
where $\atantwo$ is the quadrant-aware arctangent.
For $|b| \gg |a|$, the above simplifies to
\begin{align}
\Gamma(a+\im b)\approx
\sqrt{2\pi}
\,|b|^{a-1/2}
\,e^{-\pi|b|/2}
\,e^{\im [b\ln|b|-b+\sgn(b)(2a-1)\pi/4]}.
\label{eq:Stirling_Gamma_Im>Re}
\end{align}
which in turn implies that for $|b| \gg |a|, |a+j|$, with $a,b,j \in \RR$, 
\begin{align}
\dfrac{\Gamma(a+j+\im b)}{\Gamma(a+\im b)} 
\approx (\im b)^j .
\label{eq:Stirling_Gamma_Im>Re_Ratio}
\end{align}

\clearpage
\section{Asymptotic forms of Laguerre polynomials \label{app:Laguerre}}

We review the asymptotic forms of the Laguerre polynomials, $L_{n-\ell-1}^{(2\ell+1)} (4\lambda n)$, at large $n$ and for $\lambda \in \mathbb{R}$, as derived in Ref.~\cite{erdelyi:1960} (see also Refs.~\cite{Muckenhoupt:1970,Temme:1990} and references therein). 

We first define the functions $\beta(\lambda) \in \mathbb{R}$ for $\lambda < 1$,  and $\bar{\beta}(\lambda) \in \mathbb{R}$ for $\lambda > 0$, as follows
\begin{subequations}
\label{eq:betas_def}
\label[pluralequation]{eqs:taus_def}
\begin{align}
\beta(\lambda) 
&\equiv  
\dfrac{1}{2} \left\{
\begin{array}{ll}
\sqrt{1-\lambda} + \dfrac{{\arcsinh}{\sqrt{|\lambda|}}}{\sqrt{|\lambda|}},     
&  \lambda < 0 
\\[1em]
\sqrt{1-\lambda} + \dfrac{\arcsin{\sqrt{\lambda}}}{\sqrt{\lambda}},  
&  0< \lambda <1 
\end{array}
\right\}
\label{eq:beta_def}
= 1 - \dfrac{\lambda}{6} + \mathcal{O}\left(\lambda^2\right),
\end{align}
and
\begin{align}
\bar{\beta}(\lambda) 
&\equiv  
\dfrac{3}{2(\lambda-1)} 
\left\{
\begin{array}{ll}
\sqrt{\lambda} - \dfrac{\arccos \sqrt{\lambda}}{\sqrt{1-\lambda}},
& 0< \lambda < 1
\\[1em]
\sqrt{\lambda} - \dfrac{\arccosh \sqrt{\lambda}}{\sqrt{\lambda-1}},
& 1 < \lambda
\end{array}
\right \}
\label{eq:betabar_def}
= 1 - \dfrac{3}{10}(\lambda-1) + \mathcal{O}\left((\lambda-1)^2\right)  ,
\end{align}    
\end{subequations}
where we note that
\begin{align}
(1-\lambda)^{3/2} \bar{\beta}(\lambda) = 3\left[\pi/4-\sqrt{\lambda} \beta(\lambda)\right]
\quad \text{for} \quad
0<\lambda<1 .
\label{eq:A-B_relation_0<lambda<1}
\end{align}
We discern the following cases~\cite{erdelyi:1960,Muckenhoupt:1970}:

\bigskip

\noindent
$\bm{\lambda <1}$

\noindent
The asymptotic form of $L_{n-\ell-1}^{2\ell+1} (4\lambda n)$ can be expressed in terms of Bessel functions,
\begin{subequations}
\label{eq:Laguerre_Asymptotes_Bessel}
\label[pluralequation]{eqs:Laguerre_Asymptotes_Bessel}
\begin{align}
(4|\lambda|)^{\ell+1/2} \, e^{-2\lambda n} L_{n-\ell-1}^{(2\ell+1)} (4\lambda n) 
\approx \dfrac{\beta(\lambda)^{1/2}}{(1-\lambda)^{1/4}} 
\left\{
\begin{array}{ll}
I_{2\ell+1} \left(4n\sqrt{|\lambda|} \,\beta(\lambda)\right), 
& \lambda < 0 ,
\\[1ex]
J_{2\ell+1} \left(4n\sqrt{\lambda} \, \beta(\lambda)\right) ,
& 0< \lambda < 1 ,
\end{array}
\right.
\label{eq:Laguerre_Asymptotes_Bessel_original}
\end{align}
with $J_{2\ell+1}$ and $I_{2\ell+1}$ being the Bessel and modified Bessel functions of first kind. Using the asymptotic forms of the Bessel functions for large $n$, the above yields
\begin{align}
(4|\lambda|)^{\ell+1/2} \, e^{-2\lambda n} L_{n-\ell-1}^{(2\ell+1)} (4\lambda n) 
\approx 
\dfrac{1}{[|\lambda| (1-\lambda)]^{1/4} 
(2\pi n )^{1/2}}
\left\{
\begin{array}{ll}
\dfrac{1}{2} e^{4n\sqrt{|\lambda|} \beta(\lambda)}, 
& \lambda < 0 ,
\\[1ex]
(-1)^{\ell}
\sin \left[4n\sqrt{\lambda} \beta(\lambda) - \dfrac{\pi}{4} \right] ,
& 0< \lambda < 1 .
\end{array}
\right.
\label{eq:Laguerre_Asymptotes_Bessel_simplified}
\end{align}
\end{subequations}

\noindent
$\bm{\lambda>0}$

\noindent
The asymptotic form of $L_{n-\ell-1}^{2\ell+1} (4\lambda n)$ can be expressed in terms of the Airy function,
\begin{subequations}
\label{eq:Laguerre_Asymptotes_Airy}
\label[pluralequation]{eqs:Laguerre_Asymptotes_Airy}
\begin{align}
(4\lambda)^{\ell+1/2} \, e^{-2\lambda n} L_{n-\ell-1}^{(2\ell+1)} (4\lambda n) 
\approx 
\dfrac{(-1)^{n-\ell-1}}{(2n)^{1/3}} 
\, \dfrac{\bar{\beta}(\lambda)^{1/6}}{\lambda^{1/4}}
\, {\rm Ai} \left[(2n \, \bar{\beta}(\lambda))^{2/3} (\lambda-1)\right] .
\label{eq:Laguerre_Asymptotes_Airy_original}
\end{align}
Using the asymptotic forms of the Airy function for large $n$, the above yields
\begin{align}
(4\lambda)^{\ell+1/2} \, e^{-2\lambda n} L_{n-\ell-1}^{(2\ell+1)} (4\lambda n) 
\approx (-1)^{n-\ell-1}
\left\{
\begin{array}{ll}
\dfrac{\sin \left[\dfrac{4n}{3} (1-\lambda)^{3/2} \bar{\beta}(\lambda) + \dfrac{\pi}{4} \right]}
{[\lambda (1-\lambda)]^{1/4} (2\pi n )^{1/2}},
& 0< \lambda < 1  ,
\\[1em]
\dfrac{1}{(18 n)^{1/3} \Gamma(2/3)} ,
& \lambda =1,
\\[1.5em]
\dfrac{e^{-(4n/3) (\lambda-1)^{3/2} \bar{\beta}(\lambda)}}
{2 [\lambda (\lambda-1)]^{1/4} (2\pi n )^{1/2}} , 
& 1< \lambda .
\end{array}
\right.
\label{eq:Laguerre_Asymptotes_Airy_simplified}
\end{align}
\end{subequations}
Considering \cref{eq:A-B_relation_0<lambda<1}, the asymptotic forms \eqref{eq:Laguerre_Asymptotes_Bessel_simplified} and \eqref{eq:Laguerre_Asymptotes_Airy_simplified} are in agreement for $0<\lambda<1$. 

\clearpage
\section{Asymptotic forms of confluent hypergeometric function  \label{app:1F1}}

\subsection{Method of steepest descent}

We briefly review the steepest descent method for oscillatory integrals. Consider
\begin{align}
I(\zeta)=\int_{\mathcal C} g(z)\,e^{\im \zeta\,\phi(z)}\,\dd z,
\qquad \zeta\gg 1,
\label{eq:SD_general}
\end{align}
where $g$ and $\phi$ admit analytic continuation to a complex neighborhood of the contour ${\cal C}$. Stationary points $z_0$ satisfy $\phi'(z_0)=0$. It is convenient to write the exponential as $e^{\zeta F(z)}$ with $F(z)\equiv \im\phi(z)$. A steepest--descent path through $z_0$ is a curve along which $\Im F$ is constant and $\Re F$ decreases away from the saddle. Equivalently, since $F=\im\phi$, this amounts to $\Re\phi(z)$ constant and $\Im\phi(z)$ increasing away from $z_0$, so that $e^{\im\zeta\phi(z)}$ is exponentially damped away from the saddle.

Assuming that ${\cal C}$ can be deformed to pass through $z_0$ along a steepest--descent direction without crossing singularities or branch cuts, and that $F''(z_0)\neq 0$, we expand around the saddle,
$F(z)=F(z_0)+(1/2) F''(z_0)\,(z-z_0)^2+\cdots$. 
Choosing a local parameterization $z-z_0=s \, e^{\im\theta}$ with $s\in\RR$ such that
$\Re\qty(F''(z_0)e^{2\im\theta})<0$, 
the leading saddle contribution is
\begin{align}
I(\zeta)
\approx
g(z_0)\,e^{\zeta F(z_0)}\,
\sqrt{\frac{2\pi}{-\zeta\,F''(z_0)}} ,
\label{eq:SD_gaussian_general}
\end{align}
where the branch of the square root is fixed by the chosen steepest-descent direction (equivalently by $\theta$), and we must sum over all contributing saddles. For real saddles with $\phi''(z_0)> 0$ or $<0$, we may take $\theta=+\pi/4$ or $-\pi/4$, respectively, to ensure 
$\Re\qty(F''(z_0)e^{2\im\theta})<0$, 
and \cref{eq:SD_gaussian_general} reduces to
\begin{align}
I(\zeta)
\approx
g(z_0)\,e^{\im\zeta\phi(z_0)}
\,e^{\im\frac{\pi}{4}\sgn\qty(\phi''(z_0))}
\sqrt{\frac{2\pi}{\zeta\,|\phi''(z_0)|}} .
\label{eq:SD_gaussian_real}
\end{align}
When a stationary point coincides with an endpoint or is degenerate ($F''(z_0)=0$), the Gaussian approximation fails.

\subsection{Application to confluent hypergeometric function}

We are interested in studying the $\zetaB \gg 1+\ell$ behavior of
\begin{align}
{}_1F_1 (
1+\ell+\im \lambda \zetaB;
2\ell+2;
-\im 4\zetaB) . 
\label{eq:1F1_def}
\end{align}
The integral representation of the confluent hypergeometric is~\cite[9.211.1]{Integrals_GradshteynRyzhik}
\begin{align}
{}_1F_1 (a;~b;~z) = 
\dfrac{2^{1-b} \, e^{z/2} \, \Gamma(b)}
{\Gamma(a) \, \Gamma(b-a)}
\int_{-1}^1
(1+t)^{a-1} 
\, (1-t)^{b-a-1}
\, e^{z t/2}
\, \dd t,
\quad
\Re(b) > \Re (a) >0 .
\label{eq:1F1_IntegralRep}
\end{align}
With this, \eqref{eq:1F1_def} becomes
\begin{align}
{}_1F_1 (
1+\ell+\im \lambda \zetaB;
2\ell+2;
-\im 4\zetaB) 
\ = \ 
\dfrac{e^{-\im 2\zetaB}}{2^{2\ell+1}}
\dfrac{\Gamma(2\ell+2)}{|\Gamma(1+\ell+\im \lambda\zetaB)|^2}
\times
I_\ell (\lambda,\zetaB).
\label{eq:1F1_factors}
\end{align}
Here
\begin{align}
I_\ell (\lambda,\zetaB) \equiv
\int_{-1}^1 
e^{\im \zetaB \phi_\lambda(t)}
\, (1-t^2)^{\ell}
\, \dd t 
=
\int_0^1 
\qty(
e^{+\im \zetaB \phi_\lambda(t)} +
e^{-\im \zetaB \phi_\lambda(t)} )
\, (1-t^2)^{\ell}
\, \dd t ,
\label{eq:1F1_Integral}
\end{align}
with
\begin{align}
\phi_\lambda (t) \equiv  
\lambda \ln \dfrac{1+t}{1-t} - 2t ,
\qquad
\phi_\lambda' (t) =
\dfrac{2\lambda}{1-t^2} -2,
\qquad
\phi_\lambda'' (t) =
\dfrac{4\lambda t}{(1-t^2)^2} .
\label{eq:1F1_phase}
\end{align}
In \cref{eq:1F1_Integral}, we took into account that $\phi_\lambda(-t)=-\phi_\lambda(t)$ and symmetrized the integrand.
The phase $\phi_\lambda(t)$ has two stationary points,
\begin{align}
t_\pm = \pm\sqrt{1-\lambda}.
\label{eq:1F1_StationaryPoints}
\end{align}
We discern the cases $0<\lambda<1$, $\lambda=1$, $\lambda>1$ and $\lambda<0$.

\subsection*{$\bm{0 <\lambda < 1}$}
The stationary points are real and lie within the interval of integration, $t_\pm \in [-1,1]$; restricting the integration to $t \in [0,1]$ due to symmetry, it suffices to consider only $t_+$. Define
\begin{align}
\phi_\lambda^s \equiv 
\phi_\lambda (t_+) = 
\lambda \ln 
\dfrac{1+\sqrt{1-\lambda}}{1-\sqrt{1-\lambda}}
-2\sqrt{1-\lambda}  ,
\qquad
\phi_\lambda' (t_+) = 0,
\qquad
\phi_\lambda '' (t_+) = 
\dfrac{4\sqrt{1-\lambda}}{\lambda} .
\end{align}
Then, the Gaussian saddle-point approximation \eqref{eq:SD_gaussian_real} yields 
\begin{align}
I_\ell(\lambda,\zetaB) 
&\approx 
(1-t_+^2)^{\ell}
\, e^{\im \zetaB \phi_\lambda^s}
\, e^{\im (\pi/4) \sgn (\phi_\lambda''(t_+))}
\sqrt{\dfrac{2\pi}{\zetaB|\phi_\lambda '' (t_+)|}}
+ \cc 
\nn \\
&=
\dfrac{\lambda^{\ell+1/2}}{(1-\lambda)^{1/4}}
\left(\dfrac{2\pi}{\zetaB} \right)^{1/2}
\ \cos\left( 
\zetaB \, \phi_\lambda^s
+\dfrac{\pi}{4}
\right) .
\label{eq:Iell_0<lambda<1}
\end{align}

\subsection*{$\bm{\lambda = 1}$}
At $\lambda=1$, the stationary points \eqref{eq:1F1_StationaryPoints} coalesce to $t_\pm=0$. In this case the usual Gaussian saddle-point approximation does not apply, since
$\phi_{\lambda=1}(t_\pm)=\phi_{\lambda=1}'(t_\pm)=0=\phi_{\lambda=1}''(t_\pm)=0$, while $\phi_{\lambda=1}'''(t_\pm)\neq 0$. Expanding around zero,
\begin{align}
\phi_{\lambda=1}(t)
= \dfrac{2}{3} t^3+\mathcal{O}(t^5). 
\label{eq:1F1_phase_lambda1_expand}
\end{align}
With this, the integral \eqref{eq:1F1_Integral} can be approximated as
\begin{align}
I_\ell (1,\zetaB) 
&\approx
\int_{-1}^1 
e^{+\im (2\zetaB/3) t^3}
\, (1-t^2)^{\ell}
\, \dd t    
\approx
\int_{-\infty}^\infty 
e^{+\im (2\zetaB/3) t^3} 
\, \dd t   
=
\dfrac{\Gamma(1/3)}{12^{1/6}}
\, \zetaB^{-1/3} .
\label{eq:Iell_lambda=1}
\end{align}
Here we dropped the ${\cal O}(t^2)$ corrections in $(1-t^2)^\ell$ and extended the integration to $t\in(-\infty,\infty)$.
Since in the dominant region $t\sim \zetaB^{-1/3}$, the neglected terms give relative corrections of order ${\cal O}(\zetaB^{-2/3})$,
while the contribution from $|t|\gg \zetaB^{-1/3}$ is suppressed due to rapid oscillations. The final result was obtained by rescaling the
integrand, and using 
\begin{align}
\int_{-\infty}^{\infty} e^{\im u^3}\,\dd u
= \dfrac{\sqrt{3}}{3} 
\Gamma\left(\dfrac{1}{3}\right) .
\label{eq:cubic_integral}
\end{align}

\subsection*{$\bm{\lambda > 1~~\text{or}~~\lambda < 0}$}

In the regimes $\lambda>1$ and $\lambda<0$, the stationary points \eqref{eq:1F1_StationaryPoints} lie off the integration contour.
Moreover, the logarithm in \eqref{eq:1F1_phase} produces branch points at $t=\pm 1$, so a direct contour deformation in the $t$-plane
requires keeping track of the associated cuts. A convenient way to proceed is to remove these endpoints by the change of variables
\begin{align}
t=\tanh (u),
\qquad 
\ln\frac{1+t}{1-t}=2u,
\qquad 
1-t^2=\sech^2 (u),
\qquad 
\dd t=\sech^2 (u)\,\dd u,
\label{eq:t_tanh_u}
\end{align}
which maps $t\in(-1,1)$ to $u\in(-\infty,\infty)$. With this, the integral \eqref{eq:1F1_Integral} becomes
\begin{align}
I_\ell(\lambda,\zetaB)
=
\int_{-\infty}^{\infty}
\sech^{2\ell+2}(u)
\, e^{\im\,2\zetaB\,\varphi_\lambda(u)}\,
\dd u,
\label{eq:Iell_u}
\end{align}
where
\begin{align}
\varphi_\lambda(u)\equiv \lambda u-\tanh (u),
\qquad
\varphi_\lambda'(u)=\lambda-\sech^2 (u),
\qquad
\varphi_\lambda''(u)=2\sech^2(u)\,\tanh (u).
\label{eq:varphi_def}
\end{align}
The integrand in \eqref{eq:Iell_u} is analytic in the open region $|{\Im u}|<\pi/2$ and decays as 
$\sech^{2\ell+2}(u) \sim e^{-2(\ell+1)|u|}$
for $\Re u\to\pm\infty$. Hence the contour can be deformed within this region without boundary contributions from infinity.
The nearest singularities (poles of $\tanh u$ and $\sech u$) are located at $u=\im(\pi/2+\pi n)$ with $n \in \ZZ$. We shall now treat the two intervals, $\lambda > 1$ and $\lambda<0$, separately. 

\paragraph{$\lambda>1$:}
The saddles solve $\varphi_\lambda'\qty(u_s)=0$, i.e.\ $\sech^2 u_s=\lambda$. For $\lambda>1$ the closest saddles lie on
the imaginary axis,
\begin{align}
u_s=\pm \im v_s,
\qquad
\cos\qty(v_s)=\frac{1}{\sqrt{\lambda}},
\qquad
v_s=\arccos\qty(1/\sqrt{\lambda})
\ \in\ (0,\pi/2).
\label{eq:lambda>1_saddle}
\end{align}
Using $\tanh\qty(\im v_s)=\im\tan\qty(v_s)=\im\sqrt{\lambda-1}$ and $\sech\qty(\im v_s)=\sec\qty(v_s)=\sqrt{\lambda}$, we find that for the saddle point $u_s=\im v_s$,
\begin{subequations}
\begin{align}
\varphi_\lambda\qty(\im v_s)
&=\im\qty(\lambda v_s-\sqrt{\lambda-1})
\equiv \im\,A(\lambda),
\qquad
A(\lambda)>0,
\label{eq:A_lambda_def}
\\
\varphi_\lambda'\qty(\im v_s)
&=0,
\\
\varphi_\lambda''\qty(\im v_s)
&=\im\,2\lambda\sqrt{\lambda-1}\, .
\end{align}
\end{subequations}
Hence $\exp\qty(\im 2\zetaB\,\varphi_\lambda\qty(\im v_s))=\exp\qty(-2\zetaB A(\lambda))$. Deforming the contour within the analytic strip $|{\Im u}|<\pi/2$ to pass through this saddle point, and using \cref{eq:SD_gaussian_general}, we obtain the contribution
\begin{align}
I_\ell\qty(\lambda,\zetaB)
\approx
\lambda^{\ell+\frac12}
\sqrt{\frac{\pi}{2\zetaB\sqrt{\lambda-1}}}\;
\exp\qty[-2\zetaB\qty(\lambda\arccos\qty(1/\sqrt{\lambda})-\sqrt{\lambda-1})].
\label{eq:Iell_lambda>1}
\end{align}
On the other hand, the saddle at $u_s=-\im v_s$ would give a growing contribution for all $\zetaB >0$, proportional to $\exp\qty(+2\zetaB A(\lambda))$. 
Since on the original contour $u\in\RR$, 
$|e^{\im 2\zetaB\varphi_\lambda(u)}|=1$, therefore
$|I_\ell (\lambda,\zetaB)| \leqslant 
\int_{-\infty}^{\infty}
\sech^{2\ell+2}(u) \, \dd u<\infty$ 
for all $\zetaB$, we deduce that the integration contour cannot be deformed into a path of steepest descent passing through the saddle at $u_s=-\im v_s$ without encountering singularities. \Cref{eq:Iell_lambda>1} thus remains the sole contribution.

\paragraph{$\lambda<0$:}
The saddles satisfy $\sech^2 (u_s)=\lambda<0$, thus lie on the lines $\Im\qty(u)=\pm\pi/2$. We parametrize the lower
ones as
\begin{align}
u_s=\pm x_s-\im \pi/2,
\qquad
\sinh x_s=\frac{1}{\sqrt{|\lambda|}},
\qquad
x_s=\operatorname{asinh}\qty(1/\sqrt{|\lambda|}) ,
\label{eq:Saddles_lambda<0}
\end{align}
with
\begin{subequations}
\label{eq:Saddles_lambda<0_Phase}
\label[pluralequation]{eqs:Saddles_lambda<0_Phase}
\begin{align}
\varphi_{\lambda}\qty(\pm x_s-\im \pi/2)
&=\mp\qty(|\lambda| x_s+\sqrt{1+|\lambda|})+\im\,|\lambda|\,\pi/2,
\\
\varphi_{\lambda}'\qty(\pm x_s-\im \pi/2)
&=0,
\\
\varphi_{\lambda}''\qty(\pm x_s-\im \pi/2)
&=\mp 2|\lambda|\sqrt{1+|\lambda|}\, .
\end{align}
\end{subequations}
Here $\Im\varphi_\lambda\qty(u_s)=+|\lambda|\pi/2$, so $\exp\qty(\im 2\zetaB\,\varphi_\lambda\qty(u_s))\sim e^{-\pi|\lambda|\zetaB}$.
The upper saddles at $u=\pm x_s+\im\pi/2$ have $\Im\varphi_\lambda=-|\lambda|\pi/2$ and would yield
$\sim e^{+\pi|\lambda|\zetaB}$; as above, this is incompatible with the finiteness of $|I_\ell (\lambda,\zetaB)|$ deduced from integration along the real axis. We thus deform the contour within $|{\Im u}|\leqslant\pi/2$ to pass through the lower saddles while avoiding the pole at $u=-\im\pi/2$ (at $\Re u=0$). Summing the Gaussian contributions according to \cref{eq:SD_gaussian_general}, gives
\begin{align}
I_\ell\qty(\lambda,\zetaB)
\approx
(-1)^{\ell+1}
|\lambda|^{\ell+\frac12}
\sqrt{\frac{2\pi}{\zetaB\sqrt{1+|\lambda|}}}\;
e^{-\pi|\lambda|\zetaB}\;
\cos\qty(
2\zetaB\qty[|\lambda| x_s+\sqrt{1+|\lambda|}]
+\pi/4).
\label{eq:Iell_lambda<0}
\end{align}

\subsection*{Summary}

Using $|\Gamma(1+\ell + \im \zetaS)|^2 = (\ell!)^2 e^{-\pi\zetaS} S_\ell(\zetaS)$ with $\zetaS=\lambda\zetaB$, where $S_\ell$ is the $\ell$-wave Sommerfeld factor, we find
\begin{align}
\qty[\dfrac{2^{2\ell+1} \, \ell!}{(2\ell+1)!}]^2
e^{-2\pi \zetaS}
\, S_\ell (\zetaS)
~~ 
|{}_1F_1 (
1+\ell+\im \lambda \zetaB;
2\ell+2;
-\im 4\zetaB)|^2
=
\dfrac{|I_\ell(\lambda,\zetaB)|^2}
{(\ell!)^2 \, S_\ell(\zetaS)}.
\label{eq:combo_Iell_identity}
\end{align}
Collecting \cref{eq:Iell_0<lambda<1,eq:Iell_lambda=1,eq:Iell_lambda>1,eq:Iell_lambda<0}, and averaging oscillatory factors ($\cos^2\to 1/2$) when present, we obtain
\begin{multline}
\qty[\dfrac{2^{2\ell+1} \, \ell!}{(2\ell+1)!}]^2
e^{-2\pi \zetaS}
\, S_\ell (\zetaS)
~~|
{}_1F_1 (
1+\ell+\im \lambda \zetaB;
2\ell+2;
-\im 4\zetaB)
|^2
\approx
\\[1em] 
\approx
\frac{1}{(\ell!)^2\,S_\ell(\zetaS)}
\times
\begin{cases}
\dfrac{\pi\,|\lambda|^{2\ell+1}}{\zetaB\,\sqrt{1+|\lambda|}}\;
e^{-2\pi|\lambda|\zetaB},
& \lambda <0,
\\[1.1em]
\dfrac{\pi \, \lambda^{2\ell+1}}
{\sqrt{1-\lambda}}
\, \zetaB^{-1},
& 0< \lambda < 1,
\\[1.1em]
\dfrac{\Gamma(1/3)^2}{12^{1/3}}
\, \zetaB^{-2/3},
& \lambda = 1,
\\[1.1em]
\dfrac{\pi\,\lambda^{2\ell+1}}{2\zetaB\,\sqrt{\lambda-1}}\;
\exp\qty[-4\zetaB 
\qty(\lambda \arccos(1/\sqrt{\lambda})-\sqrt{\lambda-1})],
& \lambda > 1.
\end{cases}
\label{eq:1F1_summary}
\end{multline}
For $|\zetaS| \gtrsim 1+\ell$, we may set 
$(\ell!)^2 S_\ell(\zetaS) 
\simeq 2\pi |\zetaS|^{2\ell+1} 
e^{2\pi \min(\zetaS,0)} $ 
as in \cref{eq:SommerfeldFactor_LargeZetaS}. Doing so, \cref{eq:1F1_summary} simplifies to
\begin{multline}
\qty[\dfrac{2^{2\ell+1} \, \ell!}{(2\ell+1)!}]^2
e^{-2\pi \zetaS}
\, S_\ell (\zetaS)
~~|
{}_1F_1 (
1+\ell+\im \lambda \zetaB;
2\ell+2;
-\im 4\zetaB)
|^2
\approx
\\[1em]
\approx
\frac{1}{2\,\zetaB^{2\ell+2}}
\times
\begin{cases}
\dfrac{1}{\sqrt{1+|\lambda|}},
& \lambda < 0,
\\[1em]
\dfrac{1}{\sqrt{1-\lambda}},
& 0< \lambda < 1,
\\[1em]
\dfrac{\Gamma(1/3)^2}{12^{1/3}\,\pi}
\ \zetaB^{1/3},
& \lambda = 1,
\\[1em]
\dfrac{
\exp\qty[-4\zetaB 
\qty(\lambda \arccos(1/\sqrt{\lambda})-\sqrt{\lambda-1})]
}{2\sqrt{\lambda-1}},
& \lambda > 1.
\end{cases}
\label{eq:1F1_summary_largeZetaS}
\end{multline}
%

\clearpage

\section{Evaluation of diagonal $\WW_\ell$ elements for $\alpha_{\cal S}=0$ 
\label{app:NonAnalyticStructure_alphaS=0}}

The principal goal of this section is to evaluate the following integral that determines the diagonal elements of the $\WW_\ell$ matrix (cf.~\cref{eq:Wnn_Clausen,eq:Wnn_Inell}),
\begin{align}\label{eq:Clausen_Integral}
{\cal I}_{n\ell}(u)
\equiv
\PV
\int_0^{\infty} 
\dd t
&\ \dfrac{t^{2\ell+6}}{t^2-u}
\ \dfrac{1}{(1+t^2)^{2\ell+4}}
 \nn\\ 
&\times 
{}_3F_2
\left(
1+\ell - n, 1+\ell + n, \ell + 1;
\ell + \frac32, 2\ell + 2;
\left(\dfrac{2t}{1+t^2}\right)^2
\right)
.
\end{align}
In particular, we will demonstrate that the quantity,
\begin{align}\label{eq:PolyDef}
{\cal P}_{n\ell}(u) \equiv
\frac{1}{{\cal N}_{n\ell}}
(1+u)^{2n + 2}\ 	{\cal I}_{n\ell}(u)
\end{align}
is a polynomial of order $2n + 1$. ${\cal N}_{n\ell}$ is a normalization factor which we define as
\begin{align}
{\cal N}_{n\ell}
\equiv 
\frac{\pi}{2^{4\ell+5}}\frac{\Gamma(n-\ell)}{n\Gamma(n+\ell + 1)}
\left[\frac{\Gamma(2\ell+2)}{\Gamma(\ell+1)}\right]^2
.
\end{align}
In general,
\begin{align}
{\cal P}_{n\ell}(u)
=
\sum_{j = 0}^{\infty} C_{n\ell; j}\ u^j
\end{align}
such that the coefficients $C_{n\ell;j}$ are determined by
\begin{align}\label{eq:Coeff_Derv_Def}
C_{n\ell;j} 
\equiv 
\eval{
\frac{1}{j!}\dv[j]{u}\ {\cal P}_{n\ell}(u)
}_{u = 0}
.
\end{align}
From \cref{eq:Clausen_Integral,eq:PolyDef,eq:Coeff_Derv_Def} one can show
\begin{align}
C_{n\ell;j}
=
\frac{1}{{\cal N}_{n\ell}}
\sum_{i = 0}^{j}{2n + 2 \choose j - i}
\int_0^\infty \frac{\dd t\ t^{2(\ell + 2 - i)}}{(1 + t^2)^{2\ell + 4}}{}_3F_2
\left(
\begin{matrix}
1+\ell - n,& 1+\ell + n,&  \ell + 1\\
{} & \ell + \frac32,& 2\ell + 2
\end{matrix}
; \left(\dfrac{2t}{1+t^2}\right)^2
\right)
.
\end{align}
Expanding the hypergeometric function via \cref{eq:HypergeometricSeries}, integrating term-by-term, then resumming, we find
\begin{empheq}[box=\myshadebox]{align}
\label{eq:Coeffs_5F4}
C_{n\ell;j}	
=
\dfrac{1}{{\cal N}_{n\ell}}
\sum_{i = 0}^{j}
&
\ \binom{2n+2}{j-i}
\ \dfrac
{\Gamma(\ell  + i + \frac32)\Gamma(\ell - i + \frac52)}
{2\Gamma(2\ell + 4)}\nn 
\\[1ex]
&\times
{}_5F_4
\left(
\begin{matrix}
	1 + \ell-n, & 1 + \ell + n, & \ell + 1, &  \ell+ i + \frac32, & \ell  - i + \frac52\\
	{} & 2\ell+ 2, & \ell + 2, & \ell + \frac32, & \ell + \frac52
\end{matrix}; 1
\right)
.
\end{empheq}
The above result fully determines the coefficients of ${\cal P}_{n\ell}(u)$ and is often the most convenient expression for generating these values. On the other hand, we can expand ${}_5F_4(\cdots; 1)$ using \cref{eq:HypergeometricSeries} and sum over $i$ yielding
\begin{align}\label{eq:Coeffs_3F2}
C_{n\ell;j}
=
n\ 
\binom{2n+2}{j}
\sum_{m=0}^{\,n-\ell-1}
\frac{
(-1)^m\ 
\Gamma(1+\ell + n + m)\ 
}{
(1 + \ell + m)
\Gamma(n-\ell-m)
\Gamma(2\ell+2 + m)\,
\Gamma(m + 1)
}\times \nn \\[1ex]
\times\ 
{}_3F_2\!\left(
\begin{matrix}
-j,& \ell+m+\frac32,& 1\\[1ex]
{} & 2n-j+3,& -\ell-m-\frac32
\end{matrix}
;1
\right)    
.
\end{align}
From here it becomes immediately apparent that the coefficients vanish for both $j < 0$ and $j > 2n + 2$. For $j = 2n + 2$, we note that the hypergeometric function in \cref{eq:Coeffs_3F2} reduces to,
\begin{align}
{}_2F_1\!\left(
\begin{matrix}
	-(2n+2),\ \ell+m+\tfrac32\\[2pt]
	-\ell-m-\tfrac32
\end{matrix}
;1
\right)
=
\frac{(-2\ell - 2m - 3)_{2n + 2}}{(-\ell-m-\frac32)_{2n + 2}}
=
0
\end{align}
where in the first equality we used Gauss's summation theorem,~\cref{eq:Gauss_Theorem}, and the second equality is a consequence of the definition of the Pochhammer symbol and the fact that the sum is over $m\in [0, n - \ell - 1]$. Having demonstrated that ${\cal P}_{n\ell}(u)$ is a polynomial of order $2n+1$, we will now examine specific coefficients which allow us to examine the limiting behavior of ${\cal I}_{n\ell}(u)$. 

\subsection*{Limiting behavior: $u\to 0$}

Starting from \cref{eq:Coeffs_5F4} we find
\begin{align}
C_{n\ell; 0}
=
\frac{1}{{\cal N}_{n\ell}}
\frac{\Gamma(\ell + \frac32)\Gamma(\ell + \frac52)}{2\Gamma(2\ell + 4)}
{}_3F_2
\left(
\begin{matrix}
	1 + \ell-n, & 1 + \ell + n, & \ell + 1\\
	{} & 2\ell+ 2, & \ell + 2, 
\end{matrix}; 1
\right)
.
\end{align}
Using \cref{eq:Saalschutz_Theorem},
\begin{align}
{}_3F_2
\left(
\begin{matrix}
	1 + \ell-n, & 1 + \ell + n, & \ell + 1\\
	{} & 2\ell+ 2, & \ell + 2, 
\end{matrix}; 1
\right) = \frac{\Gamma(2 \ell + 3) \Gamma(n - \ell)}{2 n \Gamma(n + \ell + 1)},
\end{align}
from which one obtains
\begin{align}
C_{n\ell; 0} = 1
.
\end{align}
In a similar fashion, it is possible to show that
\begin{align}
C_{n\ell;1}
&=
2n + 3
.
\end{align}
It follows then that,
\begin{subequations}
\begin{align}
{\cal P}_{n\ell}(u\to 0) &\approx 1 + (2n + 3) u + \order*{u^2},\\[1ex]
{\cal I}_{n\ell}(u\to 0) &\approx {\cal N}_{n\ell} (1 + u) + \order*{u^2}
.
\end{align}    
\end{subequations}

\subsection*{Limiting behavior: $u\to \infty$}

In order to understand the large $u$ behavior, we examine the ($2n+1$)-th coefficient. To do so, it is more advantageous to start from \cref{eq:Coeffs_3F2}. Within the sum, note that~\cite[7.4.4.29]{Prudnikov:1990int}
\begin{align}
{}_3F_2\!\left(
\begin{matrix}
-(2n + 1),& \ell+m+\frac32,& 1\\[1ex]
{} & 2,& -\ell-m-\frac32
\end{matrix}
;1
\right)
&=
-\frac{2\ell + 2m + 5}{(2n + 2)(2\ell + 2m + 1)}
.
\end{align}
Inserting this result into \cref{eq:Coeffs_3F2} with $j = 2n + 1$ gives
\begin{align}
C_{n\ell;2n + 1}
=
-n
\sum_{m=0}^{\,n-\ell-1}
\frac{
(-1)^m\ 
\Gamma(1+\ell + n + m)\ 
}{
\Gamma(n-\ell-m)
\Gamma(2\ell+2 + m)\,
\Gamma(m + 1)
}
\times
\frac{2\ell + 2m + 5}{(1 + \ell + m)(2\ell + 2m + 1)}
.
\end{align}
We can then use partial fraction decomposition to write,
\begin{align}
\frac{2\ell + 2m + 5}{(1 + \ell + m)(2\ell + 2m + 1)}
=
-\frac{3}{1 + \ell + m} + \frac{8}{2 \ell + 2 m + 1}
.
\end{align}
We can now sum over $m$ to find,
\begin{align}
C_{n\ell;2n + 1}
=
-
\frac{8n - 6\ell -3}{2\ell + 1},
\end{align}
where we use the fact that,
\begin{align}
{}_3F_2
\left(
\begin{matrix}
1 + \ell - n,& 1 + \ell + n,& \ell + \frac12\\[1ex]
{} & \ell + \frac32& 2\ell + 2
\end{matrix}
;1
\right)
=
\frac{\Gamma(2\ell + 2)\Gamma(n - \ell)}{\Gamma(n + \ell + 1)},
\end{align}
by \cref{eq:Saalschutz_Theorem}. We can therefore conclude that,
\begin{subequations}
\begin{align}
{\cal P}_{n\ell}(u\to\infty) 
&\approx -\frac{8n-6\ell-3}{2\ell + 1}\ u^{2n + 1},
\\[1ex]
{\cal I}_{n\ell}(u\to\infty) 
&\approx -
 \frac{{\cal N}_{n\ell}}{u}
\left(
\frac{8n-6\ell-3}{2\ell + 1}
\right)
.
\end{align}
\end{subequations}
%

\clearpage
\addcontentsline{toc}{section}{References}
\bibliography{Bibliography}

\end{document}